\documentclass{achemso}

\usepackage{graphicx,xcolor,amsmath,bm,epstopdf,hyperref}
\usepackage[T1]{fontenc} 
\usepackage[export]{adjustbox}
\usepackage[ignoreunlbld,norefs,nocites]{refcheck}

\DeclareGraphicsExtensions{.pdf,.eps,.png,.jpg,.mps}

\mciteErrorOnUnknownfalse

\newcommand{\bcite}[1]{$^{[\text{\nocite{#1}\citenum{#1}}]}$}

\newcommand{\fig}{Figure~}
\newcommand{\eq}{Equation~}

\newcommand{\tab}{Table~}

\author{Marcin Szyniszewski} \affiliation{Department of Physics,
  Lancaster University, Lancaster LA1 4YB, United Kingdom}
\alsoaffiliation{Department of Physics and Astronomy, University
  College London, London WC1E 6BT, United Kingdom}

\author{Elaheh Mostaani} \affiliation{Department of Physics, Lancaster
  University, Lancaster LA1 4YB, United Kingdom}
\alsoaffiliation{Cambridge Graphene Centre, University of Cambridge, 9
  J.\ J.\ Thomson Avenue, Cambridge CB3 0FA, United Kingdom}

\author{Angelika Knothe} \affiliation{National Graphene Institute,
  University of Manchester, Booth Street East, Manchester M13 9PL,
  United Kingdom} \alsoaffiliation{Institut f\"{u}r Theoretische
  Physik, Universit\"{a}t Regensburg, D-93040 Regensburg, Germany}

\author{Vladimir Enaldiev} \affiliation{National Graphene Institute,
  University of Manchester, Booth Street East, Manchester M13 9PL,
  United Kingdom}

\author{Andrea C.\ Ferrari} \affiliation{Cambridge Graphene Centre,
  University of Cambridge, 9 J.\ J.\ Thomson Avenue, Cambridge CB3
  0FA, United Kingdom}

\author{Vladimir I.\ Fal'ko} \affiliation{National Graphene Institute,
  University of Manchester, Booth Street East, Manchester M13 9PL,
  United Kingdom}

\author{Neil D.\ Drummond} \affiliation{Department of Physics,
  Lancaster University, Lancaster LA1 4YB, United Kingdom}
\email{n.drummond@lancaster.ac.uk}

\title{Adhesion and Reconstruction of Graphene/Hexagonal Boron Nitride
  Heterostructures: A Quantum Monte Carlo Study}

\begin{document}

\begin{abstract}
We investigate interlayer adhesion and relaxation at interfaces
between graphene and hexagonal boron nitride (hBN) monolayers in van
der Waals heterostructures. The adhesion potential between graphene
and hBN is calculated as a function of local lattice offset using
diffusion quantum Monte Carlo methods, which provide an accurate
treatment of van der Waals interactions. Combining the adhesion
potential with elasticity theory, we determine the relaxed structures
of graphene and hBN layers at interfaces, finding no metastable
structures.  The adhesion potential is well described by simple
Lennard-Jones pair potentials that we parameterize using our quantum
Monte Carlo data.  Encapsulation of graphene between near-aligned
crystals of hBN gives rise to a moir\'{e} pattern whose period is
determined by the misalignment angle between the hBN crystals
superimposed over the moir\'{e} superlattice previously studied in
graphene on an hBN substrate. We model minibands in such
supermoir\'{e} superlattices and find them to be sensitive to a
180$^\circ$ rotation of one of the encapsulating hBN crystals. We find
that monolayer and bilayer graphene placed on a bulk hBN substrate and
bulk hBN/graphene/bulk hBN systems do not relax to adopt a common
lattice constant.  The energetic balance is much closer for
free-standing monolayer graphene/hBN bilayers and hBN/graphene/hBN
trilayers.  The layers in an alternating stack of graphene and hBN are
predicted to strain to adopt a common lattice constant and hence we
obtain a new, stable three-dimensional crystal with a unique
electronic structure.
\end{abstract}



Layered material heterostructures (LMHs) comprise vertically stacked
layered materials (LMs), held together by weak interlayer van der
Waals (vdW) forces.\cite{BONACCORSO2012,Geim2013,ACF2015, Backes_2020}
For example, a LMH can be made by mechanically transferring monolayer
graphene (1L-G) onto an atomically flat surface of a single-crystal
hexagonal boron nitride (hBN) substrate.\cite{Dean2010} Monolayers
(1L) of hBN have the same honeycomb structure as 1L-G, and are
insulators free of dangling bonds.  Hence they are one of the most
preferred substrates for graphene (G) devices because they largely
preserve 1L-G's electronic properties.\cite{Dean2010,Molitor2011} In
1L-G/1L-hBN heterostructures with near-aligned lattice vectors,
quasiperiodic hexagonal moir\'{e} patterns\cite{Xue2011, Sachs2011,
  Yankowitz2012} with periods of up to 140 {\AA} due to the small
mismatch $\delta=1-a_\text{G}/a_\text{hBN}\sim1.68$\% between the
hexagonal lattice constants $a_\text{hBN}=2.504$ {\AA} of
hBN\cite{Bhimanapati2016} and $a_\text{G}=2.462$ {\AA} of
1L-G\cite{Kelly1981, Dresselhaus1988} lead to peculiar low-energy
properties of electrons and holes. These moir\'{e} superlattice (SL)
effects in 1L-G/1L-hBN are the result of an interplay between weak
hybridization of electronic states near the Brillouin-zone (BZ)
corners in 1L-G with 1L-hBN orbitals and lattice relaxation of 1L-G,
which deforms locally to reduce its adhesion potential to
1L-hBN\@.\cite{Yankowitz2012}

The encapsulation of G by bulk hBN (B-hBN) has become a widely used
experimental technique. However, even when the lattice vectors of G
and hBN are aligned, a diverse range of atomic configurations may
arise in heterostructures of these materials, affecting their
electronic properties. The impact of misalignment between layers on
their electronic properties poses further challenges for experimental
analysis. To address these issues and gain a deeper understanding of
structural behavior at G/hBN interfaces, we have undertaken a study
involving the combination of aligned 1L-G with 1L-hBN\@. Furthermore,
our research contributes to overcoming the challenges in the field of
graphene twistronics in hBN/G/hBN multilayered structures, an area of
ongoing exploration in the scientific
literature.\cite{Wang2019,finney_tunable_2019,Wang2019a,PhysRevB.101.224107,
Yang_2020,Andelkovic_2020,Sun_2021,Moulsdale_2022,YWang2022,Chen_2023,Hu_2023}
In particular, we describe multiscale modeling of moir\'{e} SLs in
1L-G/B-hBN and in 1L-G encapsulated between two aligned B-hBN
crystals, describing the electronic structure of 1L-G in the resulting
supermoir\'{e} pattern created by the top and bottom hBN in a
B-hBN/1L-G/B-hBN LMH\@.

We have studied adhesion between 1L-G and 1L-hBN using the diffusion
Monte Carlo (DMC) method\cite{Ceperley1980,Foulkes2001} as implemented
in the \textsc{casino} code,\cite{Needs2020} previously used in
studies of the binding properties of bilayer graphene
(2L-G)\@.\cite{Mostaani2015} We use the binding energies (BEs)
$E_\text{bind}$ (see \eq\ref{eq:BE_def} in the Methods section)
computed for several 1L-G/1L-hBN stacking configurations as shown in
\tab\ref{fig:BE_data} and interlayer distances $d$ to parameterize a
symmetry-based interpolation formula describing $E_\text{bind}(d, {\bm
  \ell})$ for an arbitrary $d$ and in-plane offset ${\bm \ell}$
between the lattices of two crystallographically aligned 1Ls.  We
complete this description of local interlayer adhesion by using
elasticity theory to describe the lattice relaxation of 1L-G adjusting
to the underlying B-hBN lattice, or to the lattices of two
almost-aligned B-hBN crystals above and below the 1L-G\@.


Previous theoretical studies\cite{Giovannetti2007,Sachs2011, Fan2011,
  Leconte2017, Lebedev2017} have addressed the interlayer adhesion in
1L-G/1L-hBN within the framework of density-functional theory (DFT),
using the local density approximation (LDA),\cite{Giovannetti2007} or
DFT-vdW methods combined with the random-phase approximation
(RPA),\cite{Sachs2011, Fan2011, Leconte2017, Lebedev2017} which
produced contradictory results for the BE of
2L-G\@.\cite{Chakarova2006,Lebedeva2011,Dappe2012,Gould2013,Podeszwa2010}

Here we determine the BE and vibrational properties of a free-standing
1L-G/1L-hBN bilayer using the DMC method. We performed calculations
for 1L-G/1L-hBN rather than 1L-G on a B-hBN substrate because (i) the
variation in BE of 1L-G on B-hBN substrate as a function of in-plane
offset is dominated by the closest 1L-hBN to 1L-G,\cite{Zhao2018} and
(ii) a 2L-LMH is much more computationally tractable than a bulk LMH for
expensive quantum Monte Carlo methods, which scale as the third power
of system size.\cite{Foulkes2001} G and hBN heterostructures are
typically studied at low temperatures; hence we performed all our
calculations at zero temperature.  Zero-point vibrational effects are
neglected, since they largely cancel out of the BE\@.


\section{Results and Discussion}

\subsection{BE}

\begin{figure}[!htbp]
\includegraphics[width=0.45\textwidth]{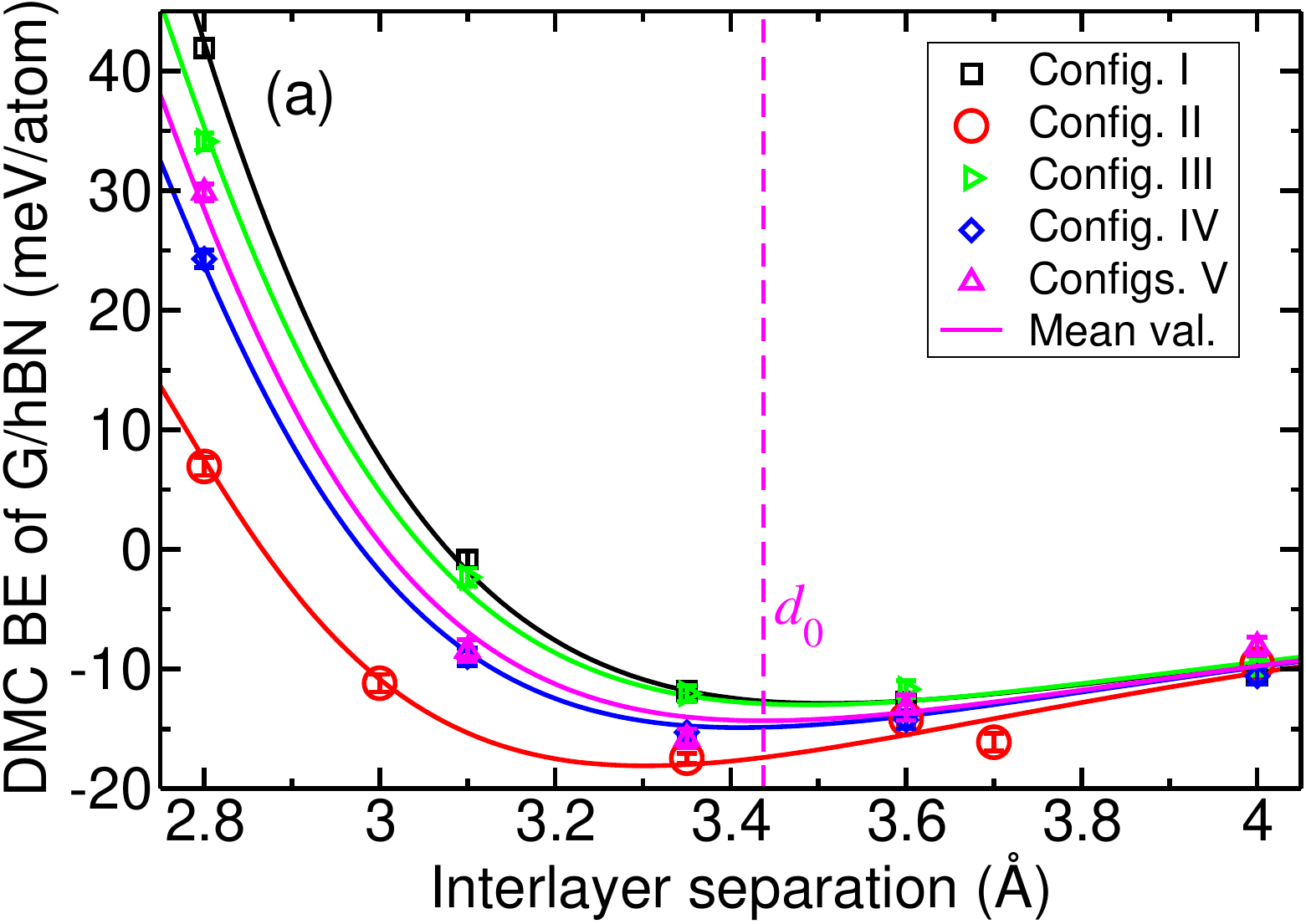}
\hspace{0.03\textwidth}
\includegraphics[width=0.45\textwidth]{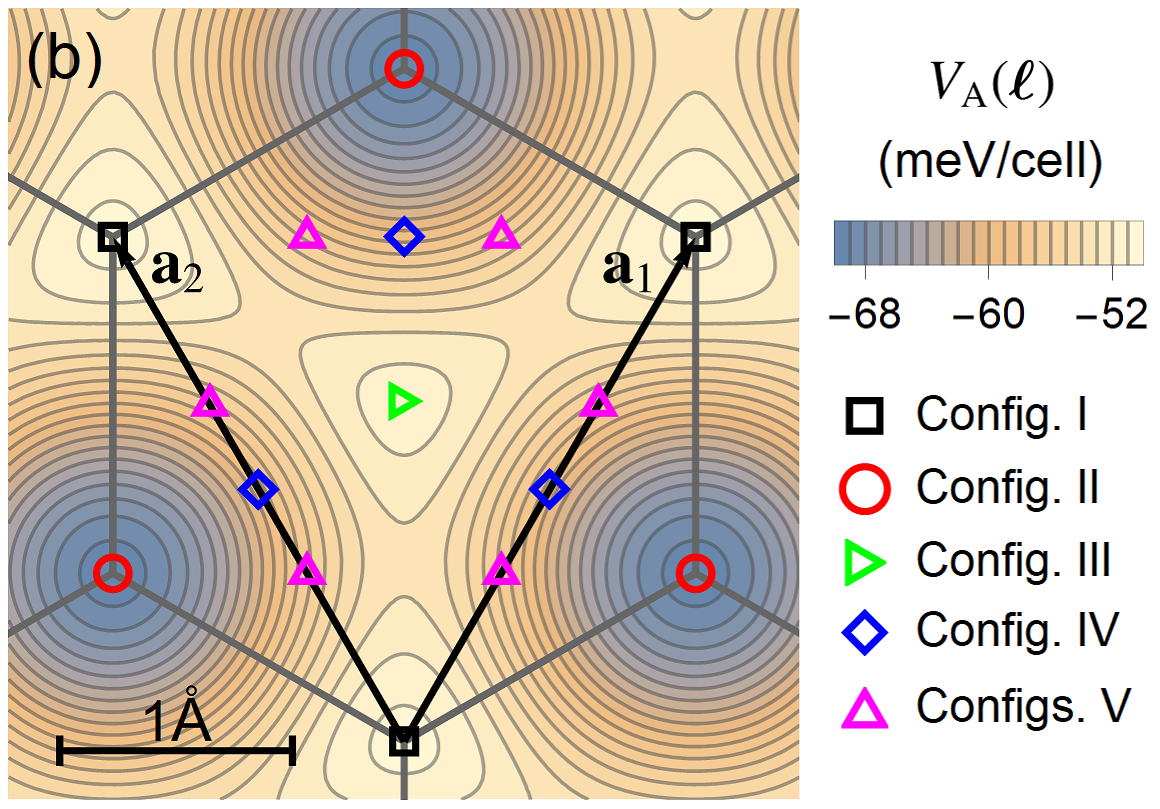}
\caption{(a)~DMC BE, $E_\text{bind}(d,{\bm \ell})$
  (\eq\ref{eq:BE_fit}), of 1L-G/1L-hBN as a function of interlayer
  separation $d$ for the five stacking configurations in
  \tab\ref{tab:G_BN_configs}. Both layers have the lattice constant of
  1L-G\@. The solid lines are a fit of \eq\ref{eq:BE_fit} to the DMC
  data. The fitted parameters are reported in Table 1 of
  the Supporting Information (SI)\@. (b)~Adhesion potential per unit
  cell $V_\text{A}({\bm \ell})\equiv 4E_\text{bind}(d_0,{\bm \ell})$
  as a function of the in-plane offset ${\bm \ell}$ of 1L-G relative
  to 1L-hBN at the layer separation $d_0$ that minimizes the
  translationally averaged BE\@. The zigzag direction of the honeycomb
  lattice lies along the $x$ axis. The even and odd parts of
  \eq\ref{eq:BE_fit} for each of the stacking configurations are in
  Figure 1 of the SI\@.
\protect\label{fig:BE_data}}
\end{figure}

\begin{table}[!htbp]
\small
\begin{center}
\caption{1L-G/1L-hBN stacking configurations and corresponding
  equilibrium separations, BEs, and breathing-mode (out-of-plane
  zone-center optical phonon) frequencies, obtained by fitting
  \eq\ref{eq:BE_fit} to DMC energy data obtained with both layers
  forced to adopt the lattice constant of G\@. C, B, and N atoms are
  shown as black, orange, and green balls, respectively. The hexagonal
  sublattices A and B are labeled in Config.~I\@. The offset ${\bm
    \ell}$ is the in-plane displacement of each C-C bond center from
  the corresponding B-N bond center. $\mathbf{a}_1$ and $\mathbf{a}_2$
  are the lattice vectors, as shown in \fig\ref{fig:BE_data}b. The DMC
  equilibrium BEs of the different configurations are correlated due
  to the use of the same DMC 1L energies in each case; hence the BE
  differences are more precise than suggested by the error bars on the
  absolute BEs. The errors on relative BEs are shown in
  \tab\ref{tab:BE_results_rel}. \protect\label{tab:G_BN_configs} }
\begin{tabular}{lccccc}
& & & \textbf{Equil.\ sep.} & \textbf{Equil.\ BE} & \textbf{Breath.-mode}  \\
\raisebox{2ex}[0pt]{\textbf{Config.}} &
\raisebox{2ex}[0pt]{\textbf{Structure}} &
\raisebox{2ex}[0pt]{\textbf{Layer offset} ${\bm \ell}$} &
\textbf{({\r A})} & \textbf{(meV/atom)} & \textbf{freq.\ (cm$^{\bm -1}$)} \\
\hline
I (AA) &
\includegraphics[valign=m,width=1.5cm]{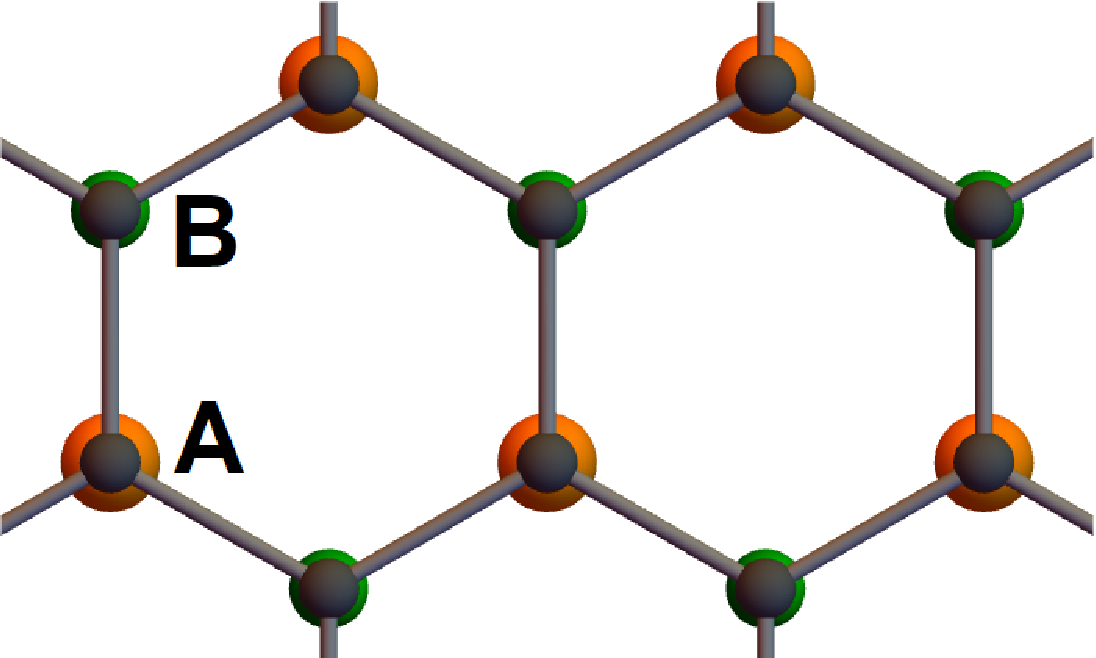}
& $\mathbf{0}$ & $3.51(1)$ & $-12.9(9)$ & $74(3)$ \\[.1cm] \hline
{\setlength\tabcolsep{0pt}\begin{tabular}{l}II (AB) \\(B on C)\end{tabular}} &
\includegraphics[valign=m,width=1.5cm]{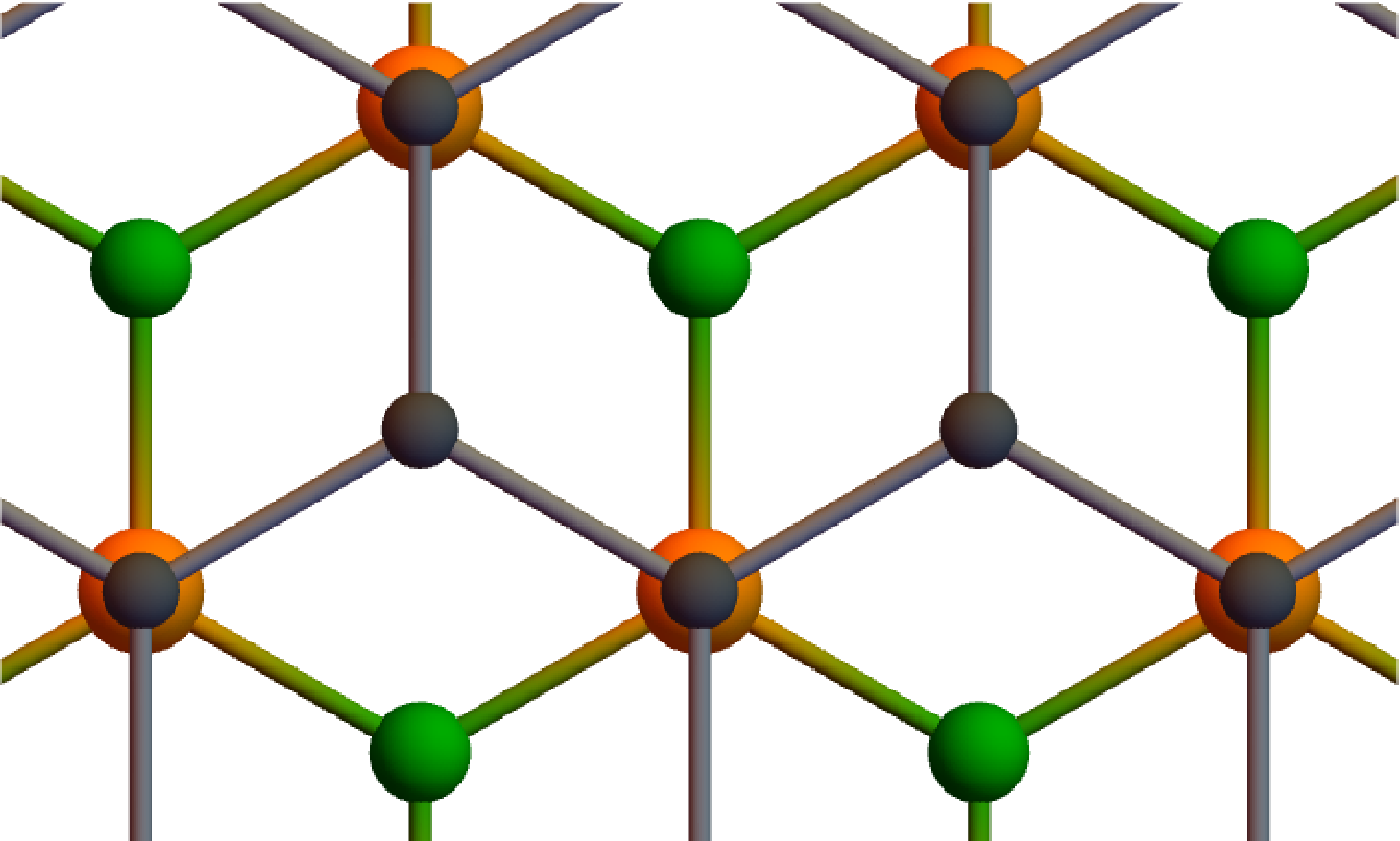}
& $\frac{2}{3}\left(\mathbf{a}_1 + \mathbf{a}_2\right)$ &
$3.30(1)$ & $-18.1(9)$ & $92(3)$ \\
\hline
{\setlength\tabcolsep{0pt}\begin{tabular}{l}III (BA)\\ (N on C)\end{tabular}} &
\includegraphics[valign=m,width=1.5cm]{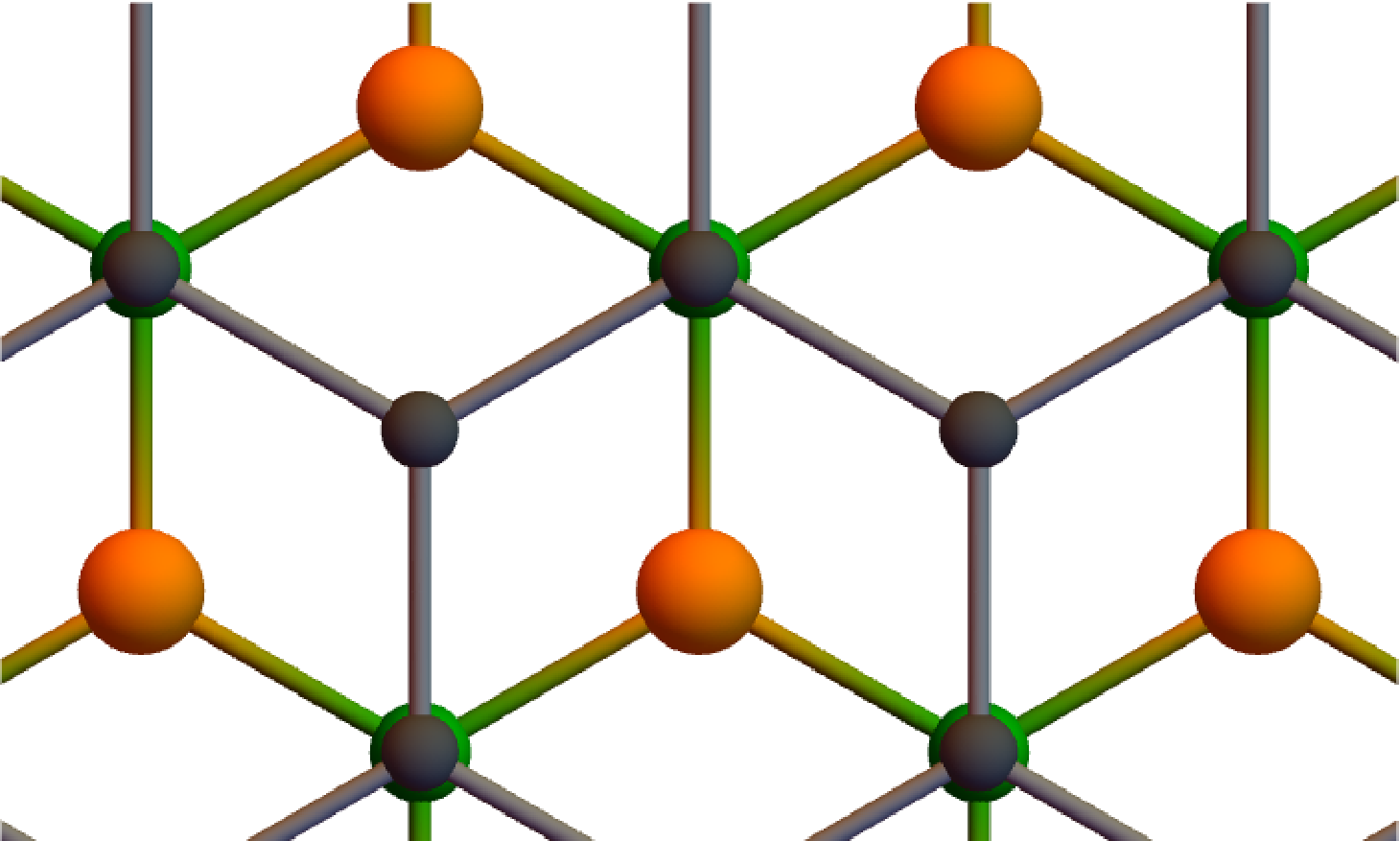}
& $\frac{1}{3}(\mathbf{a}_1 + \mathbf{a}_2)$ & $3.49(1)$ &
$-13.0(9)$ & $74(3)$ \\
\hline
IV &
\includegraphics[valign=m,width=1.5cm]{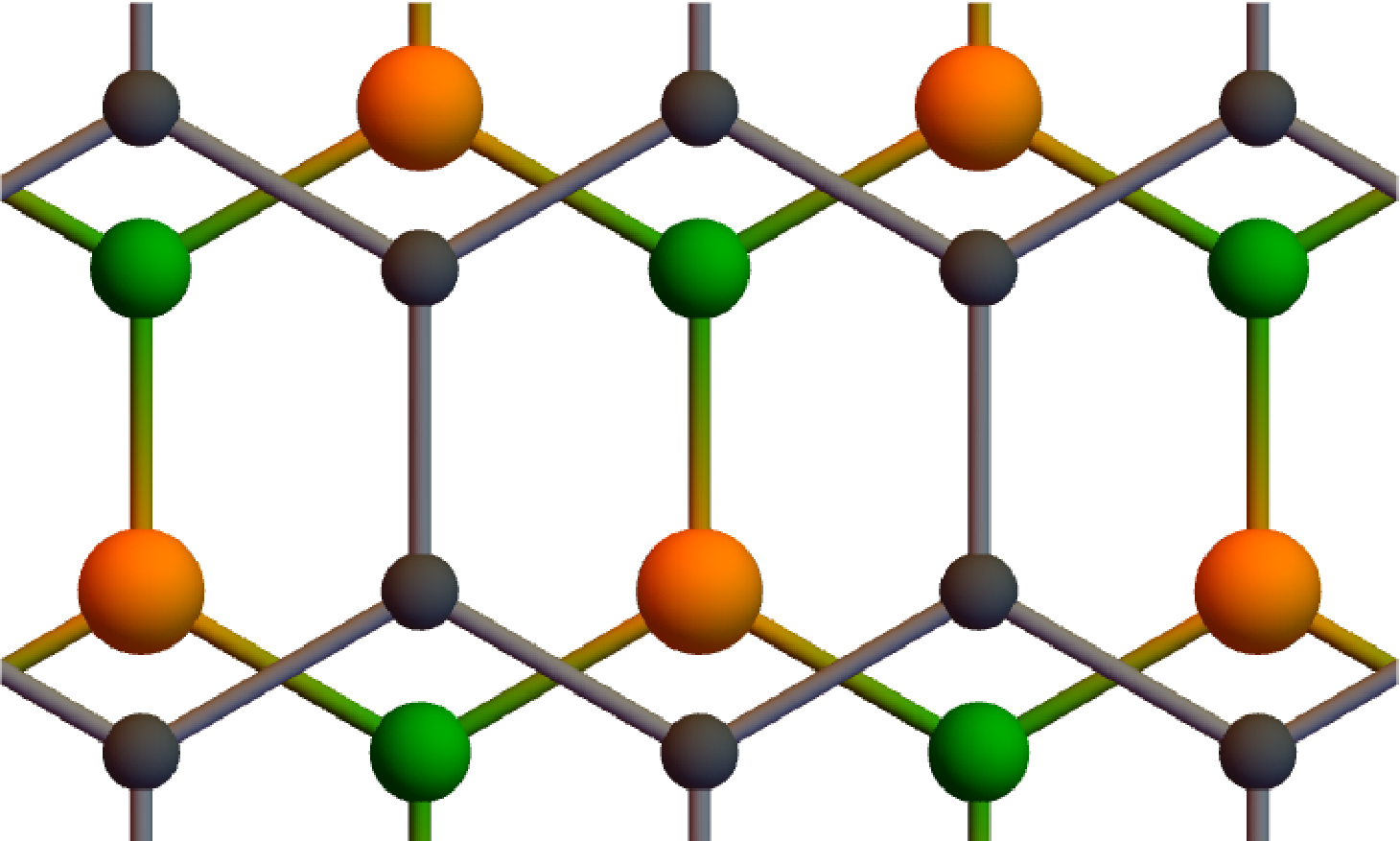}
& $\frac{1}{2}(\mathbf{a}_1+\mathbf{a}_2)$ & $3.41(1)$ &
$-14.9(9)$ & $82(3)$ \\
\hline
V (mean val.)$^a$ &
\includegraphics[valign=m,width=1.5cm]{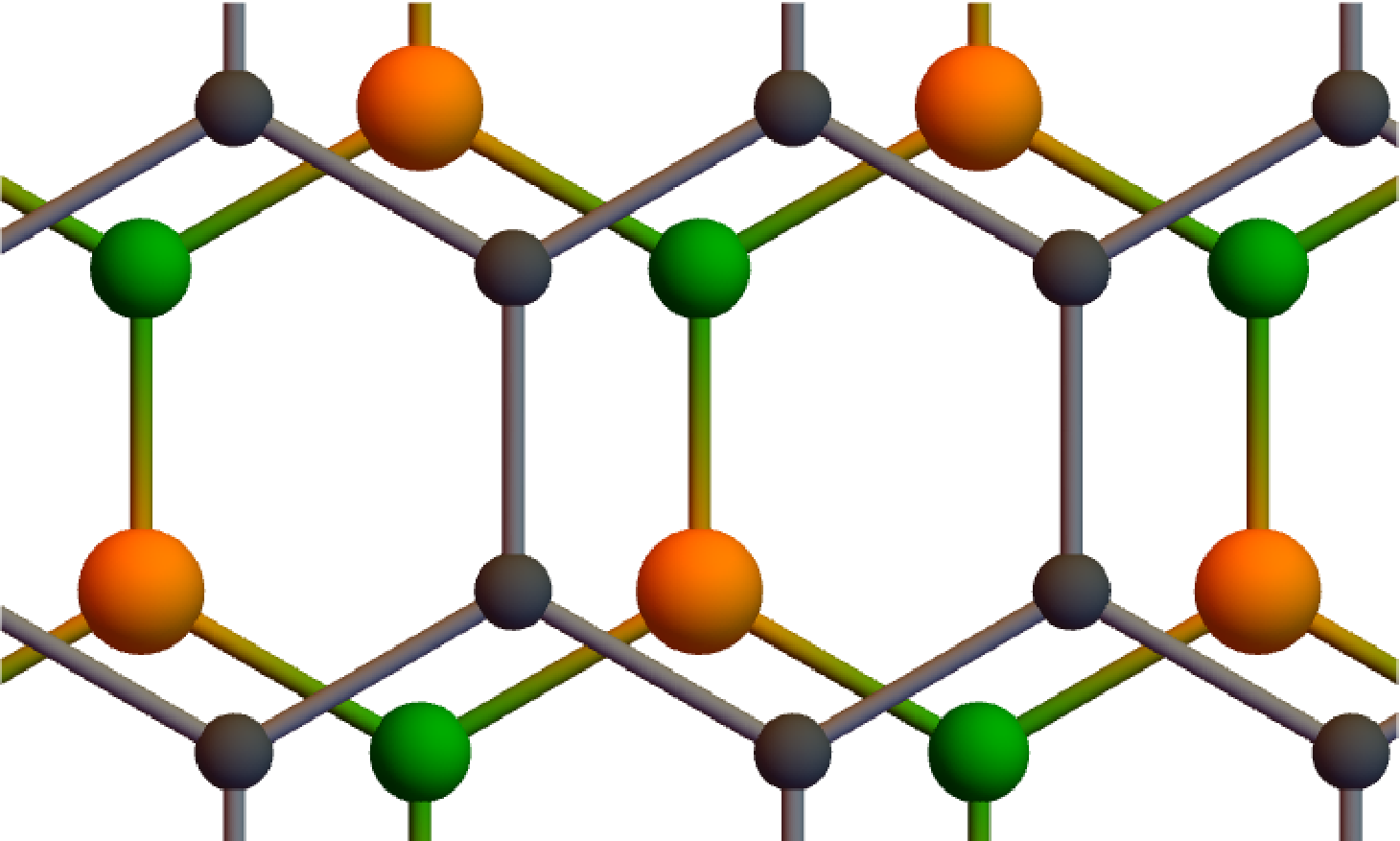}
& $\frac{1}{3}(\mathbf{a}_1-\mathbf{a}_2)$ & $3.44(1)$ &
$-14.3(9)$ & $80(3)$ \\ \hline
\end{tabular}
\end{center}
$^a$ See the Methods section for a detailed description of the
configurations used.
\end{table}

The DMC-computed BE, $E_\text{bind}(d, \bm\ell)$, is plotted in
\fig\ref{fig:BE_data} against interlayer separation for 1L-G/1L-hBN in
five different stacking configurations (\textit{i.e.}, five different
offsets ${\bm \ell}$ of 1L-G relative to 1L-hBN, with both layers
forced to adopt the 1L-G lattice constant). Because the elastic energy
required to compress the 1L-hBN to have the same lattice constant as
1L-G cancels out of the BE,\cite{SanJose2014} the precise choice of
lattice constant does not affect the BE\@. The stacking configurations
are labelled I--V, and are shown in
\tab\ref{tab:G_BN_configs}. Configuration I is AA
stacking. Configurations II and III are AB (B on C) and BA (N on C)
stacking. The BE curve of configuration II is flatter; therefore we
have studied a broader range of interlayer separations than for the
other configurations. Configuration V is a ``mean-value
configuration,'' for which the contribution to the offset-dependence
of a quantity such as the BE from the first star of reciprocal-lattice
vectors is zero.  The DMC BEs per atom were extrapolated to the
thermodynamic limit of infinite supercell size at fixed primitive-cell
geometry. We fitted our DMC BEs by the function
\begin{equation}
E_\text{bind}(d,{\bm \ell}) = \bar{E}_\text{bind}(d)+ A_{11}
e^{-\kappa_{\text{A}1}d} \sum_{m=1,3,5} \cos(\mathbf{g}_m \cdot {\bm
  \ell}) +B_{11}e^{-\kappa_{\text{B}1}d}
\sum_{m=1,3,5}\sin(\mathbf{g}_m \cdot {\bm \ell}),
\label{eq:BE_fit}
\end{equation}
where
$\bar{E}_\text{bind}(d)=A_{01}d^{-4}+A_{02}d^{-8}+A_{03}d^{-12}+A_{04}d^{-16}$,
$d$ is the interlayer separation, and ${\bm \ell}$ is the offset of
the 1L-G lattice relative to the 1L-hBN lattice, with ${\bm
  \ell_I}=\mathbf{0}$ corresponding to AA stacking (configuration
I)\@. For configuration V, $E_\text{bind}(d,{\bm \ell}_{\rm
  V})=\bar{E}_\text{bind}(d)$. The $\{A_{sj}\}$, $\{B_{sj}\}$,
$\{\kappa_{\text{A}s}\}$, and $\{\kappa_{\text{B}s}\}$ are fitting
parameters (see Table 1 of the SI for the fitted values), while
$\mathbf{g}_m$ denotes the $m$th reciprocal-lattice point of the
primitive cell, and subscript $s$ denotes the star of
reciprocal-lattice vectors. We adopt the convention that
reciprocal-lattice points are ordered by star $s$, starting with
$\mathbf{g}_0=\mathbf{0}$, and that within each star successive
reciprocal lattice points are labeled anticlockwise starting from the
$y$-axis. The labeling is illustrated in Figure 4c of the SI\@. Thus
$\mathbf{g}_1$, $\mathbf{g}_3$, and $\mathbf{g}_5$ denote the first,
third, and fifth reciprocal-lattice points in the first star of the
hexagonal reciprocal lattice. \eq\ref{eq:BE_fit} exhibits the
expected\cite{Dobson2104} configuration-independent $d^{-4}$ vdW
behavior at separation larger than a few times
$1/\min\{\kappa_{\text{A}1},\kappa_{\text{B}1}\}$, and satisfies the
translational and $D_3$ point symmetry in the offset ${\bm \ell}$.
The minimum of $\bar{E}_\text{bind}(d)$ in \eq\ref{eq:BE_fit} provides
the nominal equilibrium separation $d_0$ for a rigid pair of
misaligned LMs (because all offsets ${\bm \ell}$ are present in the
moir\'{e} supercell), while the second derivative
$\bar{E}_\text{bind}''(d_0)$ determines their layer breathing mode
(LBM) frequency $\omega_\text{BM}$. The translationally averaged DMC
breathing-mode frequency is $80(3)$ cm$^{-1}$, in good agreement with
the DFT-LDA result ($\sim77$ cm$^{-1}$; see
Figure 2 of the SI)\@. The reduced
$\chi^2$ does not change significantly when two stars of nonzero
reciprocal lattice vectors are used instead of one in the fitting
function for the BE\@. Hence, for simplicity, we have used the
one-star model in \eq\ref{eq:BE_fit}.

Our DMC results show that stacking configuration II has the lowest
energy, in agreement with DFT calculations.\cite{Giovannetti2007} In
this structure, the C 2p$_{\rm z}$ electrons are closer to the
partially positive B atoms than the partially negative N atoms.

\begin{table}[!htbp]
\begin{center}
\caption{Differences in the equilibrium BE of 1L-G/1L-hBN between the
  different stacking configurations in \tab\ref{tab:G_BN_configs}
  calculated using DFT and DMC\@. $E_Y$ is the equilibrium total
  energy per atom of 1L-G/1L-hBN in stacking configuration $Y$. Both
  layers were assumed to have the 1L-G lattice constant in the present
  work and in Refs.\ \citenum{Giovannetti2007} and \citenum{Fan2011},
  while the lattice constant of 1L-hBN was used in
  Refs.\ \citenum{Sachs2011} and \citenum{Slotman2014}. An averaged
  lattice constant was used in Ref.\ \citenum{Leconte2017}. The other
  DFT-LDA results are from the present
  work. \protect\label{tab:BE_results_rel}}
\begin{tabular}{lcccc}
& \multicolumn{4}{c}{\textbf{Relative BE (meV\,/\,atom)}} \\
\raisebox{1em}[0pt]{\textbf{Difference}} & \textbf{DFT-LDA} &
\textbf{DFT-vdW}\cite{Fan2011,Slotman2014} & \textbf{DFT-RPA} &
\textbf{DMC} \\
\hline
$E_\text{I}-E_\text{II}$ & $6$, $6$\bcite{Giovannetti2007} $\sim
9.5$\bcite{Fan2011} & $\sim 10$, $5.25$ & $5.25$\bcite{Sachs2011},
$5$\bcite{Leconte2017} & $5.2(4)$ \\
$E_\text{III}-E_\text{II}$ & $5$, $\sim 9.5$\bcite{Fan2011} & $\sim
10$, $4.25$ & $4.5$\bcite{Sachs2011}, $4$\bcite{Leconte2017} &
$5.1(5)$ \\
$E_\text{IV}-E_\text{II}$ & $3$, $5$\bcite{Giovannetti2007}, $\sim
4.5$\bcite{Fan2011} & $\sim 5$, $3.0$ & $4.5$\bcite{Sachs2011} &
$3.2(3)$ \\
$E_\text{V}-E_\text{II}$ & $1$ & & & $3.8(3)$ \\
$E_\text{I}-E_\text{III}$ & $1$, $\sim 0$\bcite{Fan2011} & $\sim 0$,
$1$ & $0.75$\bcite{Sachs2011}, $1$\bcite{Leconte2017} & $0.1(4)$ \\
\hline
\end{tabular}
\end{center}
\end{table}

\begin{table}[!htbp]
\begin{center}
\caption{Equilibrium BE of 1L-G/1L-hBN from DFT\@. Both layers were
  assumed to have the 1L-G lattice constant in the present work and in
  Refs.\ \citenum{Giovannetti2007} and \citenum{Fan2011}, while the
  1L-hBN lattice constant was used in Refs.\ \citenum{Sachs2011} and
  \citenum{Slotman2014}. An averaged lattice constant was used in
  Ref.\ \citenum{Leconte2017}. The other DFT-LDA results are from the
  present work.\protect\label{tab:BE_results}}
\begin{tabular}{lccc}
& \multicolumn{3}{c}{\textbf{BE (meV\,/\,atom)}} \\
\raisebox{2ex}[0pt]{\textbf{Config.}} & \textbf{DFT-LDA} &
\textbf{DFT-vdW}\cite{Fan2011,Slotman2014} & \textbf{DFT-RPA} \\
\hline
I & $-9$, $\sim -10$\bcite{Giovannetti2007}, $\sim
-7.5$\bcite{Fan2011} & $\sim -25$, $-28.50$ &
$-15.5$\bcite{Sachs2011}, $-18$\bcite{Leconte2017} \\
II & $-15$, $\sim -16$\bcite{Giovannetti2007}, $\sim
-17$\bcite{Fan2011} & $\sim -35$, $-33.75$ &
$-20.75$\bcite{Sachs2011}, $-23$\bcite{Leconte2017} \\
III & $-10$, $\sim -7.5$\bcite{Fan2011} & $\sim -25$, $-29.50$ &
$-16.25$\bcite{Sachs2011}, $-19$\bcite{Leconte2017} \\
IV & $-12$, $\sim -11$\bcite{Giovannetti2007}, $\sim
-12.5$\bcite{Fan2011} & $\sim -30$ & $-17.75$\bcite{Sachs2011} \\
V & $-11$ & & \\
\hline
\end{tabular}
\end{center}
\end{table}

\tab\ref{tab:BE_results} compares our DMC BEs with DFT for each
stacking configuration. DFT-LDA underestimates the BE by up to 40\%,
while DFT-vdW overbinds 1L-G/1L-hBN by up to
50\%. DFT-RPA\cite{Sachs2011} overbinds 1L-G/1L-hBN by 25\% compared
with DMC\@. Relative BEs for the different stacking configurations are
given in \tab\ref{tab:BE_results_rel}. Stacking configuration II is
$5.2(4)$ meV/atom more stable than the least stable stacking
configuration (I)\@. This is similar to the difference of $5.25$
meV/atom calculated by DFT-RPA\cite{Sachs2011} and
DFT-vdW\@.\cite{Slotman2014} The difference in the DMC BEs of
configurations I and III is negligible, as is also predicted by
DFT-vdW\cite{Slotman2014} and DFT-RPA\@.\cite{Sachs2011, Fan2011,
  Leconte2017} Even DFT-LDA correctly predicts the qualitative trend
for the relative stability of the different stacking
configurations. The DMC equilibrium separations of the different
stacking configurations are in \tab\ref{tab:G_BN_configs}, and lie in
the range 3.30--3.51 {\AA}. This is similar to the range of
separations $\sim$3.2--3.5 {\AA} and $\sim$3.35--3.55 {\AA} predicted
by DFT-vdW\cite{Fan2011, Slotman2014} and DFT-RPA;\cite{Sachs2011} see
Table 2 of the SI\@.

\subsection{Reconstruction of Perfectly Aligned G/hBN Composites into
Lattice-Matched Structures}

The total energy of 1L-G/1L-hBN has two relevant contributions: (i)
the adhesion potential $U_\text{A}$ and (ii) the elastic energy
$U_\text{E}$ due to the straining of the layers. Our DMC calculations
show that, for a 1L-G/1L-hBN LMH in which both layers have the same
lattice constant, the equilibrium BE is higher by $3.8(3)$ meV/atom in
the mean-value configuration (V) than in the most stable stacking
configuration (II); see \tab\ref{tab:BE_results_rel}. Hence the
adhesion potential could be lowered by $15(1)$ meV per 1L-G unit cell
if 1L-G were to adopt the same lattice constant as hBN, forming
stacking configuration II uniformly instead of a moir\'{e} pattern.
Where 1L-G is encapsulated between two perfectly aligned regions of
B-hBN, the adhesion potential could be lowered by as much as $30(2)$
meV per 1L-G unit cell if 1L-G were to adopt the hBN lattice constant.

The elastic energy $\Delta U_\text{E}^\text{G}(a)$ required to strain
1L-G to have lattice constant $a$ may be calculated using
Equation 5 of the SI with strain tensor
$\varepsilon=(a/a_\text{G}-1)I$, where $I$ is the $2 \times 2$
identity matrix. This gives $\Delta
U_\text{E}^\text{G}(a)=\sqrt{3}{(a-a_\text{G})}^2(\lambda_\text{G}+\mu_\text{G})$,
where $\lambda_\text{G}$ and $\mu_\text{G}$ are the Lam\'{e}
coefficients of 1L-G, whose experimentally measured values are given
in the SI\@.  Hence the elastic energy required to strain 1L-G to have
the same lattice constant as hBN is $\Delta
U_\text{E}^\text{G}(a_\text{hBN})=42$ meV per 1L-G unit cell. Thus the
reduction in adhesion potential from adopting stacking configuration
II uniformly does not compensate for the increase in the elastic
energy, so the layers retain their different, incommensurate lattice
constants, even when 1L-G is encapsulated in B-hBN and the
crystallographic directions of the lattices are aligned. This,
however, does not exclude local deformations, which adjust the two
lattices periodically following the moir\'{e} pattern of the two
slightly incommensurate crystals, leading to the weak lattice
reconstruction that we describe in the SI\@.  In these weak
reconstructions we do not find any evidence of nonglobal energy
minima, implying that there are no low-energy metastable structures at
1L-G/1L-hBN interfaces.

For a suspended 1L-G/1L-hBN bilayer, the adhesion-potential reduction
$\Delta U_\text{A}^\text{G/hBN} = -15(1)$ meV per unit cell due to
adopting the most stable stacking configuration uniformly must again
be compared with the change $\Delta U_\text{E}^\text{G/hBN}$ in the
total elastic energy.  $\Delta U_\text{E}^\text{G/hBN}$ is the sum of
the elastic energies required to stretch/compress the individual
layers to have the common lattice constant $a$ such that the elastic
energy is minimal with respect to $a$, i.e., $\Delta
U_\text{E}^\text{G/hBN} =\min_a \left(\Delta U_\text{E}^\text{G}(a) +
\Delta U_\text{E}^\text{hBN}(a)\right)=19$ meV per unit cell, where
the expression for $\Delta U_\text{E}^\text{hBN}(a)$ is analogous to
that given above for 1L-G\@. The elastic-energy penalty is $4(1)$ meV
per unit cell larger than the reduction in the adhesion potential, so
that the layers retain their incommensurate lattice constants
(just). For a suspended 1L-hBN/1L-G/1L-hBN trilayer with aligned
1L-hBN lattices, the adhesion-potential reduction $\Delta
U_\text{A}^\text{hBN/G/hBN} = -30(2)$ meV per unit cell due to
adopting a common lattice constant must be compared with the
elastic-energy penalty $\Delta U_\text{E}^\text{hBN/G/hBN} = \min_a
\left( \Delta U_\text{E}^\text{G}(a) + 2 \Delta
U_\text{E}^\text{hBN}(a)\right) = 26$ meV per unit cell. Hence the
adhesion energy gain marginally exceeds the elastic energy penalty by
$4(2)$ meV per unit cell, so that the trilayer is expected to
reconstruct to form a new 2d crystal.  However, small strains may
easily change the balance of the energies in these two free-standing
structures and hence change the fate of lattice
reconstruction. Finally, for a multilayer structure consisting of an
alternating series of 1L-G and 1L-hBN layers, $\Delta U_\text{A} =
\Delta U_\text{A}^\text{hBN/G/hBN} = -30(2)$ meV per unit cell per
1L-G/1L-hBN bilayer and $\Delta U_\text{E} = \Delta
U_\text{E}^\text{G/hBN} = 19$ meV per unit cell per 1L-G/1L-hBN
bilayer, so the balance shifts completely in favor of adhesion,
leading to a reconstructed lattice constant of $a'= 2.481$ {\AA} and
to a peculiar semimetallic band structure of this bulk composite
material.\cite{Chen_2023}

Using the fitted adhesion potential, we have parameterized a continuum
model of the local relaxation of 1L-G on a rigid hBN substrate (see
Methods), allowing us to examine the electronic properties of
B-hBN/1L-G/B-hBN\cite{Wang2019,Wang2019a,finney_tunable_2019,Yang_2020,Andelkovic_2020,Sun_2021,Moulsdale_2022,Hu_2023}
for various orientations of the hBN unit cells (see SI)\@.



\subsection{Effective Interatomic Pair Potentials Beyond Standard
Lennard-Jones Theory}

\begin{figure}[!htbp]
\begin{center}
\raisebox{-0.5\height}{\includegraphics[width=80mm]{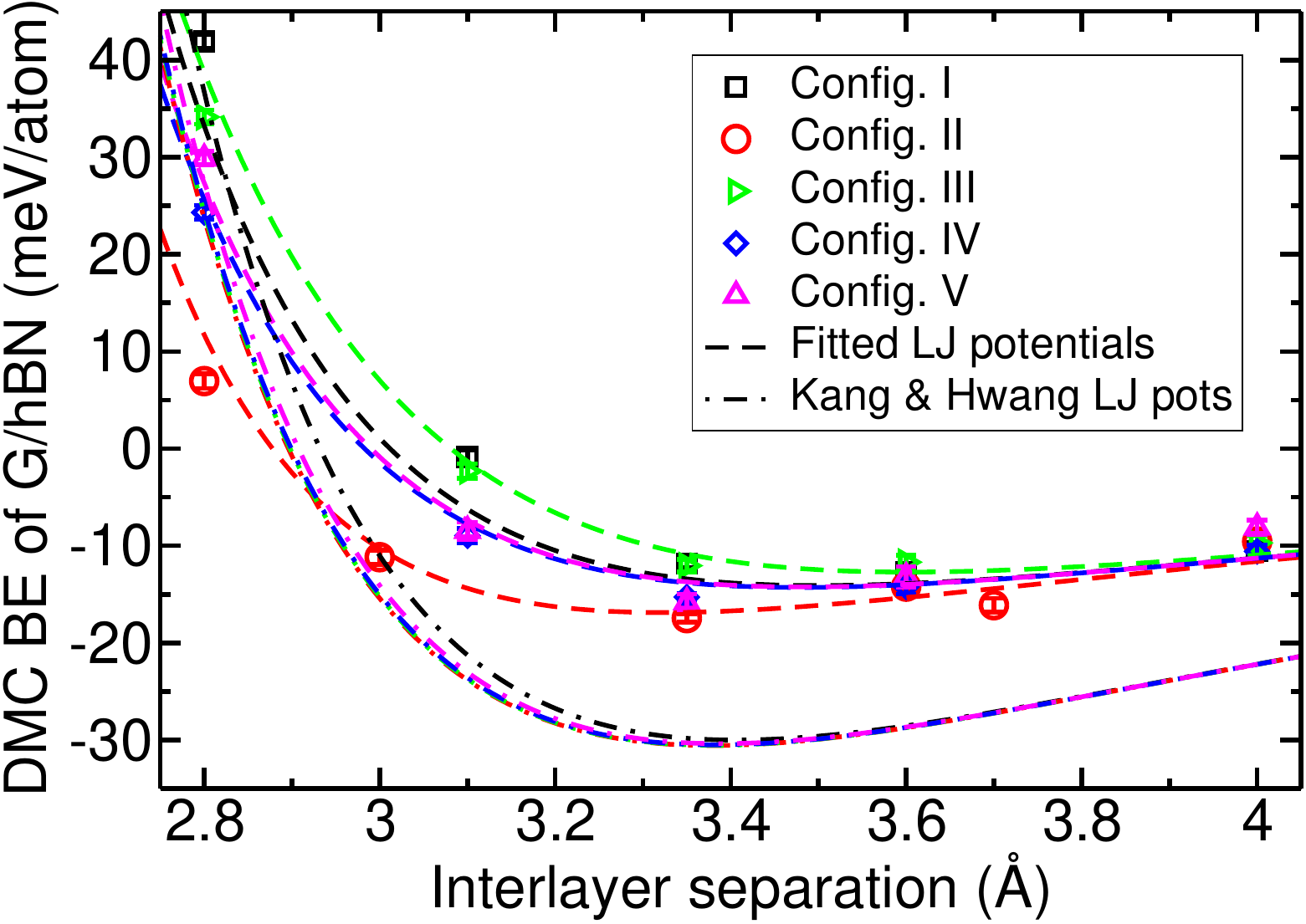}}
\hfill
\begin{minipage}[m]{60mm}
\begin{tabular}{lc}
\textbf{Parameter} & \textbf{Value} \\
\hline
$A_\text{CB}$ & $73454.7$ eV\,{\AA}$^{12}$ \\
$B_\text{CB}$ & $684.583$ eV\,{\AA}$^6$ \\
$A_\text{CN}$ & $-31772.9$ eV\,{\AA}$^{12}$ \\
$B_\text{CN}$ & $-661.272$ eV\,{\AA}$^6$ \\
\hline
\end{tabular}
\end{minipage}
\caption{DMC 1L-G/1L-hBN BEs (markers) together with BE curves
  obtained using both our fitted LJ pair potentials (dashed lines) and
  those from Ref.\ \citenum{Kang2004} (dash-dotted lines). Both layers
  are constrained to have the 1L-G lattice constant. Our fitting
  parameters are reported in the right-hand panel.  The equilibrium
  layer separations from our fitted pair potentials are 3.50, 3.32,
  3.60, 3.48, and 3.48 {\AA} for stacking configurations I--V,
  respectively, and the corresponding equilibrium BEs are $-14.1$,
  $-16.9$, $-12.7$, $-14.3$, and $-14.2$ meV/atom. These may be
  compared with the results in \tab\ref{tab:G_BN_configs}, which were
  obtained by fitting \eq\ref{eq:BE_fit} to the DMC
  data.\protect\label{fig:lj_be_curves}}
\end{center}
\end{figure}

Molecular dynamics simulations with interatomic pair potentials are
often used to model interactions between 1L-G and 1L-hBN\@.  The
simplest useful pair potential is the two-parameter Lennard-Jones (LJ)
form,\cite{Jones1924} $U(r)=A/r^{12}-B/r^6$, used to study 1L-G on
1L-hBN\cite{Neek2014,Zhang2015,Yang2018,Nguyen2019} and
6L-hBN\@.\cite{Zhang2017} However, the accuracy of LJ potentials in
1L-G/1L-hBN LMHs is questionable.  \textit{E.g.}, the parameters
$A_\text{CB}=32679$ eV\,{\AA}$^{12}$ and $B_\text{CB}=20.748$
eV\,{\AA}$^{6}$ for a C-B LJ potential and $A_\text{CN}=34545$
eV\,{\AA}$^{12}$ and $B_\text{CN}=23.709$ eV\,{\AA}$^{6}$ for a C-N LJ
potential were obtained in Ref.\ \citenum{Kang2004} and used in
Refs.\ \citenum{Neek2014} and \citenum{Nguyen2019} to study
1L-G/1L-hBN interfaces. Interlayer BE curves for a 1L-G/1L-hBN LMH
obtained with these LJ potentials are shown in
\fig\ref{fig:lj_be_curves}. These LJ potentials overbind 1L-G/1L-hBN
by around 50\% and fail to distinguish between the different stacking
configurations.

We determine the parameters $A_\text{CB}$ and $B_\text{CB}$ for LJ C-B
potentials and $A_\text{CN}$ and $B_\text{CN}$ for C-N potentials by
fitting the LJ BE to our DMC BE results for all the stacking
configurations in \fig\ref{tab:G_BN_configs}. The fitted parameters
and the resulting LJ BE curves are plotted in
\fig\ref{fig:lj_be_curves}. The fitted parameters take
counterintuitive values; \textit{e.g.}, in the C-N pair potential, the
LJ parameters are negative, so that the $r^{-12}$ term is attractive
and the $r^{-6}$ term is repulsive. Hence, over the relevant range of
interatomic distances, the C-B pair potential is attractive, while the
C-N one is repulsive. Although standard LJ potentials with attractive
$r^{-6}$ tails give a poor description of the physics of interlayer
interactions, our results demonstrate that interatomic pair potentials
of LJ form can provide a good description of 1L-G/1L-hBN adhesion
provided one is willing to dispense with transferability.

\section{Conclusions}

We have performed DMC calculations of the binding of 1L-G and 1L-hBN,
finding the BE for different stacking configurations (\textit{i.e.},
lattice offsets).  We have used these BEs to argue that G and hBN
retain different, incommensurate lattice constants for 1L-G on a B-hBN
substrate, while G and hBN adopt a common lattice constant in a
stacked heterostructure of alternating layers of 1L-G and 1L-hBN\@.
Free-standing 1L-G/1L-hBN bilayers and 1L-hBN/1L-G/1L-hBN trilayers
are close to the boundary between these two types of behavior.

As described in the SI, we have used our calculated adhesion potential
to evaluate the relaxation of 1L-G on a B-hBN substrate and 1L-G
encapsulated in B-hBN\@.  This has allowed us to examine the
electronic structure of B-hBN/1L-G/B-hBN, showing that the electronic
structure depends on the relative orientation of the B-hBN lattices
above and below the 1L-G\@.

Finally we report pair potentials for modelling interactions between
1L-G and 1L-hBN\@.

\section{Methods}

\subsection{Atomic Structure of 1L-G/1L-hBN at a Given Layer Separation}

To determine the atomic structure of 1L-G/1L-hBN for a given layer
separation, we pinned the lattice vectors, the mean interlayer
distance, and the mean lattice offset, and we relaxed the remaining
structural parameters, which describe the slight buckling of less than
$0.003$ {\AA} of the layers, within DFT\@.\cite{Clark2005} We used
ultrasoft pseudopotentials,\cite{Vanderbilt1990,Clark2005} a
plane-wave cutoff energy of $25$ Ha, a $15\times15$ Monkhorst-Pack
$\mathbf{k}$-point mesh, and Grimme's
dispersion-corrected\cite{Grimme2006} Perdew-Burke Ernzerhof
(PBE)\cite{Perdew1996} functional in a periodic simulation cell of
height $16$ {\AA}. We fixed the in-plane lattice constants of both
1L-G and 1L-hBN at the experimental 1L-G lattice constant
$a_\text{G}=2.46$ {\AA}\@.\cite{Kelly1981,Dresselhaus1988} The
resulting bucklings are energetically insignificant.  To calculate the
BE of 1L-G/1L-hBN in \fig\ref{fig:BE_data} for Config.\ V, we have
used similar configurations to the mean-value configuration (V) as
shown in \tab\ref{tab:G_BN_configs}. For Config.\ V, we performed
calculations in which the layer offsets at $d=2.8$, 3.1, 3.35, 3.6,
and 4 {\AA} are ${\bm \ell} = 0.09\mathbf{a}_1-0.04\mathbf{a}_2$,
$0.08\mathbf{a}_1-0.04\mathbf{a}_2$,
$0.04\mathbf{a}_1-0.03\mathbf{a}_2$,
$0.34\mathbf{a}_1-0.33\mathbf{a}_2$, and $1/3\mathbf{a}_1$,
respectively.  Offsets corresponding to the relaxed structures were
used in the fit of \eq\ref{eq:BE_fit}.

\subsection{Details of the Quantum Monte Carlo Approach}

The BE per atom of 1L-G/1L-hBN can be written as
\begin{equation}
E_\text{bind}(d,{\bm \ell})=E_\text{G/hBN}(d,{\bm
  \ell})-\frac{E_\text{hBN}+E_\text{G}}{2}, \label{eq:BE_def}
\end{equation}
where $E_\text{G/hBN}$, $E_\text{hBN}$, and $E_\text{G}$ are the total
energies per atom of 1L-G/1L-hBN, 1L-hBN, and 1L-G, calculated using
the fixed-node DMC method as implemented in the \textsc{casino}
code.\cite{Needs2020} We used Dirac-Fock
pseudopotentials\cite{Trail2005a, Trail2005b} to represent the atomic
cores, using the pseudopotential locality
approximation.\cite{Mitas1991} Our many-body trial wave functions
consisted of Slater determinants for spin-up and spin-down electrons,
multiplied by a Jastrow correlation factor\cite{Foulkes2001}
containing polynomial and plane-wave electron-electron terms and
polynomial electron-nucleus and electron-electron-nucleus
terms.\cite{Drummond2004} The Slater determinants contained Kohn-Sham
orbitals\cite{KohnSham1965} generated using the \textsc{castep}
plane-wave DFT code\cite{Clark2005} and re-expressed in a blip
(B-spline) basis.\cite{Alfe2004} Free parameters in our Jastrow
factors were optimized within variational quantum Monte Carlo by
unreweighted variance minimization.\cite{Umrigar1988, Drummond2005} We
generated the Kohn-Sham orbitals using the LDA functional and a
plane-wave energy cutoff of $110$ Ha. Although the DFT calculations
were performed in a three-dimensionally periodic cell, the DMC
calculations were carried out in a 2d-periodic simulation supercell
using a 2d Ewald interaction between charges.\cite{Parry1975,
  Parry1976}

To remove biases due to finite time steps and populations of walkers,
we performed pairs of DMC calculations using time steps in the ratio
$1:2.5$ with the corresponding target walker populations in the ratio
$2.5:1$, and we linearly extrapolated the resulting DMC energies to
zero time step and infinite population. Time-step errors are discussed
in the SI\@. The fixed-node error is of uncertain magnitude, but is
always positive.\cite{Reynolds_1982} The regions of high electron
density close to the nuclei make the dominant contribution to the
fixed-node error, so fixed-node errors are expected largely to cancel
when the BE is calculated.

To reduce quasirandom single-particle finite-size errors caused by
momentum quantization, we evaluated twist-averaged\cite{Lin2001} DMC
ground-state energies per atom in simulation supercells containing
$3\times 3$ and $5\times 5$ unit cells for 1L-G, 1L-hBN, and
1L-G/1L-hBN\@. To remove systematic finite-size errors due to the long
range of the Coulomb interaction and two-body correlations, we
extrapolated the twist-averaged energies to the thermodynamic limit of
infinite system size. Twist averaging was performed by fitting
\begin{equation}
E(N,\mathbf{k}_\text{s})=\bar{E}(N)
+a\left[E^\text{DFT}(N,\mathbf{k}_\text{s})-E^\text{DFT}(\infty)\right]
\end{equation}
to our DMC energies per atom $E(N,\mathbf{k}_\text{s})$ at twists
(simulation-cell Bloch vectors) $\mathbf{k}_\text{s}$ in $N$-electron
supercells, where $\bar{E}(N)$ and $a$ are fitting parameters,
$E^\text{DFT}(\infty)$ is the DFT energy per atom obtained with a fine
$50\times 50$ $\mathbf{k}$-point sampling and
$E^\text{DFT}(N,\mathbf{k}_\text{s})$ is the DFT energy with a
$\mathbf{k}$-point set corresponding to the DMC calculation. To
maximize the cancellation of errors in the BE, at each system size the
same set of twelve random twists was used for the 1Ls and for
1L-G/1L-hBN at all separations and with all stacking
configurations. The twist-averaged energies per atom $\bar{E}(N)$ were
extrapolated to the thermodynamic limit of infinite system size by
fitting
\begin{equation}
\bar{E}(N)=E(\infty)+bN^{-5/4}
\end{equation}
to our data, where the energy per atom in the thermodynamic limit
$E(\infty)$ and $b$ are fitting parameters.\cite{Drummond2008}

\subsection{Continuum Model of Relaxation of Layers}

To model the relaxation and deformation of 1L-G on a B-hBN substrate,
we use a continuum model of the displacement field in an elastic 1L-G
layer on rigid B-hBN\@.\cite{SanJose2014} The misalignment angle
between the lattice vectors of the elastic layer (1L-G) and the rigid
layer (B-hBN) is denoted by $\theta$ and is assumed to be small. Let
the unit-cell positions of the elastic layer after relaxation be
$\mathbf{r}_n+\mathbf{u}(\mathbf{r}_n)$, where $\{\mathbf{r}_n\}$ are
the unrelaxed lattice points, and $\mathbf{u}(\mathbf{r})$ is the
displacement field.  The displacement field is assumed to have the
periodicity of the moir\'{e} supercell. We can therefore write the
Fourier expansion of the displacement field as
$\mathbf{u}(\mathbf{r})=\sum_{\mathbf{q}}\mathbf{u}_{\mathbf{q}}e^{i
  \mathbf{q}\cdot\mathbf{r}}$, where the sum runs over the reciprocal
lattice points $\mathbf{q}$ of the moir\'{e} supercell.


In general, the adhesion potential per unit cell depends on
$\theta$. However, this dependence is negligible at small angles,
since the aligned case ($\theta=0$) is extremal. Hence at small
$\theta$ the adhesion potential at any point in the bilayer simply
depends on the local interlayer lattice offset ${\bm \ell}$. The
spatial variation in the interlayer distance is small, so we evaluate
the local contribution to the adhesion potential per unit cell as
$V_\text{A}({\bm \ell})=4E_\text{bind}(d_0,{\bm \ell})$. Plots of the
adhesion potential as a function of the offset of 1L-G relative to
B-hBN are in \fig\ref{fig:BE_data} and the SI (Figure 3). The
expression for the adhesion potential is in the SI\@.

In the continuum model we write the total adhesion potential and the
elastic energy in terms of the Fourier components of the displacement
field.  The equilibrium displacement field is then found by minimizing
the sum of the elastic energy and total adhesion potential with
respect to the Fourier components of the displacement field.  Full
details are given in the SI\@.  Also given in the SI is an analysis of
the consequences of lattice relaxation for the electronic structure of
1L-G\@.

\section*{Acknowledgements}

Computational resources were provided by Lancaster University's High
End Computing cluster. M.\ Szyniszewski was funded by ERC Grant
No.\ 853368 and EPSRC Grant No.\ EP/P010180/1.

\section*{Supporting Information}

Supporting Information: BE Fitting Parameters; Equilibrium Separation,
BE, and LBM Frequency; DFT Geometries; DFT Phonon Calculations;
Lam\'{e} Parameters; 2d Adhesion-Potential Parameters; Relaxation of
1L-G/B-hBN within the Continuum Model; Minimizing the Analytical
Approximation to the Total Energy within the Continuum Model;
Brute-Force Minimization of the Total Energy within the Continuum
Model; Modelling Moir\'e SL Minibands for Electrons in
B-hBN/1L-G/B-hBN\@; Band-Structure Reconstruction of B-hBN/1L-G/B-hBN:
Derivation of the Hamiltonian of Shortest Period; Time-Step Errors in
DMC Calculations (PDF)\@.

\bibliography{G_BN}

\providecommand{\latin}[1]{#1}
\makeatletter
\providecommand{\doi}
  {\begingroup\let\do\@makeother\dospecials
  \catcode`\{=1 \catcode`\}=2 \doi@aux}
\providecommand{\doi@aux}[1]{\endgroup\texttt{#1}}
\makeatother
\providecommand*\mcitethebibliography{\thebibliography}
\csname @ifundefined\endcsname{endmcitethebibliography}
  {\let\endmcitethebibliography\endthebibliography}{}
\begin{mcitethebibliography}{65}
\providecommand*\natexlab[1]{#1}
\providecommand*\mciteSetBstSublistMode[1]{}
\providecommand*\mciteSetBstMaxWidthForm[2]{}
\providecommand*\mciteBstWouldAddEndPuncttrue
  {\def\EndOfBibitem{\unskip.}}
\providecommand*\mciteBstWouldAddEndPunctfalse
  {\let\EndOfBibitem\relax}
\providecommand*\mciteSetBstMidEndSepPunct[3]{}
\providecommand*\mciteSetBstSublistLabelBeginEnd[3]{}
\providecommand*\EndOfBibitem{}
\mciteSetBstSublistMode{f}
\mciteSetBstMaxWidthForm{subitem}{(\alph{mcitesubitemcount})}
\mciteSetBstSublistLabelBeginEnd
  {\mcitemaxwidthsubitemform\space}
  {\relax}
  {\relax}

\bibitem[Bonaccorso \latin{et~al.}(2012)Bonaccorso, Lombardo, Hasan, Sun,
  Colombo, and Ferrari]{BONACCORSO2012}
Bonaccorso,~F.; Lombardo,~A.; Hasan,~T.; Sun,~Z.; Colombo,~L.; Ferrari,~A.~C.
  Production and Processing of Graphene and 2d Crystals. \emph{Mater. Today}
  \textbf{2012}, \emph{15}, 564 -- 589\relax
\mciteBstWouldAddEndPuncttrue
\mciteSetBstMidEndSepPunct{\mcitedefaultmidpunct}
{\mcitedefaultendpunct}{\mcitedefaultseppunct}\relax
\EndOfBibitem
\bibitem[Geim and Grigorieva(2013)Geim, and Grigorieva]{Geim2013}
Geim,~A.~K.; Grigorieva,~I.~V. Van der {W}aals Heterostructures. \emph{Nature}
  \textbf{2013}, \emph{499}, 419\relax
\mciteBstWouldAddEndPuncttrue
\mciteSetBstMidEndSepPunct{\mcitedefaultmidpunct}
{\mcitedefaultendpunct}{\mcitedefaultseppunct}\relax
\EndOfBibitem
\bibitem[Ferrari \latin{et~al.}(2015)Ferrari, Bonaccorso, Fal{'}ko, Novoselov,
  Roche, B{\o}ggild, Borini, Koppens, Palermo, Pugno, Garrido, Sordan, Bianco,
  Ballerini, Prato, Lidorikis, Kivioja, Marinelli, Ryh\"{a}nen, Morpurgo,
  Coleman, Nicolosi, Colombo, Fert, Garcia-Hernandez, Bachtold, Schneider,
  Guinea, Dekker, Barbone, Sun, Galiotis, Grigorenko, Konstantatos, Kis,
  Katsnelson, Vandersypen, Loiseau, Morandi, Neumaier, Treossi, Pellegrini,
  Polini, Tredicucci, Williams, Hee~Hong, Ahn, Min~Kim, Zirath, van Wees,
  van~der Zant, Occhipinti, Di~Matteo, Kinloch, Seyller, Quesnel, Feng, Teo,
  Rupesinghe, Hakonen, Neil, Tannock, L\"{o}fwander, and Kinaret]{ACF2015}
Ferrari,~A.~C. \latin{et~al.}  Science and Technology Roadmap for Graphene
  Related Two-Dimensional Crystals and Hybrid Systems. \emph{Nanoscale}
  \textbf{2015}, \emph{7}, 4598--4810\relax
\mciteBstWouldAddEndPuncttrue
\mciteSetBstMidEndSepPunct{\mcitedefaultmidpunct}
{\mcitedefaultendpunct}{\mcitedefaultseppunct}\relax
\EndOfBibitem
\bibitem[Backes \latin{et~al.}(2020)Backes, Abdelkader, Alonso,
  Andrieux-Ledier, Arenal, Azpeitia, Balakrishnan, Banszerus, Barjon, Bartali,
  Bellani, Berger, Berger, Ortega, {C}arlo Bernard, Beton, Beyer, Bianco,
  B{\o}ggild, Bonaccorso, Barin, Botas, Bueno, Carriazo, Castellanos-Gomez,
  Christian, Ciesielski, Ciuk, Cole, Coleman, Coletti, Crema, Cun, Dasler,
  Fazio, D{\'{\i}}ez, Drieschner, Duesberg, Fasel, Feng, Fina, Forti, Galiotis,
  Garberoglio, Garc{\'{\i}}a, Garrido, Gibertini, G\"{o}lzh\"{a}user,
  G{\'{o}}mez, Greber, Hauke, Hemmi, Hernandez-Rodriguez, Hirsch, Hodge,
  Huttel, Jepsen, Jimenez, Kaiser, Kaplas, Kim, Kis, Papagelis, Kostarelos,
  Krajewska, Lee, Li, Lipsanen, Liscio, Lohe, Loiseau, Lombardi, L{\'{o}}pez,
  Martin, Mart{\'{\i}}n, Mart{\'{\i}}nez, Martin-Gago, Mart{\'{\i}}nez,
  Marzari, Mayoral, McManus, Melucci, M{\'{e}}ndez, Merino, Merino, Meyer,
  Miniussi, Miseikis, Mishra, Morandi, Munuera, Mu{\~{n}}oz, Nolan, Ortolani,
  Ott, Palacio, Palermo, Parthenios, Pasternak, Patane, Prato, Prevost,
  Prudkovskiy, Pugno, Rojo, Rossi, Ruffieux, Samor{\`{\i}}, Schu{\'{e}},
  Setijadi, Seyller, Speranza, Stampfer, Stenger, Strupinski, Svirko, Taioli,
  Teo, Testi, Tomarchio, Tortello, Treossi, Turchanin, Vazquez, Villaro,
  Whelan, Xia, Yakimova, Yang, Yazdi, Yim, Yoon, Zhang, Zhuang, Colombo,
  Ferrari, and Garcia-Hernandez]{Backes_2020}
Backes,~C. \latin{et~al.}  Production and Processing of Graphene and Related
  Materials. \emph{2D Mater.} \textbf{2020}, \emph{7}, 022001\relax
\mciteBstWouldAddEndPuncttrue
\mciteSetBstMidEndSepPunct{\mcitedefaultmidpunct}
{\mcitedefaultendpunct}{\mcitedefaultseppunct}\relax
\EndOfBibitem
\bibitem[Dean \latin{et~al.}(2010)Dean, Young, Meric, Lee, Wang, Sorgenfrei,
  Watanabe, Taniguchi, Kim, Shepard, and Hone]{Dean2010}
Dean,~C.~R.; Young,~A.~F.; Meric,~I.; Lee,~C.; Wang,~L.; Sorgenfrei,~S.;
  Watanabe,~K.; Taniguchi,~T.; Kim,~P.; Shepard,~K.~L.; Hone,~J. Boron Nitride
  Substrates for High-Quality Graphene Electronics. \emph{Nat. Nanotechnol.}
  \textbf{2010}, \emph{5}, 722\relax
\mciteBstWouldAddEndPuncttrue
\mciteSetBstMidEndSepPunct{\mcitedefaultmidpunct}
{\mcitedefaultendpunct}{\mcitedefaultseppunct}\relax
\EndOfBibitem
\bibitem[Molitor \latin{et~al.}(2011)Molitor, G\"{u}ttinger, Stampfer,
  Dr\"{o}scher, Jacobsen, Ihn, and Ensslin]{Molitor2011}
Molitor,~F.; G\"{u}ttinger,~J.; Stampfer,~C.; Dr\"{o}scher,~S.; Jacobsen,~A.;
  Ihn,~T.; Ensslin,~K. Electronic Properties of Graphene Nanostructures.
  \emph{J. Phys. Condens. Mater.} \textbf{2011}, \emph{23}, 243201\relax
\mciteBstWouldAddEndPuncttrue
\mciteSetBstMidEndSepPunct{\mcitedefaultmidpunct}
{\mcitedefaultendpunct}{\mcitedefaultseppunct}\relax
\EndOfBibitem
\bibitem[Xue \latin{et~al.}(2011)Xue, Sanchez-Yamagishi, Bulmash, Jacquod,
  Deshpande, Watanabe, Taniguchi, Jarillo-Herrero, and LeRoy]{Xue2011}
Xue,~J.; Sanchez-Yamagishi,~J.; Bulmash,~D.; Jacquod,~P.; Deshpande,~A.;
  Watanabe,~K.; Taniguchi,~T.; Jarillo-Herrero,~P.; LeRoy,~B.~J. Scanning
  Tunnelling Microscopy and Spectroscopy of Ultra-Flat Graphene on Hexagonal
  Boron Nitride. \emph{Nat. Mater.} \textbf{2011}, \emph{10}, 282\relax
\mciteBstWouldAddEndPuncttrue
\mciteSetBstMidEndSepPunct{\mcitedefaultmidpunct}
{\mcitedefaultendpunct}{\mcitedefaultseppunct}\relax
\EndOfBibitem
\bibitem[Sachs \latin{et~al.}(2011)Sachs, Wehling, Katsnelson, and
  Lichtenstein]{Sachs2011}
Sachs,~B.; Wehling,~T.~O.; Katsnelson,~M.~I.; Lichtenstein,~A.~I. Adhesion and
  Electronic Structure of Graphene on Hexagonal Boron Nitride Substrates.
  \emph{Phys. Rev. B} \textbf{2011}, \emph{84}, 195414\relax
\mciteBstWouldAddEndPuncttrue
\mciteSetBstMidEndSepPunct{\mcitedefaultmidpunct}
{\mcitedefaultendpunct}{\mcitedefaultseppunct}\relax
\EndOfBibitem
\bibitem[Yankowitz \latin{et~al.}(2012)Yankowitz, Xue, Cormode,
  Sanchez-Yamagishi, Watanabe, Taniguchi, Jarillo-Herrero, Jacquod, and
  LeRoy]{Yankowitz2012}
Yankowitz,~M.; Xue,~J.; Cormode,~D.; Sanchez-Yamagishi,~J.~D.; Watanabe,~K.;
  Taniguchi,~T.; Jarillo-Herrero,~P.; Jacquod,~P.; LeRoy,~B.~J. Emergence of
  Superlattice {D}irac Points in Graphene on Hexagonal Boron Nitride.
  \emph{Nat. Phys.} \textbf{2012}, \emph{8}, 382\relax
\mciteBstWouldAddEndPuncttrue
\mciteSetBstMidEndSepPunct{\mcitedefaultmidpunct}
{\mcitedefaultendpunct}{\mcitedefaultseppunct}\relax
\EndOfBibitem
\bibitem[Bhimanapati \latin{et~al.}(2016)Bhimanapati, Glavin, and
  Robinson]{Bhimanapati2016}
Bhimanapati,~G.; Glavin,~N.; Robinson,~J. In \emph{2D Materials}; Iacopi,~F.,
  Boeckl,~J.~J., Jagadish,~C., Eds.; Semiconductors and Semimetals; Elsevier,
  2016; Vol.~95; Chapter 3, pp 101 -- 147\relax
\mciteBstWouldAddEndPuncttrue
\mciteSetBstMidEndSepPunct{\mcitedefaultmidpunct}
{\mcitedefaultendpunct}{\mcitedefaultseppunct}\relax
\EndOfBibitem
\bibitem[Kelly(1981)]{Kelly1981}
Kelly,~B.~T. \emph{Physics of Graphite}; Applied Science: United Kingdom,
  1981\relax
\mciteBstWouldAddEndPuncttrue
\mciteSetBstMidEndSepPunct{\mcitedefaultmidpunct}
{\mcitedefaultendpunct}{\mcitedefaultseppunct}\relax
\EndOfBibitem
\bibitem[Dresselhaus \latin{et~al.}(1988)Dresselhaus, Dresselhaus, Sugihara,
  Spain, and Goldberg]{Dresselhaus1988}
Dresselhaus,~M.~S.; Dresselhaus,~G.; Sugihara,~K.; Spain,~I.~L.;
  Goldberg,~H.~A. \emph{Graphite Fibers and Filaments}; Springer-Verlag,
  1988\relax
\mciteBstWouldAddEndPuncttrue
\mciteSetBstMidEndSepPunct{\mcitedefaultmidpunct}
{\mcitedefaultendpunct}{\mcitedefaultseppunct}\relax
\EndOfBibitem
\bibitem[Wang \latin{et~al.}(2019)Wang, Zihlmann, Liu, Makk, Watanabe,
  Taniguchi, Baumgartner, and Sch\"{o}nenberger]{Wang2019}
Wang,~L.; Zihlmann,~S.; Liu,~M.-H.; Makk,~P.; Watanabe,~K.; Taniguchi,~T.;
  Baumgartner,~A.; Sch\"{o}nenberger,~C. New Generation of Moir\'{e}
  Superlattices in Doubly Aligned h{BN}/Graphene/h{BN} Heterostructures.
  \emph{Nano Lett.} \textbf{2019}, \emph{19}, 2371--2376\relax
\mciteBstWouldAddEndPuncttrue
\mciteSetBstMidEndSepPunct{\mcitedefaultmidpunct}
{\mcitedefaultendpunct}{\mcitedefaultseppunct}\relax
\EndOfBibitem
\bibitem[Finney \latin{et~al.}(2019)Finney, Yankowitz, Muraleetharan, Watanabe,
  Taniguchi, Dean, and Hone]{finney_tunable_2019}
Finney,~N.~R.; Yankowitz,~M.; Muraleetharan,~L.; Watanabe,~K.; Taniguchi,~T.;
  Dean,~C.~R.; Hone,~J. Tunable Crystal Symmetry in Graphene-Boron Nitride
  Heterostructures with Coexisting Moir\'{e} Superlattices. \emph{Nat.
  Nanotechnol.} \textbf{2019}, \emph{14}, 1029--1034\relax
\mciteBstWouldAddEndPuncttrue
\mciteSetBstMidEndSepPunct{\mcitedefaultmidpunct}
{\mcitedefaultendpunct}{\mcitedefaultseppunct}\relax
\EndOfBibitem
\bibitem[Wang \latin{et~al.}(2019)Wang, Wang, Yin, T{\'o}v{\'a}ri, Yang, Lin,
  Holwill, Birkbeck, Perello, Xu, Zultak, Gorbachev, Kretinin, Taniguchi,
  Watanabe, Morozov, An{\dj}elkovi{\'c}, Milovanovi{\'c}, Covaci, Peeters,
  Mishchenko, Geim, Novoselov, Fal'ko, Knothe, and Woods]{Wang2019a}
Wang,~Z. \latin{et~al.}  Composite Super-Moir\'e Lattices in Double-Aligned
  Graphene Heterostructures. \emph{Sci. Adv.} \textbf{2019}, \emph{5},
  eaay8897\relax
\mciteBstWouldAddEndPuncttrue
\mciteSetBstMidEndSepPunct{\mcitedefaultmidpunct}
{\mcitedefaultendpunct}{\mcitedefaultseppunct}\relax
\EndOfBibitem
\bibitem[Zhu \latin{et~al.}(2020)Zhu, Cazeaux, Luskin, and
  Kaxiras]{PhysRevB.101.224107}
Zhu,~Z.; Cazeaux,~P.; Luskin,~M.; Kaxiras,~E. Modeling Mechanical Relaxation in
  Incommensurate Trilayer van der {W}aals Heterostructures. \emph{Phys. Rev. B}
  \textbf{2020}, \emph{101}, 224107\relax
\mciteBstWouldAddEndPuncttrue
\mciteSetBstMidEndSepPunct{\mcitedefaultmidpunct}
{\mcitedefaultendpunct}{\mcitedefaultseppunct}\relax
\EndOfBibitem
\bibitem[Yang \latin{et~al.}(2020)Yang, Li, Yin, Xu, Mullan, Taniguchi,
  Watanabe, Geim, Novoselov, and Mishchenko]{Yang_2020}
Yang,~Y.; Li,~J.; Yin,~J.; Xu,~S.; Mullan,~C.; Taniguchi,~T.; Watanabe,~K.;
  Geim,~A.~K.; Novoselov,~K.~S.; Mishchenko,~A. \textit{In Situ} Manipulation
  of van der {W}aals Heterostructures for Twistronics. \emph{Sci. Adv.}
  \textbf{2020}, \emph{6}, eabd3655\relax
\mciteBstWouldAddEndPuncttrue
\mciteSetBstMidEndSepPunct{\mcitedefaultmidpunct}
{\mcitedefaultendpunct}{\mcitedefaultseppunct}\relax
\EndOfBibitem
\bibitem[An{\dj}elkovi{\'{c}} \latin{et~al.}(2020)An{\dj}elkovi{\'{c}},
  Milovanovi{\'{c}}, Covaci, and Peeters]{Andelkovic_2020}
An{\dj}elkovi{\'{c}},~M.; Milovanovi{\'{c}},~S.~P.; Covaci,~L.; Peeters,~F.~M.
  Double Moir{\'e} with a Twist: Supermoir{\'e} in Encapsulated Graphene.
  \emph{Nano Lett.} \textbf{2020}, \emph{20}, 979--988\relax
\mciteBstWouldAddEndPuncttrue
\mciteSetBstMidEndSepPunct{\mcitedefaultmidpunct}
{\mcitedefaultendpunct}{\mcitedefaultseppunct}\relax
\EndOfBibitem
\bibitem[Sun \latin{et~al.}(2021)Sun, Zhang, Liu, Zhu, Huang, Yuan, Wang,
  Watanabe, Taniguchi, Li, Zhu, Mao, Yang, Kang, Liu, Ye, Han, and
  Zhang]{Sun_2021}
Sun,~X. \latin{et~al.}  Correlated States in Doubly-Aligned
  h{BN}/Graphene/h{BN} Heterostructures. \emph{Nat. Commun.} \textbf{2021},
  \emph{12}, 7196\relax
\mciteBstWouldAddEndPuncttrue
\mciteSetBstMidEndSepPunct{\mcitedefaultmidpunct}
{\mcitedefaultendpunct}{\mcitedefaultseppunct}\relax
\EndOfBibitem
\bibitem[Moulsdale \latin{et~al.}(2022)Moulsdale, Knothe, and
  Fal'ko]{Moulsdale_2022}
Moulsdale,~C.; Knothe,~A.; Fal'ko,~V. Kagome Network of Miniband-Edge States in
  Double-Aligned Graphene-Hexagonal Boron Nitride Structures. \emph{Phys. Rev.
  B} \textbf{2022}, \emph{105}, L201112\relax
\mciteBstWouldAddEndPuncttrue
\mciteSetBstMidEndSepPunct{\mcitedefaultmidpunct}
{\mcitedefaultendpunct}{\mcitedefaultseppunct}\relax
\EndOfBibitem
\bibitem[Wang \latin{et~al.}(2022)Wang, Jiang, Xiao, Cai, Zhang, Wang, Ma, Han,
  Huang, Watanabe, Taniguchi, Guo, Wang, Mayorov, and Yu]{YWang2022}
Wang,~Y.; Jiang,~S.; Xiao,~J.; Cai,~X.; Zhang,~D.; Wang,~P.; Ma,~G.; Han,~Y.;
  Huang,~J.; Watanabe,~K.; Taniguchi,~T.; Guo,~Y.; Wang,~L.; Mayorov,~A.~S.;
  Yu,~G. Ferroelectricity in h{BN} Intercalated Double-Layer Graphene.
  \emph{Front. Phys.} \textbf{2022}, \emph{17}, 43504\relax
\mciteBstWouldAddEndPuncttrue
\mciteSetBstMidEndSepPunct{\mcitedefaultmidpunct}
{\mcitedefaultendpunct}{\mcitedefaultseppunct}\relax
\EndOfBibitem
\bibitem[Chen \latin{et~al.}(2023)Chen, Zollner, Moulsdale, Fal'ko, and
  Knothe]{Chen_2023}
Chen,~X.; Zollner,~K.; Moulsdale,~C.; Fal'ko,~V.~I.; Knothe,~A. Semimetallic
  and Semiconducting Graphene-h{BN} Multilayers with Parallel or Reverse
  Stacking. \emph{Phys. Rev. B} \textbf{2023}, \emph{107}, 125402\relax
\mciteBstWouldAddEndPuncttrue
\mciteSetBstMidEndSepPunct{\mcitedefaultmidpunct}
{\mcitedefaultendpunct}{\mcitedefaultseppunct}\relax
\EndOfBibitem
\bibitem[Hu \latin{et~al.}(2023)Hu, Tan, Al~Ezzi, Chattopadhyay, Gou, Zheng,
  Wang, Chen, Thottathil, Luo, Watanabe, Taniguchi, Wee, Adam, and
  Ariando]{Hu_2023}
Hu,~J.; Tan,~J.; Al~Ezzi,~M.~M.; Chattopadhyay,~U.; Gou,~J.; Zheng,~Y.;
  Wang,~Z.; Chen,~J.; Thottathil,~R.; Luo,~J.; Watanabe,~K.; Taniguchi,~T.;
  Wee,~A. T.~S.; Adam,~S.; Ariando,~A. Controlled Alignment of Supermoir{\'e}
  Lattice in Double-Aligned Graphene Heterostructures. \emph{Nat. Commun.}
  \textbf{2023}, \emph{14}, 4142\relax
\mciteBstWouldAddEndPuncttrue
\mciteSetBstMidEndSepPunct{\mcitedefaultmidpunct}
{\mcitedefaultendpunct}{\mcitedefaultseppunct}\relax
\EndOfBibitem
\bibitem[Ceperley and Alder(1980)Ceperley, and Alder]{Ceperley1980}
Ceperley,~D.~M.; Alder,~B.~J. Ground State of the Electron Gas by a Stochastic
  Method. \emph{Phys. Rev. Lett.} \textbf{1980}, \emph{45}, 566--569\relax
\mciteBstWouldAddEndPuncttrue
\mciteSetBstMidEndSepPunct{\mcitedefaultmidpunct}
{\mcitedefaultendpunct}{\mcitedefaultseppunct}\relax
\EndOfBibitem
\bibitem[Foulkes \latin{et~al.}(2001)Foulkes, Mitas, Needs, and
  Rajagopal]{Foulkes2001}
Foulkes,~W. M.~C.; Mitas,~L.; Needs,~R.~J.; Rajagopal,~G. Quantum {M}onte
  {C}arlo Simulations of Solids. \emph{Rev. Mod. Phys.} \textbf{2001},
  \emph{73}, 33--83\relax
\mciteBstWouldAddEndPuncttrue
\mciteSetBstMidEndSepPunct{\mcitedefaultmidpunct}
{\mcitedefaultendpunct}{\mcitedefaultseppunct}\relax
\EndOfBibitem
\bibitem[Needs \latin{et~al.}(2020)Needs, Towler, Drummond,
  L\'{o}pez~R\'{\i}os, and Trail]{Needs2020}
Needs,~R.~J.; Towler,~M.~D.; Drummond,~N.~D.; L\'{o}pez~R\'{\i}os,~P.;
  Trail,~J.~R. Variational and Diffusion Quantum {M}onte {C}arlo Calculations
  with the {CASINO} Code. \emph{J. Chem. Phys.} \textbf{2020}, \emph{152},
  154106\relax
\mciteBstWouldAddEndPuncttrue
\mciteSetBstMidEndSepPunct{\mcitedefaultmidpunct}
{\mcitedefaultendpunct}{\mcitedefaultseppunct}\relax
\EndOfBibitem
\bibitem[Mostaani \latin{et~al.}(2015)Mostaani, Drummond, and
  Fal'ko]{Mostaani2015}
Mostaani,~E.; Drummond,~N.~D.; Fal'ko,~V.~I. Quantum {M}onte {C}arlo
  Calculation of the Binding Energy of Bilayer Graphene. \emph{Phys. Rev.
  Lett.} \textbf{2015}, \emph{115}, 115501\relax
\mciteBstWouldAddEndPuncttrue
\mciteSetBstMidEndSepPunct{\mcitedefaultmidpunct}
{\mcitedefaultendpunct}{\mcitedefaultseppunct}\relax
\EndOfBibitem
\bibitem[Giovannetti \latin{et~al.}(2007)Giovannetti, Khomyakov, Brocks, Kelly,
  and van~den Brink]{Giovannetti2007}
Giovannetti,~G.; Khomyakov,~P.~A.; Brocks,~G.; Kelly,~P.~J.; van~den Brink,~J.
  Substrate-Induced Band Gap in Graphene on Hexagonal Boron Nitride: \textit{Ab
  Initio} Density Functional Calculations. \emph{Phys. Rev. B} \textbf{2007},
  \emph{76}, 073103\relax
\mciteBstWouldAddEndPuncttrue
\mciteSetBstMidEndSepPunct{\mcitedefaultmidpunct}
{\mcitedefaultendpunct}{\mcitedefaultseppunct}\relax
\EndOfBibitem
\bibitem[Fan \latin{et~al.}(2011)Fan, Zhao, Wang, Zhang, and Zhang]{Fan2011}
Fan,~Y.; Zhao,~M.; Wang,~Z.; Zhang,~X.; Zhang,~H. Tunable Electronic Structures
  of Graphene/Boron Nitride Heterobilayers. \emph{Appl. Phys. Lett.}
  \textbf{2011}, \emph{98}, 083103\relax
\mciteBstWouldAddEndPuncttrue
\mciteSetBstMidEndSepPunct{\mcitedefaultmidpunct}
{\mcitedefaultendpunct}{\mcitedefaultseppunct}\relax
\EndOfBibitem
\bibitem[Leconte \latin{et~al.}(2017)Leconte, Jung, Leb\`egue, and
  Gould]{Leconte2017}
Leconte,~N.; Jung,~J.; Leb\`egue,~S.; Gould,~T. Moir\'e-Pattern Interlayer
  Potentials in van der {W}aals Materials in the Random-Phase Approximation.
  \emph{Phys. Rev. B} \textbf{2017}, \emph{96}, 195431\relax
\mciteBstWouldAddEndPuncttrue
\mciteSetBstMidEndSepPunct{\mcitedefaultmidpunct}
{\mcitedefaultendpunct}{\mcitedefaultseppunct}\relax
\EndOfBibitem
\bibitem[Lebedev \latin{et~al.}(2017)Lebedev, Lebedeva, Popov, and
  Knizhnik]{Lebedev2017}
Lebedev,~A.~V.; Lebedeva,~I.~V.; Popov,~A.~M.; Knizhnik,~A.~A. Stacking in
  Incommensurate Graphene/Hexagonal-Boron-Nitride Heterostructures Based on
  \textit{Ab Initio} Study of Interlayer Interaction. \emph{Phys. Rev. B}
  \textbf{2017}, \emph{96}, 085432\relax
\mciteBstWouldAddEndPuncttrue
\mciteSetBstMidEndSepPunct{\mcitedefaultmidpunct}
{\mcitedefaultendpunct}{\mcitedefaultseppunct}\relax
\EndOfBibitem
\bibitem[Chakarova-K\"ack \latin{et~al.}(2006)Chakarova-K\"ack, Schr\"oder,
  Lundqvist, and Langreth]{Chakarova2006}
Chakarova-K\"ack,~S.~D.; Schr\"oder,~E.; Lundqvist,~B.~I.; Langreth,~D.~C.
  Application of van der {W}aals Density Functional to an Extended System:
  Adsorption of Benzene and Naphthalene on Graphite. \emph{Phys. Rev. Lett.}
  \textbf{2006}, \emph{96}, 146107\relax
\mciteBstWouldAddEndPuncttrue
\mciteSetBstMidEndSepPunct{\mcitedefaultmidpunct}
{\mcitedefaultendpunct}{\mcitedefaultseppunct}\relax
\EndOfBibitem
\bibitem[Lebedeva \latin{et~al.}(2011)Lebedeva, Knizhnik, Popov, Lozovik, and
  Potapkin]{Lebedeva2011}
Lebedeva,~I.~V.; Knizhnik,~A.~A.; Popov,~A.~M.; Lozovik,~Y.~E.; Potapkin,~B.~V.
  Interlayer Interaction and Relative Vibrations of Bilayer Graphene.
  \emph{Phys. Chem. Chem. Phys.} \textbf{2011}, \emph{13}, 5687--5695\relax
\mciteBstWouldAddEndPuncttrue
\mciteSetBstMidEndSepPunct{\mcitedefaultmidpunct}
{\mcitedefaultendpunct}{\mcitedefaultseppunct}\relax
\EndOfBibitem
\bibitem[Dappe \latin{et~al.}(2012)Dappe, Bolcatto, Ortega, and
  Flores]{Dappe2012}
Dappe,~Y.~J.; Bolcatto,~P.~G.; Ortega,~J.; Flores,~F. Dynamical Screening of
  the van der {W}aals Interaction between Graphene Layers. \emph{J. Phys.
  Condens. Mater.} \textbf{2012}, \emph{24}, 424208\relax
\mciteBstWouldAddEndPuncttrue
\mciteSetBstMidEndSepPunct{\mcitedefaultmidpunct}
{\mcitedefaultendpunct}{\mcitedefaultseppunct}\relax
\EndOfBibitem
\bibitem[Gould \latin{et~al.}(2013)Gould, Leb{\`{e}}gue, and Dobson]{Gould2013}
Gould,~T.; Leb{\`{e}}gue,~S.; Dobson,~J.~F. Dispersion Corrections in Graphenic
  Systems: a Simple and Effective Model of Binding. \emph{J. Phys. Condens.
  Mater.} \textbf{2013}, \emph{25}, 445010\relax
\mciteBstWouldAddEndPuncttrue
\mciteSetBstMidEndSepPunct{\mcitedefaultmidpunct}
{\mcitedefaultendpunct}{\mcitedefaultseppunct}\relax
\EndOfBibitem
\bibitem[Podeszwa(2010)]{Podeszwa2010}
Podeszwa,~R. Interactions of Graphene Sheets Deduced from Properties of
  Polycyclic Aromatic Hydrocarbons. \emph{J. Chem. Phys.} \textbf{2010},
  \emph{132}, 044704\relax
\mciteBstWouldAddEndPuncttrue
\mciteSetBstMidEndSepPunct{\mcitedefaultmidpunct}
{\mcitedefaultendpunct}{\mcitedefaultseppunct}\relax
\EndOfBibitem
\bibitem[Zhao and Liu(2018)Zhao, and Liu]{Zhao2018}
Zhao,~Z.-Y.; Liu,~Q.-L. Study of the Layer-Dependent Properties of {M}o{S}$_2$
  Nanosheets with Different Crystal Structures by {DFT} Calculations.
  \emph{Catal. Sci. Technol.} \textbf{2018}, \emph{8}, 1867--1879\relax
\mciteBstWouldAddEndPuncttrue
\mciteSetBstMidEndSepPunct{\mcitedefaultmidpunct}
{\mcitedefaultendpunct}{\mcitedefaultseppunct}\relax
\EndOfBibitem
\bibitem[San-Jose \latin{et~al.}(2014)San-Jose, Guti\'errez-Rubio, Sturla, and
  Guinea]{SanJose2014}
San-Jose,~P.; Guti\'errez-Rubio,~A.; Sturla,~M.; Guinea,~F. Spontaneous Strains
  and Gap in Graphene on Boron Nitride. \emph{Phys. Rev. B} \textbf{2014},
  \emph{90}, 075428\relax
\mciteBstWouldAddEndPuncttrue
\mciteSetBstMidEndSepPunct{\mcitedefaultmidpunct}
{\mcitedefaultendpunct}{\mcitedefaultseppunct}\relax
\EndOfBibitem
\bibitem[Dobson \latin{et~al.}(2014)Dobson, Gould, and Vignale]{Dobson2104}
Dobson,~J.~F.; Gould,~T.; Vignale,~G. How Many-Body Effects Modify the van der
  {W}aals Interaction between Graphene Sheets. \emph{Phys. Rev. X}
  \textbf{2014}, \emph{4}, 021040\relax
\mciteBstWouldAddEndPuncttrue
\mciteSetBstMidEndSepPunct{\mcitedefaultmidpunct}
{\mcitedefaultendpunct}{\mcitedefaultseppunct}\relax
\EndOfBibitem
\bibitem[Slotman \latin{et~al.}(2014)Slotman, de~Wijs, Fasolino, and
  Katsnelson]{Slotman2014}
Slotman,~G.~J.; de~Wijs,~G.~A.; Fasolino,~A.; Katsnelson,~M.~I. Phonons and
  Electron-Phonon Coupling in Graphene-h-{BN} Heterostructures. \emph{Ann.
  Phys. (Berl.)} \textbf{2014}, \emph{526}, 381--386\relax
\mciteBstWouldAddEndPuncttrue
\mciteSetBstMidEndSepPunct{\mcitedefaultmidpunct}
{\mcitedefaultendpunct}{\mcitedefaultseppunct}\relax
\EndOfBibitem
\bibitem[E.(1924)]{Jones1924}
E.,~J.~J. The Determination of Molecular Fields. {II}. From the Equation of
  State of a Gas. \emph{Proc. R. Soc. A} \textbf{1924}, \emph{106},
  463--477\relax
\mciteBstWouldAddEndPuncttrue
\mciteSetBstMidEndSepPunct{\mcitedefaultmidpunct}
{\mcitedefaultendpunct}{\mcitedefaultseppunct}\relax
\EndOfBibitem
\bibitem[Neek-Amal and Peeters(2014)Neek-Amal, and Peeters]{Neek2014}
Neek-Amal,~M.; Peeters,~F.~M. Graphene on Boron-Nitride: Moir\'{e} Pattern in
  the van der {W}aals Energy. \emph{Appl. Phys. Lett.} \textbf{2014},
  \emph{104}, 041909\relax
\mciteBstWouldAddEndPuncttrue
\mciteSetBstMidEndSepPunct{\mcitedefaultmidpunct}
{\mcitedefaultendpunct}{\mcitedefaultseppunct}\relax
\EndOfBibitem
\bibitem[Zhang \latin{et~al.}(2015)Zhang, Hong, and Yue]{Zhang2015}
Zhang,~J.; Hong,~Y.; Yue,~Y. Thermal Transport across Graphene and Single Layer
  Hexagonal Boron Nitride. \emph{J. Appl. Phys.} \textbf{2015}, \emph{117},
  134307\relax
\mciteBstWouldAddEndPuncttrue
\mciteSetBstMidEndSepPunct{\mcitedefaultmidpunct}
{\mcitedefaultendpunct}{\mcitedefaultseppunct}\relax
\EndOfBibitem
\bibitem[Yang \latin{et~al.}(2018)Yang, Zhang, Zhang, and Zeng]{Yang2018}
Yang,~H.; Zhang,~Z.; Zhang,~J.; Zeng,~X.~C. Machine Learning and Artificial
  Neural Network Prediction of Interfacial Thermal Resistance between Graphene
  and Hexagonal Boron Nitride. \emph{Nanoscale} \textbf{2018}, \emph{10},
  19092--19099\relax
\mciteBstWouldAddEndPuncttrue
\mciteSetBstMidEndSepPunct{\mcitedefaultmidpunct}
{\mcitedefaultendpunct}{\mcitedefaultseppunct}\relax
\EndOfBibitem
\bibitem[Nguyen(2019)]{Nguyen2019}
Nguyen,~H. T.~T. Graphene Layer of Hybrid Graphene/Hexagonal Boron Nitride
  Model Upon Heating. \emph{Carbon Lett.} \textbf{2019}, \emph{29},
  521--528\relax
\mciteBstWouldAddEndPuncttrue
\mciteSetBstMidEndSepPunct{\mcitedefaultmidpunct}
{\mcitedefaultendpunct}{\mcitedefaultseppunct}\relax
\EndOfBibitem
\bibitem[Zhang \latin{et~al.}(2017)Zhang, Hu, Chen, and Li]{Zhang2017}
Zhang,~Z.; Hu,~S.; Chen,~J.; Li,~B. Hexagonal Boron Nitride: A Promising
  Substrate for Graphene with High Heat Dissipation. \emph{Nanotechnology}
  \textbf{2017}, \emph{28}, 225704\relax
\mciteBstWouldAddEndPuncttrue
\mciteSetBstMidEndSepPunct{\mcitedefaultmidpunct}
{\mcitedefaultendpunct}{\mcitedefaultseppunct}\relax
\EndOfBibitem
\bibitem[Kang and Hwang(2004)Kang, and Hwang]{Kang2004}
Kang,~J.~W.; Hwang,~H.~J. Comparison of {C$_{60}$} Encapsulations into Carbon
  and Boron Nitride Nanotubes. \emph{J. Phys. Condens. Mater.} \textbf{2004},
  \emph{16}, 3901--3908\relax
\mciteBstWouldAddEndPuncttrue
\mciteSetBstMidEndSepPunct{\mcitedefaultmidpunct}
{\mcitedefaultendpunct}{\mcitedefaultseppunct}\relax
\EndOfBibitem
\bibitem[Clark \latin{et~al.}(2005)Clark, Segall, Pickard, Hasnip, Probert,
  Refson, and Payne~Mike]{Clark2005}
Clark,~S.~J.; Segall,~M.~D.; Pickard,~C.~J.; Hasnip,~P.~J.; Probert,~M. I.~J.;
  Refson,~K.; Payne~Mike,~C. First Principles Methods Using {CASTEP}. \emph{Z.
  Kristallogr. Cryst. Mater.} \textbf{2005}, \emph{220}, 567--570\relax
\mciteBstWouldAddEndPuncttrue
\mciteSetBstMidEndSepPunct{\mcitedefaultmidpunct}
{\mcitedefaultendpunct}{\mcitedefaultseppunct}\relax
\EndOfBibitem
\bibitem[Vanderbilt(1990)]{Vanderbilt1990}
Vanderbilt,~D. Soft Self-Consistent Pseudopotentials in a Generalized
  Eigenvalue Formalism. \emph{Phys. Rev. B} \textbf{1990}, \emph{41},
  7892--7895\relax
\mciteBstWouldAddEndPuncttrue
\mciteSetBstMidEndSepPunct{\mcitedefaultmidpunct}
{\mcitedefaultendpunct}{\mcitedefaultseppunct}\relax
\EndOfBibitem
\bibitem[Grimme(2006)]{Grimme2006}
Grimme,~S. Semiempirical {GGA}-Type Density Functional Constructed with a
  Long-Range Dispersion Correction. \emph{J. Comp. Chem.} \textbf{2006},
  \emph{27}, 1787--1799\relax
\mciteBstWouldAddEndPuncttrue
\mciteSetBstMidEndSepPunct{\mcitedefaultmidpunct}
{\mcitedefaultendpunct}{\mcitedefaultseppunct}\relax
\EndOfBibitem
\bibitem[Perdew \latin{et~al.}(1996)Perdew, Burke, and Ernzerhof]{Perdew1996}
Perdew,~J.~P.; Burke,~K.; Ernzerhof,~M. Generalized Gradient Approximation Made
  Simple. \emph{Phys. Rev. Lett.} \textbf{1996}, \emph{77}, 3865--3868\relax
\mciteBstWouldAddEndPuncttrue
\mciteSetBstMidEndSepPunct{\mcitedefaultmidpunct}
{\mcitedefaultendpunct}{\mcitedefaultseppunct}\relax
\EndOfBibitem
\bibitem[Trail and Needs(2005)Trail, and Needs]{Trail2005a}
Trail,~J.~R.; Needs,~R.~J. Norm-Conserving {H}artree-{F}ock Pseudopotentials
  and Their Asymptotic Behavior. \emph{J. Chem. Phys.} \textbf{2005},
  \emph{122}, 014112\relax
\mciteBstWouldAddEndPuncttrue
\mciteSetBstMidEndSepPunct{\mcitedefaultmidpunct}
{\mcitedefaultendpunct}{\mcitedefaultseppunct}\relax
\EndOfBibitem
\bibitem[Trail and Needs(2005)Trail, and Needs]{Trail2005b}
Trail,~J.~R.; Needs,~R.~J. Smooth Relativistic {H}artree-{F}ock
  Pseudopotentials for {H} to {B}a and {L}u to {H}g. \emph{J. Chem. Phys.}
  \textbf{2005}, \emph{122}, 174109\relax
\mciteBstWouldAddEndPuncttrue
\mciteSetBstMidEndSepPunct{\mcitedefaultmidpunct}
{\mcitedefaultendpunct}{\mcitedefaultseppunct}\relax
\EndOfBibitem
\bibitem[Mit\'{a}\u{s} \latin{et~al.}(1991)Mit\'{a}\u{s}, Shirley, and
  Ceperley]{Mitas1991}
Mit\'{a}\u{s},~L.; Shirley,~E.~L.; Ceperley,~D.~M. Nonlocal Pseudopotentials
  and Diffusion {M}onte {C}arlo. \emph{J. Chem. Phys.} \textbf{1991},
  \emph{95}, 3467--3475\relax
\mciteBstWouldAddEndPuncttrue
\mciteSetBstMidEndSepPunct{\mcitedefaultmidpunct}
{\mcitedefaultendpunct}{\mcitedefaultseppunct}\relax
\EndOfBibitem
\bibitem[Drummond \latin{et~al.}(2004)Drummond, Towler, and
  Needs]{Drummond2004}
Drummond,~N.~D.; Towler,~M.~D.; Needs,~R.~J. {J}astrow Correlation Factor for
  Atoms, Molecules, and Solids. \emph{Phys. Rev. B} \textbf{2004}, \emph{70},
  235119\relax
\mciteBstWouldAddEndPuncttrue
\mciteSetBstMidEndSepPunct{\mcitedefaultmidpunct}
{\mcitedefaultendpunct}{\mcitedefaultseppunct}\relax
\EndOfBibitem
\bibitem[{K}ohn and {S}ham(1965){K}ohn, and {S}ham]{KohnSham1965}
{K}ohn,~W.; {S}ham,~L.~J. Self-Consistent Equations Including Exchange and
  Correlation Effects. \emph{Phys. Rev.} \textbf{1965}, \emph{140},
  A1133--A1138\relax
\mciteBstWouldAddEndPuncttrue
\mciteSetBstMidEndSepPunct{\mcitedefaultmidpunct}
{\mcitedefaultendpunct}{\mcitedefaultseppunct}\relax
\EndOfBibitem
\bibitem[Alf\`e and Gillan(2004)Alf\`e, and Gillan]{Alfe2004}
Alf\`e,~D.; Gillan,~M.~J. Efficient Localized Basis Set for Quantum {M}onte
  {C}arlo Calculations on Condensed Matter. \emph{Phys. Rev. B} \textbf{2004},
  \emph{70}, 161101\relax
\mciteBstWouldAddEndPuncttrue
\mciteSetBstMidEndSepPunct{\mcitedefaultmidpunct}
{\mcitedefaultendpunct}{\mcitedefaultseppunct}\relax
\EndOfBibitem
\bibitem[Umrigar \latin{et~al.}(1988)Umrigar, Wilson, and Wilkins]{Umrigar1988}
Umrigar,~C.~J.; Wilson,~K.~G.; Wilkins,~J.~W. Optimized Trial Wave Functions
  for Quantum {M}onte {C}arlo Calculations. \emph{Phys. Rev. Lett.}
  \textbf{1988}, \emph{60}, 1719--1722\relax
\mciteBstWouldAddEndPuncttrue
\mciteSetBstMidEndSepPunct{\mcitedefaultmidpunct}
{\mcitedefaultendpunct}{\mcitedefaultseppunct}\relax
\EndOfBibitem
\bibitem[Drummond and Needs(2005)Drummond, and Needs]{Drummond2005}
Drummond,~N.~D.; Needs,~R.~J. Variance-Minimization Scheme for Optimizing
  {J}astrow Factors. \emph{Phys. Rev. B} \textbf{2005}, \emph{72}, 085124\relax
\mciteBstWouldAddEndPuncttrue
\mciteSetBstMidEndSepPunct{\mcitedefaultmidpunct}
{\mcitedefaultendpunct}{\mcitedefaultseppunct}\relax
\EndOfBibitem
\bibitem[Parry(1975)]{Parry1975}
Parry,~D. The Electrostatic Potential in the Surface Region of an Ionic
  Crystal. \emph{Surf. Sci.} \textbf{1975}, \emph{49}, 433--440\relax
\mciteBstWouldAddEndPuncttrue
\mciteSetBstMidEndSepPunct{\mcitedefaultmidpunct}
{\mcitedefaultendpunct}{\mcitedefaultseppunct}\relax
\EndOfBibitem
\bibitem[Parry(1976)]{Parry1976}
Parry,~D. Erratum: The Electrostatic Potential in the Surface Region of an
  Ionic Crystal. \emph{Surf. Sci.} \textbf{1976}, \emph{54}, 195\relax
\mciteBstWouldAddEndPuncttrue
\mciteSetBstMidEndSepPunct{\mcitedefaultmidpunct}
{\mcitedefaultendpunct}{\mcitedefaultseppunct}\relax
\EndOfBibitem
\bibitem[Reynolds \latin{et~al.}(1982)Reynolds, Ceperley, Alder, and
  Lester]{Reynolds_1982}
Reynolds,~P.~J.; Ceperley,~D.~M.; Alder,~B.~J.; Lester,~J.,~William~A.
  Fixed-Node Quantum {M}onte {C}arlo for Molecules. \emph{J. Chem. Phys.}
  \textbf{1982}, \emph{77}, 5593--5603\relax
\mciteBstWouldAddEndPuncttrue
\mciteSetBstMidEndSepPunct{\mcitedefaultmidpunct}
{\mcitedefaultendpunct}{\mcitedefaultseppunct}\relax
\EndOfBibitem
\bibitem[Lin \latin{et~al.}(2001)Lin, Zong, and Ceperley]{Lin2001}
Lin,~C.; Zong,~F.~H.; Ceperley,~D.~M. Twist-Averaged Boundary Conditions in
  Continuum Quantum {M}onte {C}arlo Algorithms. \emph{Phys. Rev. E}
  \textbf{2001}, \emph{64}, 016702\relax
\mciteBstWouldAddEndPuncttrue
\mciteSetBstMidEndSepPunct{\mcitedefaultmidpunct}
{\mcitedefaultendpunct}{\mcitedefaultseppunct}\relax
\EndOfBibitem
\bibitem[Drummond \latin{et~al.}(2008)Drummond, Needs, Sorouri, and
  Foulkes]{Drummond2008}
Drummond,~N.~D.; Needs,~R.~J.; Sorouri,~A.; Foulkes,~W. M.~C. Finite-Size
  Errors in Continuum Quantum {M}onte {C}arlo Calculations. \emph{Phys. Rev. B}
  \textbf{2008}, \emph{78}, 125106\relax
\mciteBstWouldAddEndPuncttrue
\mciteSetBstMidEndSepPunct{\mcitedefaultmidpunct}
{\mcitedefaultendpunct}{\mcitedefaultseppunct}\relax
\EndOfBibitem
\end{mcitethebibliography}


\providecommand{\latin}[1]{#1}
\makeatletter
\providecommand{\doi}
  {\begingroup\let\do\@makeother\dospecials
  \catcode`\{=1 \catcode`\}=2 \doi@aux}
\providecommand{\doi@aux}[1]{\endgroup\texttt{#1}}
\makeatother
\providecommand*\mcitethebibliography{\thebibliography}
\csname @ifundefined\endcsname{endmcitethebibliography}
  {\let\endmcitethebibliography\endthebibliography}{}
\begin{mcitethebibliography}{26}
\providecommand*\natexlab[1]{#1}
\providecommand*\mciteSetBstSublistMode[1]{}
\providecommand*\mciteSetBstMaxWidthForm[2]{}
\providecommand*\mciteBstWouldAddEndPuncttrue
  {\def\EndOfBibitem{\unskip.}}
\providecommand*\mciteBstWouldAddEndPunctfalse
  {\let\EndOfBibitem\relax}
\providecommand*\mciteSetBstMidEndSepPunct[3]{}
\providecommand*\mciteSetBstSublistLabelBeginEnd[3]{}
\providecommand*\EndOfBibitem{}
\mciteSetBstSublistMode{f}
\mciteSetBstMaxWidthForm{subitem}{(\alph{mcitesubitemcount})}
\mciteSetBstSublistLabelBeginEnd
  {\mcitemaxwidthsubitemform\space}
  {\relax}
  {\relax}

\bibitem[Efron and Tibshirani(1994)Efron, and Tibshirani]{Efron1993}
Efron,~B.; Tibshirani,~R.~J. \emph{An Introduction to the Bootstrap}; Chapman
  and Hall/CRC, 1994\relax
\mciteBstWouldAddEndPuncttrue
\mciteSetBstMidEndSepPunct{\mcitedefaultmidpunct}
{\mcitedefaultendpunct}{\mcitedefaultseppunct}\relax
\EndOfBibitem
\bibitem[Mostaani \latin{et~al.}(2015)Mostaani, Drummond, and
  Fal'ko]{Mostaani2015}
Mostaani,~E.; Drummond,~N.~D.; Fal'ko,~V.~I. Quantum {M}onte {C}arlo
  Calculation of the Binding Energy of Bilayer Graphene. \emph{Phys. Rev.
  Lett.} \textbf{2015}, \emph{115}, 115501\relax
\mciteBstWouldAddEndPuncttrue
\mciteSetBstMidEndSepPunct{\mcitedefaultmidpunct}
{\mcitedefaultendpunct}{\mcitedefaultseppunct}\relax
\EndOfBibitem
\bibitem[Giovannetti \latin{et~al.}(2007)Giovannetti, Khomyakov, Brocks, Kelly,
  and van~den Brink]{Giovannetti2007}
Giovannetti,~G.; Khomyakov,~P.~A.; Brocks,~G.; Kelly,~P.~J.; van~den Brink,~J.
  Substrate-Induced Band Gap in Graphene on Hexagonal Boron Nitride: \textit{Ab
  Initio} Density Functional Calculations. \emph{Phys. Rev. B} \textbf{2007},
  \emph{76}, 073103\relax
\mciteBstWouldAddEndPuncttrue
\mciteSetBstMidEndSepPunct{\mcitedefaultmidpunct}
{\mcitedefaultendpunct}{\mcitedefaultseppunct}\relax
\EndOfBibitem
\bibitem[Fan \latin{et~al.}(2011)Fan, Zhao, Wang, Zhang, and Zhang]{Fan2011}
Fan,~Y.; Zhao,~M.; Wang,~Z.; Zhang,~X.; Zhang,~H. Tunable Electronic Structures
  of Graphene/Boron Nitride Heterobilayers. \emph{Appl. Phys. Lett.}
  \textbf{2011}, \emph{98}, 083103\relax
\mciteBstWouldAddEndPuncttrue
\mciteSetBstMidEndSepPunct{\mcitedefaultmidpunct}
{\mcitedefaultendpunct}{\mcitedefaultseppunct}\relax
\EndOfBibitem
\bibitem[Sachs \latin{et~al.}(2011)Sachs, Wehling, Katsnelson, and
  Lichtenstein]{Sachs2011}
Sachs,~B.; Wehling,~T.~O.; Katsnelson,~M.~I.; Lichtenstein,~A.~I. Adhesion and
  Electronic Structure of Graphene on Hexagonal Boron Nitride Substrates.
  \emph{Phys. Rev. B} \textbf{2011}, \emph{84}, 195414\relax
\mciteBstWouldAddEndPuncttrue
\mciteSetBstMidEndSepPunct{\mcitedefaultmidpunct}
{\mcitedefaultendpunct}{\mcitedefaultseppunct}\relax
\EndOfBibitem
\bibitem[Slotman \latin{et~al.}(2014)Slotman, de~Wijs, Fasolino, and
  Katsnelson]{Slotman2014}
Slotman,~G.~J.; de~Wijs,~G.~A.; Fasolino,~A.; Katsnelson,~M.~I. Phonons and
  Electron-Phonon Coupling in Graphene-h-{BN} Heterostructures. \emph{Ann.
  Phys. (Berl.)} \textbf{2014}, \emph{526}, 381--386\relax
\mciteBstWouldAddEndPuncttrue
\mciteSetBstMidEndSepPunct{\mcitedefaultmidpunct}
{\mcitedefaultendpunct}{\mcitedefaultseppunct}\relax
\EndOfBibitem
\bibitem[Leconte \latin{et~al.}(2017)Leconte, Jung, Leb\`egue, and
  Gould]{Leconte2017}
Leconte,~N.; Jung,~J.; Leb\`egue,~S.; Gould,~T. Moir\'e-Pattern Interlayer
  Potentials in van der {W}aals Materials in the Random-Phase Approximation.
  \emph{Phys. Rev. B} \textbf{2017}, \emph{96}, 195431\relax
\mciteBstWouldAddEndPuncttrue
\mciteSetBstMidEndSepPunct{\mcitedefaultmidpunct}
{\mcitedefaultendpunct}{\mcitedefaultseppunct}\relax
\EndOfBibitem
\bibitem[Feynman(1939)]{Feynman1939}
Feynman,~R.~P. Forces in Molecules. \emph{Phys. Rev.} \textbf{1939}, \emph{56},
  340--343\relax
\mciteBstWouldAddEndPuncttrue
\mciteSetBstMidEndSepPunct{\mcitedefaultmidpunct}
{\mcitedefaultendpunct}{\mcitedefaultseppunct}\relax
\EndOfBibitem
\bibitem[Blakslee \latin{et~al.}(1970)Blakslee, Proctor, Seldin, Spence, and
  Weng]{Blakslee_1970}
Blakslee,~O.~L.; Proctor,~D.~G.; Seldin,~E.~J.; Spence,~G.~B.; Weng,~T. Elastic
  Constants of Compression-Annealed Pyrolytic Graphite. \emph{J. Appl. Phys.}
  \textbf{1970}, \emph{41}, 3373--3382\relax
\mciteBstWouldAddEndPuncttrue
\mciteSetBstMidEndSepPunct{\mcitedefaultmidpunct}
{\mcitedefaultendpunct}{\mcitedefaultseppunct}\relax
\EndOfBibitem
\bibitem[Fayos(1999)]{Fayos_1999}
Fayos,~J. Possible {3D} Carbon Structures as Progressive Intermediates in
  Graphite to Diamond Phase Transition. \emph{J. Solid State Chem.}
  \textbf{1999}, \emph{148}, 278--285\relax
\mciteBstWouldAddEndPuncttrue
\mciteSetBstMidEndSepPunct{\mcitedefaultmidpunct}
{\mcitedefaultendpunct}{\mcitedefaultseppunct}\relax
\EndOfBibitem
\bibitem[Lee \latin{et~al.}(2008)Lee, Wei, Kysar, and Hone]{Lee_2008}
Lee,~C.; Wei,~X.; Kysar,~J.~W.; Hone,~J. Measurement of the Elastic Properties
  and Intrinsic Strength of Monolayer Graphene. \emph{Science} \textbf{2008},
  \emph{321}, 385--388\relax
\mciteBstWouldAddEndPuncttrue
\mciteSetBstMidEndSepPunct{\mcitedefaultmidpunct}
{\mcitedefaultendpunct}{\mcitedefaultseppunct}\relax
\EndOfBibitem
\bibitem[Falin \latin{et~al.}(2017)Falin, Cai, Santos, Scullion, Qian, Zhang,
  Yang, Huang, Watanabe, Taniguchi, Barnett, Chen, Ruoff, and Li]{Falin2017}
Falin,~A.; Cai,~Q.; Santos,~E. J.~G.; Scullion,~D.; Qian,~D.; Zhang,~R.;
  Yang,~Z.; Huang,~S.; Watanabe,~K.; Taniguchi,~T.; Barnett,~M.~R.; Chen,~Y.;
  Ruoff,~R.~S.; Li,~L.~H. Mechanical Properties of Atomically Thin Boron
  Nitride and the Role of Interlayer Interactions. \emph{Nat. Commun.}
  \textbf{2017}, \emph{8}, 15815\relax
\mciteBstWouldAddEndPuncttrue
\mciteSetBstMidEndSepPunct{\mcitedefaultmidpunct}
{\mcitedefaultendpunct}{\mcitedefaultseppunct}\relax
\EndOfBibitem
\bibitem[San-Jose \latin{et~al.}(2014)San-Jose, Guti\'errez-Rubio, Sturla, and
  Guinea]{SanJose2014}
San-Jose,~P.; Guti\'errez-Rubio,~A.; Sturla,~M.; Guinea,~F. Spontaneous Strains
  and Gap in Graphene on Boron Nitride. \emph{Phys. Rev. B} \textbf{2014},
  \emph{90}, 075428\relax
\mciteBstWouldAddEndPuncttrue
\mciteSetBstMidEndSepPunct{\mcitedefaultmidpunct}
{\mcitedefaultendpunct}{\mcitedefaultseppunct}\relax
\EndOfBibitem
\bibitem[Hestenes and Stiefel(1952)Hestenes, and Stiefel]{Hestenes1952}
Hestenes,~M.~R.; Stiefel,~E. Methods of Conjugate Gradients for Solving Linear
  Systems. \emph{J. Res. Natl. Bur. Stand.} \textbf{1952}, \emph{49},
  409--435\relax
\mciteBstWouldAddEndPuncttrue
\mciteSetBstMidEndSepPunct{\mcitedefaultmidpunct}
{\mcitedefaultendpunct}{\mcitedefaultseppunct}\relax
\EndOfBibitem
\bibitem[Wang \latin{et~al.}(2019)Wang, Zihlmann, Liu, Makk, Watanabe,
  Taniguchi, Baumgartner, and Sch\"{o}nenberger]{Wang2019}
Wang,~L.; Zihlmann,~S.; Liu,~M.-H.; Makk,~P.; Watanabe,~K.; Taniguchi,~T.;
  Baumgartner,~A.; Sch\"{o}nenberger,~C. New Generation of Moir\'{e}
  Superlattices in Doubly Aligned h{BN}/Graphene/h{BN} Heterostructures.
  \emph{Nano Lett.} \textbf{2019}, \emph{19}, 2371--2376\relax
\mciteBstWouldAddEndPuncttrue
\mciteSetBstMidEndSepPunct{\mcitedefaultmidpunct}
{\mcitedefaultendpunct}{\mcitedefaultseppunct}\relax
\EndOfBibitem
\bibitem[Wang \latin{et~al.}(2019)Wang, Wang, Yin, T{\'o}v{\'a}ri, Yang, Lin,
  Holwill, Birkbeck, Perello, Xu, Zultak, Gorbachev, Kretinin, Taniguchi,
  Watanabe, Morozov, An{\dj}elkovi{\'c}, Milovanovi{\'c}, Covaci, Peeters,
  Mishchenko, Geim, Novoselov, Fal'ko, Knothe, and Woods]{Wang2019a}
Wang,~Z. \latin{et~al.}  Composite Super-Moir\'e Lattices in Double-Aligned
  Graphene Heterostructures. \emph{Sci. Adv.} \textbf{2019}, \emph{5},
  eaay8897\relax
\mciteBstWouldAddEndPuncttrue
\mciteSetBstMidEndSepPunct{\mcitedefaultmidpunct}
{\mcitedefaultendpunct}{\mcitedefaultseppunct}\relax
\EndOfBibitem
\bibitem[Finney \latin{et~al.}(2019)Finney, Yankowitz, Muraleetharan, Watanabe,
  Taniguchi, Dean, and Hone]{finney_tunable_2019}
Finney,~N.~R.; Yankowitz,~M.; Muraleetharan,~L.; Watanabe,~K.; Taniguchi,~T.;
  Dean,~C.~R.; Hone,~J. Tunable Crystal Symmetry in Graphene-Boron Nitride
  Heterostructures with Coexisting Moir\'{e} Superlattices. \emph{Nat.
  Nanotechnol.} \textbf{2019}, \emph{14}, 1029--1034\relax
\mciteBstWouldAddEndPuncttrue
\mciteSetBstMidEndSepPunct{\mcitedefaultmidpunct}
{\mcitedefaultendpunct}{\mcitedefaultseppunct}\relax
\EndOfBibitem
\bibitem[Yang \latin{et~al.}(2020)Yang, Li, Yin, Xu, Mullan, Taniguchi,
  Watanabe, Geim, Novoselov, and Mishchenko]{Yang_2020}
Yang,~Y.; Li,~J.; Yin,~J.; Xu,~S.; Mullan,~C.; Taniguchi,~T.; Watanabe,~K.;
  Geim,~A.~K.; Novoselov,~K.~S.; Mishchenko,~A. \textit{In Situ} Manipulation
  of van der {W}aals Heterostructures for Twistronics. \emph{Sci. Adv.}
  \textbf{2020}, \emph{6}, eabd3655\relax
\mciteBstWouldAddEndPuncttrue
\mciteSetBstMidEndSepPunct{\mcitedefaultmidpunct}
{\mcitedefaultendpunct}{\mcitedefaultseppunct}\relax
\EndOfBibitem
\bibitem[An{\dj}elkovi{\'{c}} \latin{et~al.}(2020)An{\dj}elkovi{\'{c}},
  Milovanovi{\'{c}}, Covaci, and Peeters]{Andelkovic_2020}
An{\dj}elkovi{\'{c}},~M.; Milovanovi{\'{c}},~S.~P.; Covaci,~L.; Peeters,~F.~M.
  Double Moir{\'e} with a Twist: Supermoir{\'e} in Encapsulated Graphene.
  \emph{Nano Lett.} \textbf{2020}, \emph{20}, 979--988\relax
\mciteBstWouldAddEndPuncttrue
\mciteSetBstMidEndSepPunct{\mcitedefaultmidpunct}
{\mcitedefaultendpunct}{\mcitedefaultseppunct}\relax
\EndOfBibitem
\bibitem[Sun \latin{et~al.}(2021)Sun, Zhang, Liu, Zhu, Huang, Yuan, Wang,
  Watanabe, Taniguchi, Li, Zhu, Mao, Yang, Kang, Liu, Ye, Han, and
  Zhang]{Sun_2021}
Sun,~X. \latin{et~al.}  Correlated States in Doubly-Aligned
  h{BN}/Graphene/h{BN} Heterostructures. \emph{Nat. Commun.} \textbf{2021},
  \emph{12}, 7196\relax
\mciteBstWouldAddEndPuncttrue
\mciteSetBstMidEndSepPunct{\mcitedefaultmidpunct}
{\mcitedefaultendpunct}{\mcitedefaultseppunct}\relax
\EndOfBibitem
\bibitem[Moulsdale \latin{et~al.}(2022)Moulsdale, Knothe, and
  Fal'ko]{Moulsdale_2022}
Moulsdale,~C.; Knothe,~A.; Fal'ko,~V. Kagome Network of Miniband-Edge States in
  Double-Aligned Graphene-Hexagonal Boron Nitride Structures. \emph{Phys. Rev.
  B} \textbf{2022}, \emph{105}, L201112\relax
\mciteBstWouldAddEndPuncttrue
\mciteSetBstMidEndSepPunct{\mcitedefaultmidpunct}
{\mcitedefaultendpunct}{\mcitedefaultseppunct}\relax
\EndOfBibitem
\bibitem[Hu \latin{et~al.}(2023)Hu, Tan, Al~Ezzi, Chattopadhyay, Gou, Zheng,
  Wang, Chen, Thottathil, Luo, Watanabe, Taniguchi, Wee, Adam, and
  Ariando]{Hu_2023}
Hu,~J.; Tan,~J.; Al~Ezzi,~M.~M.; Chattopadhyay,~U.; Gou,~J.; Zheng,~Y.;
  Wang,~Z.; Chen,~J.; Thottathil,~R.; Luo,~J.; Watanabe,~K.; Taniguchi,~T.;
  Wee,~A. T.~S.; Adam,~S.; Ariando,~A. Controlled Alignment of Supermoir{\'e}
  Lattice in Double-Aligned Graphene Heterostructures. \emph{Nat. Commun.}
  \textbf{2023}, \emph{14}, 4142\relax
\mciteBstWouldAddEndPuncttrue
\mciteSetBstMidEndSepPunct{\mcitedefaultmidpunct}
{\mcitedefaultendpunct}{\mcitedefaultseppunct}\relax
\EndOfBibitem
\bibitem[Wallbank \latin{et~al.}(2013)Wallbank, Patel, Mucha-Kruczy\'{n}ski,
  Geim, and Fal'ko]{Wallbank_2013}
Wallbank,~J.~R.; Patel,~A.~A.; Mucha-Kruczy\'{n}ski,~M.; Geim,~A.~K.;
  Fal'ko,~V.~I. Generic Miniband Structure of Graphene on a Hexagonal
  Substrate. \emph{Phys. Rev. B} \textbf{2013}, \emph{87}, 245408\relax
\mciteBstWouldAddEndPuncttrue
\mciteSetBstMidEndSepPunct{\mcitedefaultmidpunct}
{\mcitedefaultendpunct}{\mcitedefaultseppunct}\relax
\EndOfBibitem
\bibitem[Novoselov \latin{et~al.}(2005)Novoselov, Geim, Morozov, Jiang,
  Katsnelson, Grigorieva, Dubonos, and Firsov]{Novoselov2005}
Novoselov,~K.~S.; Geim,~A.~K.; Morozov,~S.~V.; Jiang,~D.; Katsnelson,~M.~I.;
  Grigorieva,~I.~V.; Dubonos,~S.~V.; Firsov,~A.~A. Two-Dimensional Gas of
  Massless {D}irac Fermions in Graphene. \emph{Nature} \textbf{2005},
  \emph{438}, 197--200\relax
\mciteBstWouldAddEndPuncttrue
\mciteSetBstMidEndSepPunct{\mcitedefaultmidpunct}
{\mcitedefaultendpunct}{\mcitedefaultseppunct}\relax
\EndOfBibitem
\bibitem[Cosma \latin{et~al.}(2014)Cosma, Wallbank, Cheianov, and
  Fal'ko]{Cosma_2014}
Cosma,~D.~A.; Wallbank,~J.~R.; Cheianov,~V.; Fal'ko,~V.~I. Moir\'{e} Pattern as
  a Magnifying Glass for Strain and Dislocations in van der {W}aals
  Heterostructures. \emph{Faraday Discuss.} \textbf{2014}, \emph{173},
  137--143\relax
\mciteBstWouldAddEndPuncttrue
\mciteSetBstMidEndSepPunct{\mcitedefaultmidpunct}
{\mcitedefaultendpunct}{\mcitedefaultseppunct}\relax
\EndOfBibitem
\end{mcitethebibliography}

\newpage

\section*{For Table of Contents Only}

\begin{center}
\includegraphics[clip]{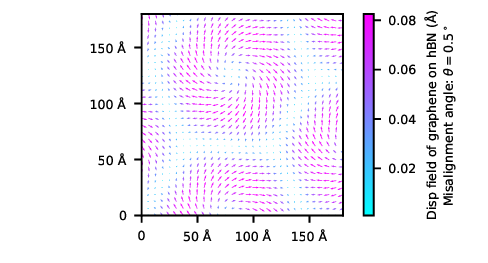}
\end{center}

\end{document}


\section{BE Fitting Parameters}

\begin{table}[!htbp]
\begin{center}
\caption{Fitting parameters in Equation 1 of the main text for
  the BE of 1L-G/1L-hBN\@. \protect\label{sup_tab:fit}}
\begin{tabular}{lrl}
\textbf{Parameter} & \multicolumn{2}{c}{\textbf{Value}} \\
\hline
$A_{01}$ & $-1.604928$ & eV\,{\r A}$\bm{^4}$\,/\,atom \\

$A_{02}$ & $-490.4703$ & eV\,{\r A}$\bm{^8}$\,/\,atom \\

$A_{03}$ & $74950.57$ & eV\,{\r A}$\bm{^{12}}$\,/\,atom \\

$A_{04}$ & $-1975507$& eV\,{\r A}$\bm{^{16}}$\,/\,atom \\

$A_{11}$ & $57.39082$ & eV\,/\,atom \\

$B_{11}$ & $-16.12002$& eV\,/\,atom \\

$\kappa_{\text{A}1}$ & $3.363638$ & {\r A}$\bm{^{-1}}$ \\

$\kappa_{\text{B}1}$ & $2.860883$ & {\r A}$\bm{^{-1}}$ \\
\hline
\end{tabular}
\end{center}
\end{table}

\begin{figure}[!htbp]
\centerline{\includegraphics[width=100mm]{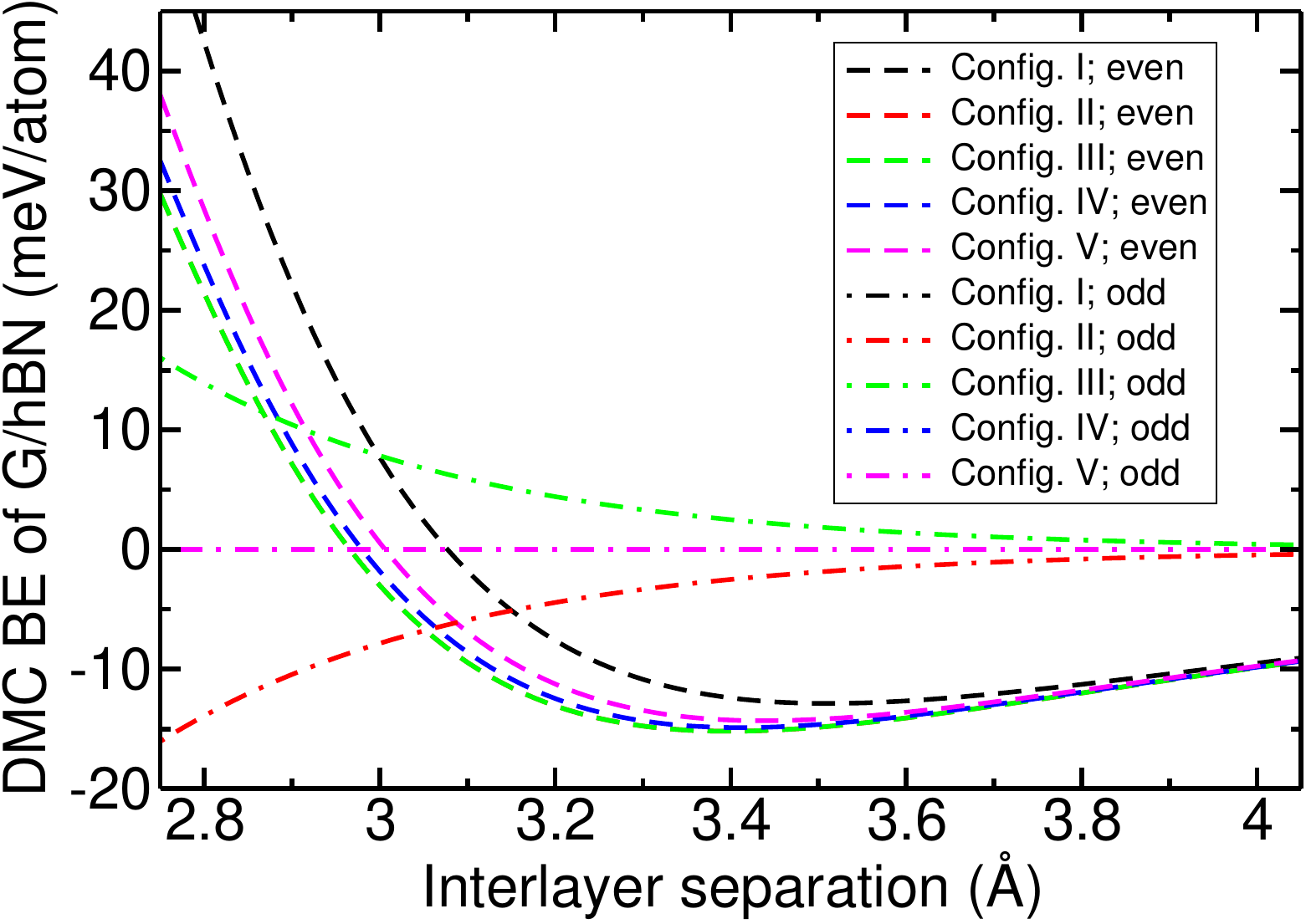}}
\caption{Even and odd parts of Equation 1 of the main text for each of
  the stacking configurations. The even BE parts of configurations II
  and III lie on top of each other, while the odd parts for
  configurations I, IV, and V are zero. The resultant BE curves are in
  Figure 1 of the main text.
  \protect\label{sup_fig:BE_symcomp}}
\end{figure}

The fitting parameters in Equation 1 of the main text are
shown in \tab\ref{sup_tab:fit}, while the even and odd parts of the BE
are shown in \fig\ref{sup_fig:BE_symcomp}.

\section{Equilibrium Separation, BE, and LBM Frequency}

The LBM frequency of an incommensurate 1L-G/1L-hBN bilayer is
$\omega_\text{BM}=2\sqrt{\bar{E}_\text{bind}''(d_0)/\mu}$, where $d_0$
is the equilibrium separation that minimizes the translationally
averaged BE $\bar{E}_\text{bind}(d)$ (the first term of Equation 1 of
the main text), and $\mu=
2m_\text{C}(m_\text{B}+m_\text{N})/(2m_\text{C}
+m_\text{B}+m_\text{N})$ is the reduced mass of a primitive cell of
1L-G/1L-hBN\@. $m_\text{C}$, $m_\text{B}$, $m_\text{N}$ are the C, B,
and N atomic masses. To evaluate the statistical error bars on $d_0$,
$\bar{E}(d_0)$, and $\omega_\text{BM}$ we used a bootstrap Monte Carlo
procedure,\cite{Efron1993} in which we repeatedly fitted Equation 1 of
the main text to resampled DMC data drawn from Gaussian distributions
centered on the mean DMC energy at each layer separation, with the
standard deviation being the standard error in the mean energy at that
layer separation. The mean and standard error in the mean of $d_0$,
$\bar{E}(d_0)$, and $\omega_\text{BM}$ are then
accumulated.\cite{Mostaani2015}

\section{DFT Geometries}

DFT-calculated interlayer distances for 1L-G/1L-hBN are reported in
\tab\ref{sup_tab:eq_sep}.

\begin{table}[!htbp]
\begin{center}
\caption{Equilibrium interlayer separation of 1L-G/1L-hBN calculated
  by DFT\@. Both layers were assumed to have the 1L-G lattice constant
  in the present work and in Refs.\ \citenum{Giovannetti2007} and
  \citenum{Fan2011}, while the lattice constant of 1L-hBN was used in
  Refs.\ \citenum{Sachs2011} and \citenum{Slotman2014}. An averaged
  lattice constant was used in Ref.\ \citenum{Leconte2017}. The other
  DFT-LDA results without citation are from the present
  work. \protect\label{sup_tab:eq_sep}}
\begin{tabular}{lccc}
& \multicolumn{3}{c}{\textbf{Equilibrium interlayer separation ({\r A})}} \\
\raisebox{1em}[0pt]{\textbf{Config.}} & \textbf{DFT-LDA} &
\textbf{DFT-vdW}\cite{Fan2011,Slotman2014} & \textbf{DFT-RPA} \\
\hline
I & $3.5$, $3.50$\bcite{Giovannetti2007} & $3.5$, $3.49$ &
$3.55$\bcite{Sachs2011}, $3.46$\bcite{Leconte2017} \\
II & $3.2$, $3.22$\bcite{Giovannetti2007} & $3.2$, $3.30$ &
$3.35$\bcite{Sachs2011}, $3.27$\bcite{Leconte2017} \\
III &$3.4$, $3.40$\bcite{Giovannetti2007} & $3.4$ & $\sim
3.4$\bcite{Sachs2011}, $3.43$\bcite{Leconte2017} \\
IV  & $3.3$ &  $3.4$, $3.45$ & $3.5$\bcite{Sachs2011} \\
V   & $3.4$ & & \\
\hline
\end{tabular}
\end{center}
\end{table}

\section{DFT Phonon Calculations}

DFT phonon dispersions were calculated using ultrasoft
pseudopotentials and a plane-wave cutoff energy of $25$ Ha within the
LDA\@.  We relaxed the lattice constant of the four-atom cell of
1L-G/1L-hBN\@. We used a $5\times 5$ supercell with a $35\times 35$
Monkhorst-Pack $\mathbf{k}$-point mesh, and displaced the atoms by
$\pm 0.04$ {\AA} to evaluate the matrix of force constants within the
finite-displacement method. The initial equilibrium atomic positions
were relaxed until the Hellmann-Feynman forces\cite{Feynman1939} were
$<5\times 10^{-5}$ eV\,{\AA}$^{-1}$.

\begin{figure}[!htbp]
\begin{center}
\begin{minipage}{0.46\textwidth}
\includegraphics[width=76mm]{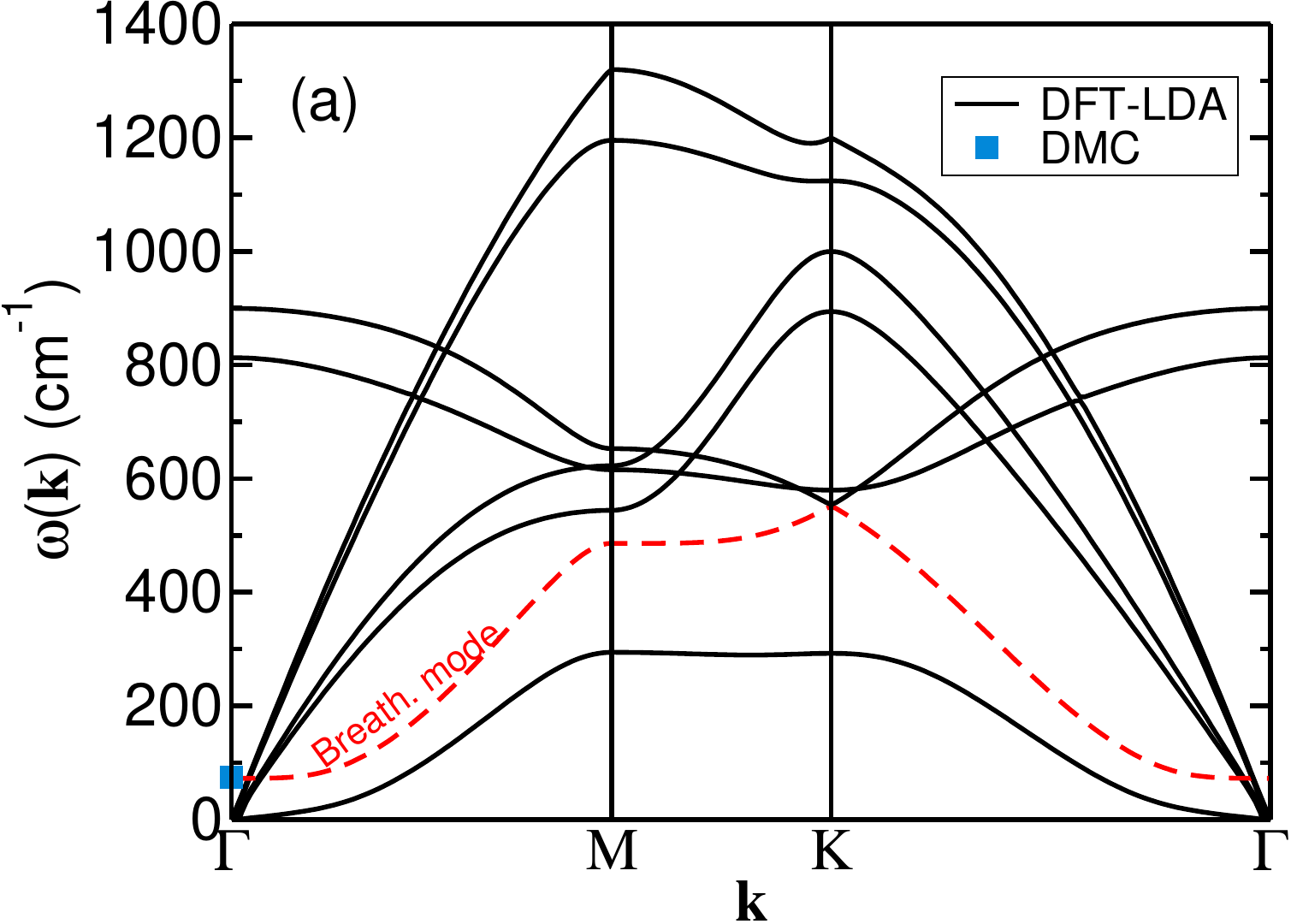}
\end{minipage}
\hspace{.6cm}
\begin{minipage}{0.46\textwidth}
\includegraphics[width=76mm]{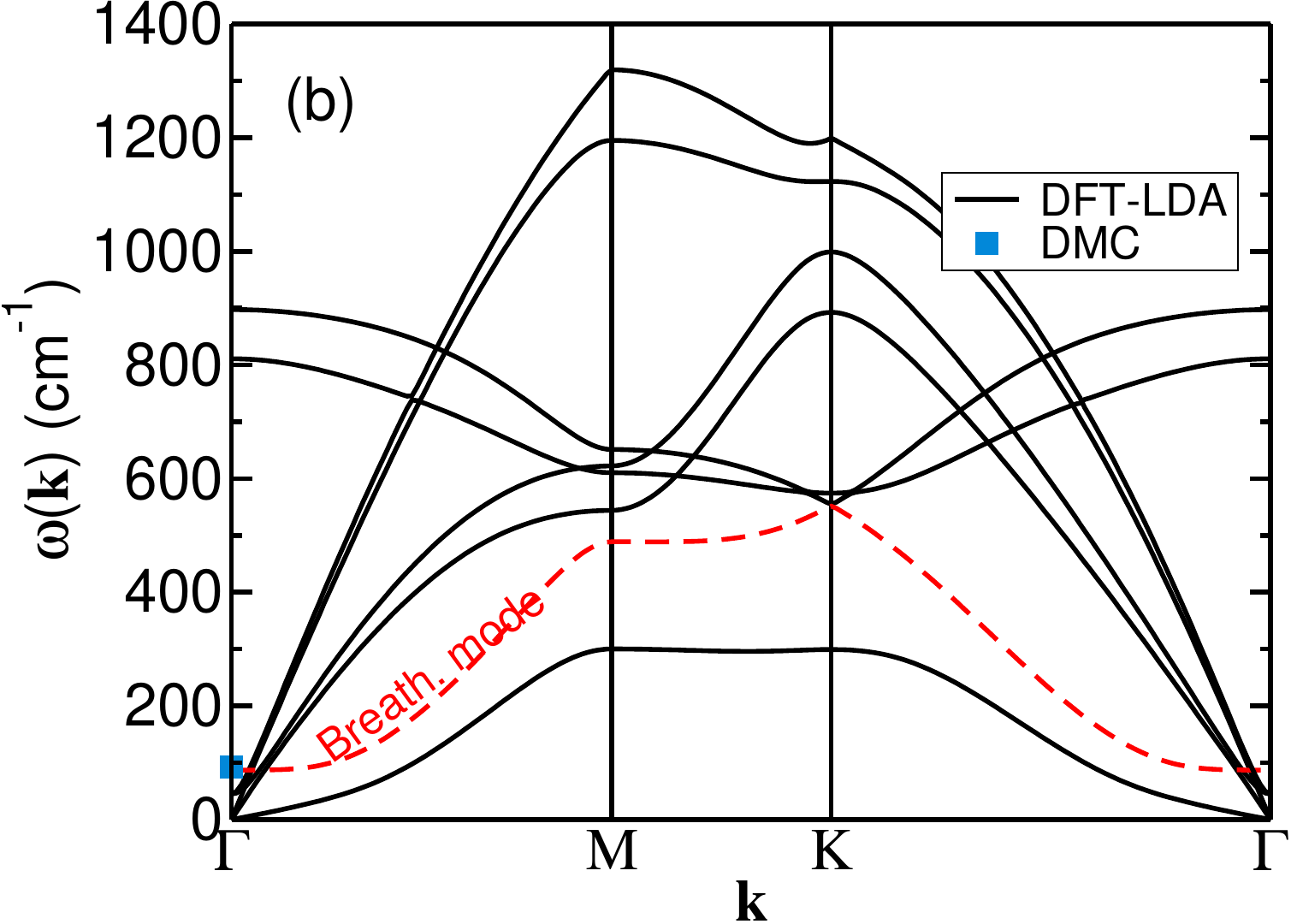}
\end{minipage}
\caption{DFT-LDA phonon dispersions of 1L-G/1L-hBN for stacking
  configurations (a) I and (b) II\@. The branches that go to the LBM
  frequency at $\Gamma$ are shown by dashed red curves and the DMC LBM
  frequencies at $\Gamma$ point are shown by blue squares.
  \protect\label{sup_fig:phonon_dipspersion}}
\end{center}
\end{figure}

\fig\ref{sup_fig:phonon_dipspersion} plots the DFT-LDA phonon
dispersion curves for 1L-G/1L-hBN in a four-atom cell at the relaxed
common in-plane lattice constant $2.47$ {\AA} and the relaxed
equilibrium separations of $3.5$ and $3.2$ {\AA} for stacking
configurations I and II, respectively. Since we are considering low
frequencies close to acoustic branches, we do not calculate Kohn
anomaly effects in 1L-G\@.

\section{Lam\'{e} Parameters}

The Young's modulus of bulk graphite is $Y=1.02(3)$ TPa, and the
in-plane Poisson's ratio is $\nu=0.165$.\cite{Blakslee_1970} The
lattice constants of Bernal-stacked graphite are $a=2.461$ {\AA} and
$c=6.708$ {\AA} at 300 K\@.\cite{Fayos_1999} The 2d Young's modulus of
1L-G is $Y_\text{2D}=Yc/2=340$ N\,m$^{-1}= 21.4$ eV\,{\AA}$^{-2}$. The
first Lam\'{e} parameter is
$\lambda=Y_\text{2D}\nu/[(1+\nu)(1-2\nu)]=4.5$ eV\,{\AA}$^{-2}$.  The
second Lam\'{e} parameter is $\mu=Y_\text{2D}/[2(1+\nu)]=9.2$
eV\,{\AA}$^{-2}$.  For thin graphene films, the Young's modulus was
measured to be $1.0(1)$ TPa,\cite{Lee_2008} in agreement with the bulk
graphite value.

The Lam\'e parameters of 1L-hBN are $\lambda_\text{hBN} = 4.0$
eV\,{\AA}$^{-2}$ and $\mu_\text{hBN} = 7.4$ eV\,{\AA}$^{-2}$, as
determined by atomic force microscopy.\cite{Falin2017}

\section{2d Adhesion-Potential Parameters}

\begin{figure}[!htbp]
\centerline{\includegraphics[width=100mm]{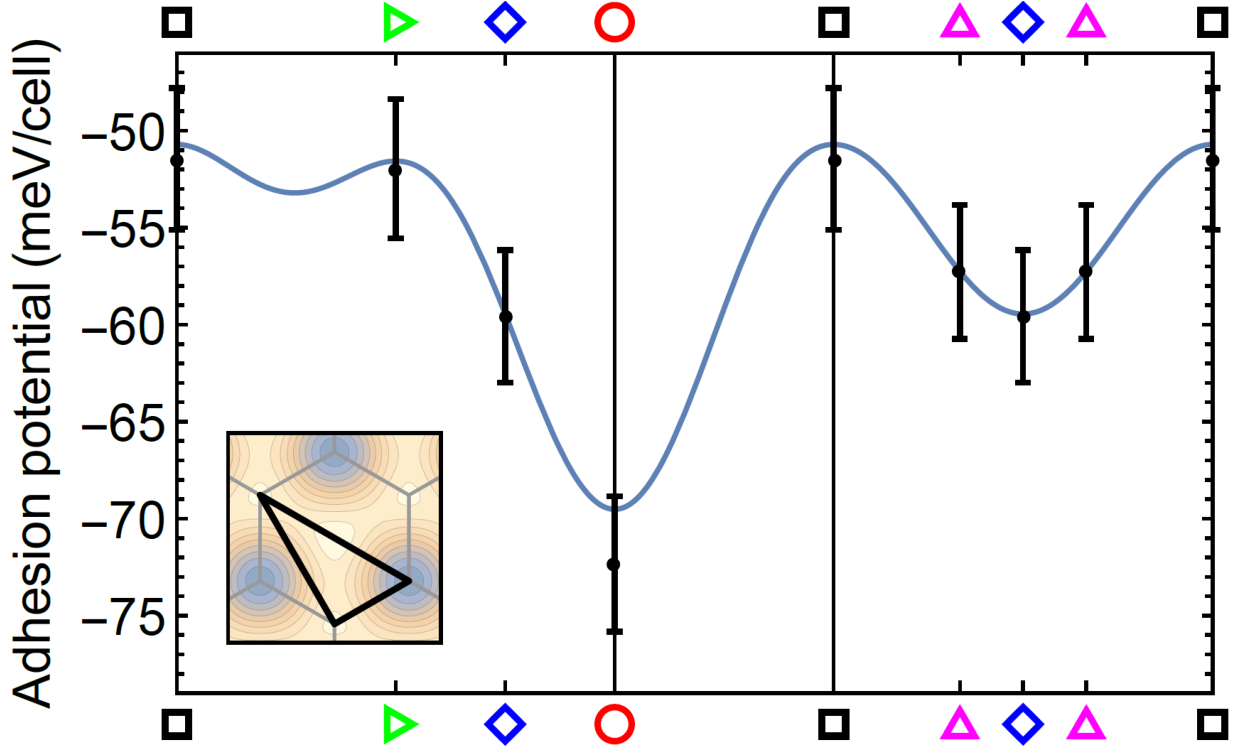}}
\caption{Adhesion potential $V_\text{A}$ of 1L-G/B-hBN as a function
  of offset ${\bm \ell}$ of 1L-G relative to B-hBN around the
  triangular path through the unit cell shown in the inset. The DMC
  results are presented in black and $V_\text{A}({\bm \ell})$ is in
  blue. The symbols indicate the stacking configurations in Table 1
  and Figure 1 of the main text. Both layers have the 1L-G lattice
  constant. The error bars show the standard error in the mean of the
  BE at the minimum of the BE curve for each stacking
  configuration. The error bars on relative adhesion potentials are
  smaller than suggested by the error bars on the absolute values, due
  to the fact that the 1L energies cancel out of differences in
  adhesion
  potential. \protect\label{sup_fig:adhesion_potential_bslike}}
\end{figure}

To approximate the adhesion potential per unit cell $V_\text{A}({\bm
  \ell})$ for 1L-G/B-hBN with in-plane offset ${\bm \ell}$ and a
common lattice parameter, we use a truncated Fourier expansion:
\begin{eqnarray}
V_\text{A} ({\bm \ell}) = \sum_m v_m e^{i \mathbf{g}_m \cdot {\bm
    \ell}} & \approx & 4E_\text{bind}(d_0,{\bm \ell}) \nonumber \\ &
\equiv & v_{\text{s}0} + v_{\text{s}1} \sum_{m=1,3,5}
\cos(\mathbf{g}_m \cdot {\bm \ell})
+v_{\text{as}1}\sum_{m=1,3,5}\sin(\mathbf{g}_m \cdot {\bm
  \ell}), \label{sup_eq:adpot_params}
\end{eqnarray}
where $\mathbf{g}_m$ is a reciprocal lattice point and $E_\text{bind}$
is given in Equation 1 of the main text. From the fit of that
equation, we find $v_{\text{s}0}=-57(3)$, $v_{\text{s}1}=2.2(3)$, and
$v_{\text{as}1}=-3.5(4)$ meV per primitive unit cell.  The adhesion
potential is plotted in \fig\ref{sup_fig:adhesion_potential_bslike}.

\section{Relaxation of 1L-G/B-hBN within the Continuum Model}

\begin{figure}[!htbp]
\centerline{\includegraphics[width=120mm]{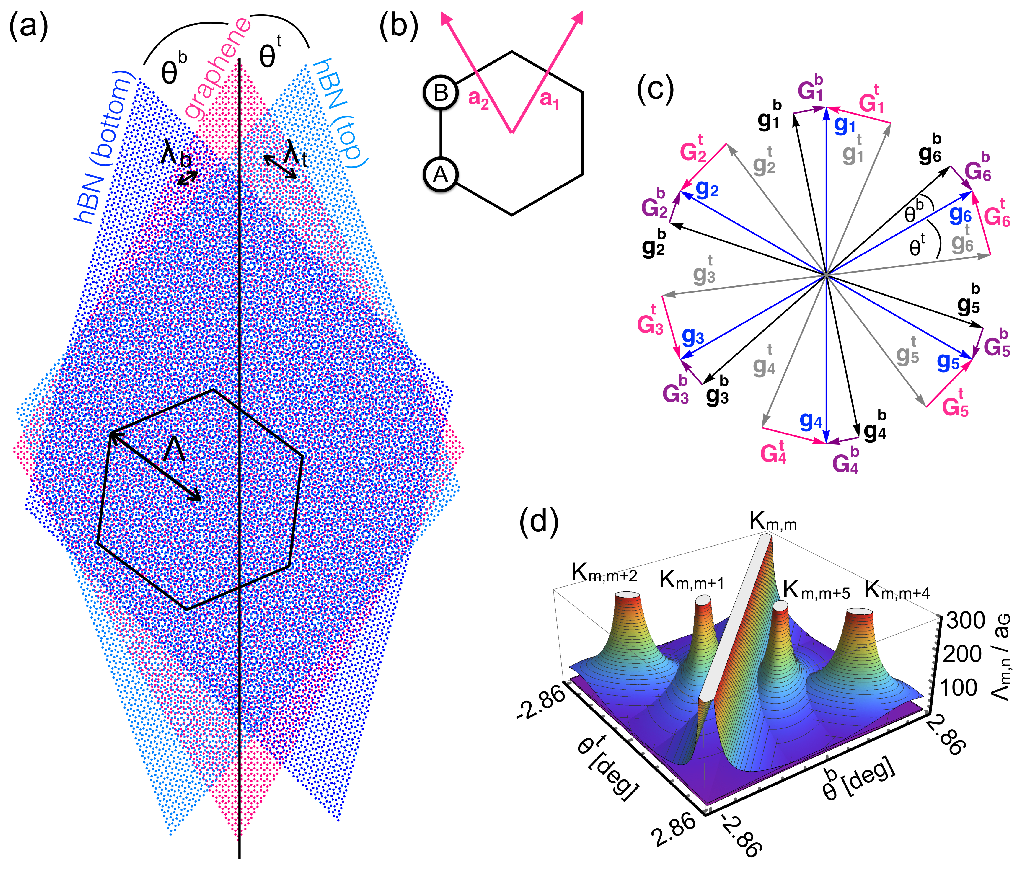}}
\caption{(a) Beating between the moir\'{e} patterns (of periods
  $\lambda_\text{t}$ and $\lambda_\text{b}$) from the top and bottom
  1L-G/B-hBN interfaces yields a supermoir\'{e} pattern of period
  $\Lambda$ in the B-hBN/1L-G/B-hBN LMH\@. (b) Real-space lattice
  vectors of the 1L-G lattice. (c) Combining reciprocal lattice
  vectors to give the reciprocal lattice vectors
  $\mathbf{K}_{nm}=\mathbf{G}^\text{t}_{n}-\mathbf{G}^\text{b}_{m }$
  of the supermoir\'{e} pattern. (d) Misalignment angles where the
  supermoir\'{e} reciprocal vectors $\mathbf{K}_{nm}$ vanish, and the
  corresponding supermoir\'{e} lattice vector ${\bm \Lambda}_{nm}$
  diverges. \protect\label{sup_fig:SLLattVecL}}
\end{figure}

When we allow an elastic layer to relax on a rigid substrate, the
adhesion potential $V_\text{A}$ provides the energy of a unit cell at
position $\mathbf{r}$, because $V_\text{A}(\mathbf{r})$ has the
periodicity of the rigid lattice, and hence $V_\text{A}
(\mathbf{r})=V_\text{A}({\bm\ell}_\mathbf{r})$, where
${\bm\ell}_\mathbf{r}$ is the position of the unit cell of the elastic
layer relative to the closest rigid lattice point. In other words,
${\bm\ell}_\mathbf{r}$ is simply the local offset of the elastic
lattice relative to the rigid lattice. Let $\mathbf{g}_m$ and
$\mathbf{g}^\text{b}_m$ be the reciprocal lattice points of the
unrelaxed elastic layer and the rigid (``bottom'') layer,
respectively. We may write the Fourier expansion of the adhesion
potential as $V_\text{A}({\bm \ell})=\sum_m v_m
e^{i\mathbf{g}^\text{b}_m \cdot {\bm\ell}}$ (see
\eq\ref{sup_eq:adpot_params}). The difference of each corresponding
pair of reciprocal-lattice points $\mathbf{g}_m$ and
$\mathbf{g}_m^\text{b}$ defines a reciprocal lattice point
$\mathbf{G}_m^\text{b}=\mathbf{g}_m-\mathbf{g}_m^\text{b}$ of the
moir\'{e} supercell; see \fig\ref{sup_fig:SLLattVecL}c.  The total
adhesion potential per unit cell of the elastic layer can be written
as
\begin{eqnarray}
U_\text{A} = \frac{1}{N} \sum_{n=1}^N V_\text{A}
(\mathbf{r}_n+\mathbf{u}(\mathbf{r}_n)) & = & \frac{1}{N} \sum_{n=1}^N
\sum_m v_m e^{i \mathbf{g}^\text{b}_m \cdot \mathbf{r}_n} e^{i
  \mathbf{g}^\text{b}_m \cdot \mathbf{u} (\mathbf{r}_n)} \nonumber
\\ & = & \frac{1}{N} \sum_{n=1}^N \sum_m v_m e^{- i
  \mathbf{G}_m^\text{b} \cdot \mathbf{r}_n} e^{i \mathbf{g}^\text{b}_m
  \cdot \mathbf{u} (\mathbf{r}_n)} \label{sup_eq:adh-full} \\ & \approx &
v_0 + i \sum_m v_m \mathbf{g}^\text{b}_m \cdot
\mathbf{u}_{\mathbf{G}^\text{b}_m}. \label{sup_eq:adh-linear}
\end{eqnarray}
\eq\ref{sup_eq:adh-full} uses
$e^{i\mathbf{g}^\text{b}_m\cdot\mathbf{r}_n}=e^{i(\mathbf{g}^\text{b}_m
  -\mathbf{g}_m)\cdot\mathbf{r}_n}=e^{-i\mathbf{G}_m^\text{b}\cdot
  \mathbf{r}_n}$. \eq\ref{sup_eq:adh-linear} assumes the displacement
field to be small compared to the rigid lattice constant, which is the
case if
\begin{equation}
  |\mathbf{u}|\ll a^\text{b}/(2\pi),
  \label{sup_eq:small_field}
\end{equation}
where $a^\text{b}$ is the lattice parameter of the rigid layer.

The elastic energy per unit cell is
\begin{eqnarray}
U_\text{E} & = & \frac{1}{N} \int_\text{crystal} \frac{1}{2}
\left[\lambda \text{Tr}(\varepsilon)^2 + 2\mu\text{Tr}(\varepsilon^2)
  \right] \, d^2\mathbf{r}
\label{sup_eq:elastic_real} \\
& = & \frac{1}{2} \sum_{\mathbf{q}} \mathbf{u}_{\mathbf{q}}^\dagger
W_{\mathbf{q}} \mathbf{u}_{\mathbf{q}},
\end{eqnarray}
where $\varepsilon_{ij}=(\partial u_i/\partial x_j+\partial
u_j/\partial x_i)/2$ is the 2d pure strain field, $\lambda$ and $\mu$
are the Lam\'{e} coefficients of the elastic layer, and $N$ is the
number of unit cells in the elastic layer. The dynamical matrix at
wavevector $\mathbf{q}$ is\cite{SanJose2014}
\begin{equation}
W_{\mathbf{q}}=A \left[(\lambda+\mu)\mathbf{q}\mathbf{q}^\text{T}+\mu q^2 I\right],
\end{equation}
where $A=|\mathbf{a}_1\times\mathbf{a}_2|$ is the area of an unrelaxed
primitive unit cell and $I$ is the $2\times 2$ identity matrix.

\begin{figure}[!htbp]
\centerline{\includegraphics[width=120mm]{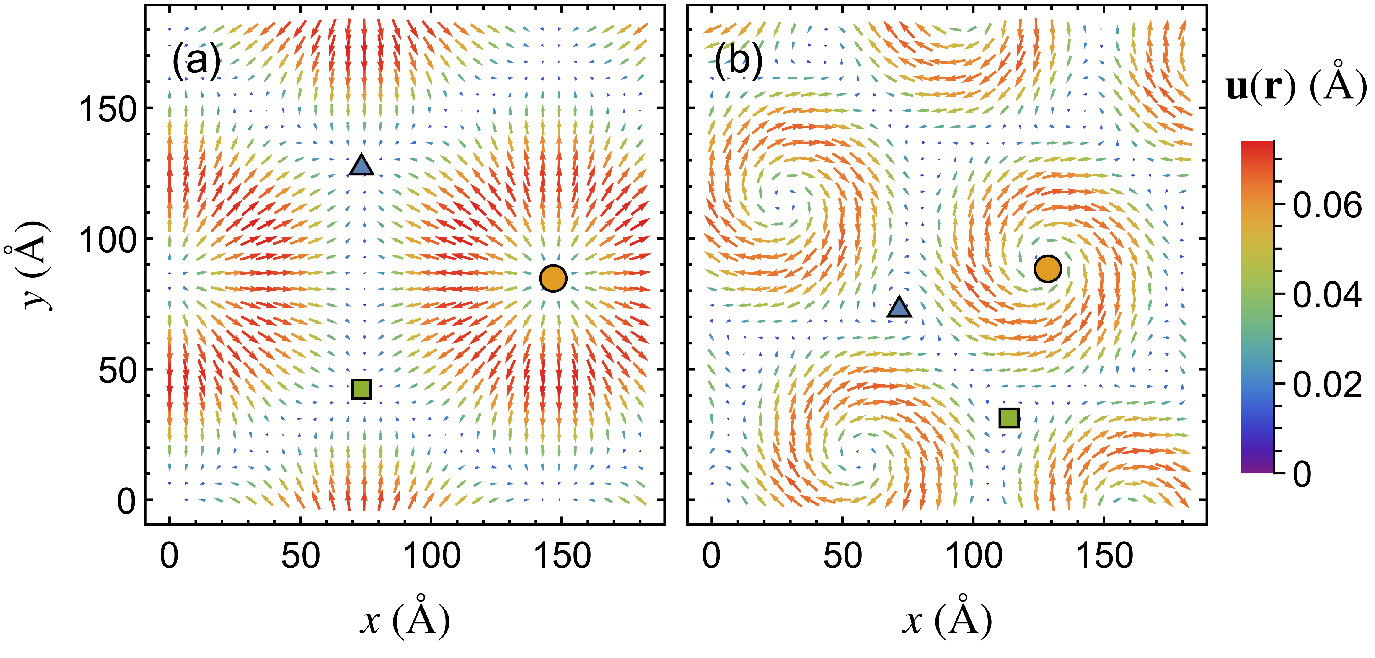}}
\caption{Displacement field $\mathbf{u}(\mathbf{r})$ of 1L-G unit
  cells in 1L-G/B-hBN LMHs, obtained by numerically minimizing the
  total energy of \eq\ref{sup_eq:adh-full} within the continuum
  model. The misalignment angle between the two layers is (a)
  $\theta=0$; (b) $\theta=1^\circ$. The symbols indicate the stacking
  configurations shown in Table 1 of the main
  text. \protect\label{sup_fig:displacement}}
\end{figure}

By minimizing the total energy $U=U_\text{E}+U_\text{A}$ with respect
to the Fourier components of the displacement field, we find the
Fourier components of the displacement field (which has the
periodicity of the moir\'e superlattice) to be
\begin{equation}
\mathbf{u}_{\mathbf{G}^\text{b}_m} = iv_m^{\ast}
W_{\mathbf{G}^\text{b}_m}^{-
  1}\mathbf{g}^\text{b}_m.\label{sup_eq:analytic_soln}
\end{equation}
(See the section below for the derivation of
\eq\ref{sup_eq:analytic_soln}.)  This result holds for any reciprocal
lattice vector $m$. We do not assume any particular symmetry or
structure in the derivation of \eq\ref{sup_eq:analytic_soln}, so that
\eq\ref{sup_eq:analytic_soln} is valid for any situation where one LM
is allowed to relax after being transferred onto a substrate of
another LM with a sufficiently similar lattice constant. Using a
numerical evaluation of the right-hand side of
\eq\ref{sup_eq:adh-full}, we have investigated the continuum model
without the approximation of small displacement fields.  See the
section below for details.  For bilayers with greatly enhanced
variations in the adhesion potential as a function of lattice offset
or with greatly reduced elastic parameters, this method finds multiple
energy minima. However, for 1L-G/B-hBN LMHs, we only find the one,
global minimum of the energy over the range of displacement fields
($\ll a_\text{hBN}$) for which the continuum model applies. For
aligned lattice vectors, this global minimum is $0.82$ $\mu$eV per
primitive cell lower than the energy predicted by the analytical
approximation of \eq\ref{sup_eq:analytic_soln}, which finds that the
1L-G relaxation lowers the energy by $0.37$ meV per primitive
cell. Physically, the stiffness of the two layers is too great to be
overcome by the variations in the adhesion potential, which only
weakly perturbs the 1L-G atomic structure. The displacement field that
minimizes the energy is shown in \fig\ref{sup_fig:displacement}.

\section{Minimizing the Analytical Approximation to the Total Energy
within the Continuum Model}

We assume finite-temperature effects to be negligible. Under zero
external stress, the relaxed displacement field is that which
minimizes the total energy. Under the assumption of a small
displacement field the total energy can be written as:
\begin{equation}
U = U_\text{E}+U_\text{A} \approx \frac{1}{2}
\sum_{\mathbf{q}}\mathbf{u}_\mathbf{q}^\dagger
W_{\mathbf{q}}\mathbf{u}_{\mathbf{q}}+v_0+i\sum_m v_m
\mathbf{g}^\text{b}_m \cdot\mathbf{u}_{\mathbf{G}^\text{b}_m},
\end{equation}
as shown in \eq\ref{sup_eq:adh-linear}.  We therefore require the
derivatives of the energy with respect to the independent complex
Fourier components of the displacement field to be zero,
\textit{i.e.}, $\nabla_{\mathbf{u}_\mathbf{q}} U =
\mathbf{0}^\text{T}$ and $\nabla_{\mathbf{u}_\mathbf{q}^\dagger} U
=\mathbf{0}$.  $W_\mathbf{q}$ is real and symmetric, and
$W_\mathbf{q}=W_{-\mathbf{q}}$. Since $\mathbf{u}(\mathbf{r})$ is
real, $\mathbf{u}^\ast_\mathbf{q}=\mathbf{u}_{-\mathbf{q}}$. So
\begin{eqnarray}
\nabla_{\mathbf{u}_{\mathbf{q}}} U & = & \nabla_{\mathbf{u}_{\mathbf{q}}} \left( \frac{1}{2}\sum_{\mathbf{q}}
\mathbf{u}_\mathbf{q}^\dagger W_{\mathbf{q}} \mathbf{u}_{\mathbf{q}}
\right) + i \sum_m v_m (\mathbf{g}^\text{b}_m)^\text{T}
\delta_{\mathbf{q},\mathbf{G}^\text{b}_m} \nonumber\\ & = &
\nabla_{\mathbf{u}_{\mathbf{q}}} \left( \frac{1}{2}
\mathbf{u}_\mathbf{q}^\dagger W_{\mathbf{q}} \mathbf{u}_{\mathbf{q}} +
\frac{1}{2} \mathbf{u}_{-\mathbf{q}}^\dagger W_{- \mathbf{q}}
\mathbf{u}_{- \mathbf{q}} \right) + i \sum_m v_m
(\mathbf{g}^\text{b}_m)^\text{T}
\delta_{\mathbf{q},\mathbf{G}^\text{b}_m} \nonumber\\ & = &
\mathbf{u}^{\dagger}_{\mathbf{q}} W_{\mathbf{q}} + i \sum_m v_m
(\mathbf{g}^\text{b}_m)^\text{T}
\delta_{\mathbf{q},\mathbf{G}^\text{b}_m}
\end{eqnarray}
and, similarly,
\begin{eqnarray}
  \nabla_{\mathbf{u}_\mathbf{q}^\dagger} U & = &
  \nabla_{u_\mathbf{q}^\dagger} \left( \frac{1}{2}
  \mathbf{u}_\mathbf{q}^\dagger W_\mathbf{q} \mathbf{u}_\mathbf{q} +
  \frac{1}{2} \mathbf{u}_{-\mathbf{q}}^\dagger W_{-\mathbf{q}}
  \mathbf{u}_{-\mathbf{q}} \right)+i
  \nabla_{\mathbf{u}_\mathbf{q}^\dagger} \sum_m v^*_m
  (-\mathbf{g}_m^\text{b}) \cdot \mathbf{u}_{-\mathbf{G}^\text{b}_m}
  \nonumber \\ & = & \nabla_{u_\mathbf{q}^\dagger}
  \left(\mathbf{u}_\mathbf{q}^\dagger W_\mathbf{q}
  \mathbf{u}_\mathbf{q} \right)+i
  \nabla_{\mathbf{u}_\mathbf{q}^\dagger} \sum_m -v_m^\ast
  \mathbf{u}_{\mathbf{G}^\text{b}_m}^\dagger \mathbf{g}_m^\text{b}
  \nonumber \\ & = & W_\mathbf{q}\mathbf{u}_\mathbf{q}-i\sum_m
  v_m^\ast \mathbf{g}_m^\text{b}
  \delta_{\mathbf{q},\mathbf{G}^\text{b}_m}.
\end{eqnarray}
Therefore, the displacement field that minimizes the total energy of
1L-G/B-hBN satisfies
\begin{equation}
  \mathbf{u}_{\mathbf{q}}^\dagger = - i \sum_m v_m
  (\mathbf{g}^\text{b}_m)^\text{T} W_{\mathbf{q}}^{-1}
  \delta_{\mathbf{q},\mathbf{G}^\text{b}_m} \text{~~~~~and~~~~~}
  \mathbf{u}_\mathbf{q}=i\sum_m v_m^\ast W_\mathbf{q}^{-1}
  \mathbf{g}^\text{b}_m\delta_{\mathbf{q}, \mathbf{G}^\text{b}_m},
\end{equation}
so that $\mathbf{u}_\mathbf{q}$ and $\mathbf{u}_\mathbf{q}^\dagger$
are an adjoint pair.  This gives us \eq\ref{sup_eq:analytic_soln}.

\section{Brute-Force Minimization of the Total Energy within the Continuum
Model}

We wrote a program to evaluate the total energy per unit cell within
the continuum model, \eq\ref{sup_eq:adh-full}, by brute-force
summation over 1L-G lattice sites, with the Fourier components of the
displacement field determined by numerical minimization of the total
energy using the conjugate-gradients method.\cite{Hestenes1952}
Initial real and imaginary parts of the Fourier components of the
displacement field in the $x$- and $y$-directions are chosen randomly
from a uniform distribution. The condition
$\mathbf{u}(-\mathbf{q})=\mathbf{u}^*(\mathbf{q})$ is imposed to
ensure that the displacement field is real, but no other constraints
are imposed on the Fourier coefficients.

For aligned 1L-G and B-hBN lattice vectors, the total energy was
evaluated by summing over $N=173889$ 1L-G unit cells, corresponding to
a $7\times 7$ array of moir\'{e} supercells. The Fourier expansion of
the displacement field included moir\'{e} supercell reciprocal lattice
vectors up to a magnitude of $1$ {\AA}$^{-1}$. This ensures that we
can describe features on the scale of a single 1L-hBN unit cell
(\textit{i.e.}, the limit of validity of the continuum model). The
procedure was repeated for 240 different initial
$\mathbf{u}_{\mathbf{G}^\text{b}_m}$, randomly chosen with magnitudes
up to $10$ {\AA}\@.

\section{Modelling Moir\'e SL Minibands for Electrons in B-hBN/1L-G/B-hBN}

For 1L-G encapsulated between two B-hBN
crystals,\cite{Wang2019,Wang2019a,finney_tunable_2019,Yang_2020,Andelkovic_2020,Sun_2021,Moulsdale_2022,Hu_2023}
B-hBN/1L-G/B-hBN, the displacement field $\mathbf{u}^\text{t/b}=\sum_m
\mathbf{u}_{\mathbf{G}^\text{t/b}_m}\,e^{-i\mathbf{G}^\text{t/b}_m\cdot\mathbf{r}}$
can be approximated as a sum of the displacements arising from the top
(t) B-hBN at twist angle $\theta^\text{t}$ and bottom (b) B-hBN at
twist angle $\theta^\text{b}$, each described using
\eq\ref{sup_eq:analytic_soln}. Such deformations lead to a complex
moir\'e SL pattern experienced by the electrons in the 1L-G, which
differs for B-hBN layers with parallel and antiparallel orientations
of their unit cells (the latter case has
$\theta^\text{t}=\theta^b+180^{\circ}$).

A highly aligned 1L-G/B-hBN interface produces a perturbation $\delta
\mathcal{H}$ to Dirac electrons, with Hamiltonian
$\mathcal{H}=v_\text{F} \mathbf{p}\cdot{\bm
  \sigma}+\delta\mathcal{H}$,\cite{Wallbank_2013} where
$v_\text{F}=10^6$ m\,s$^{-1}$ is the 1L-G Fermi
velocity,\cite{Novoselov2005} $\mathbf{p}$ is the momentum operator
and ${\bm \sigma}$ is the vector of Pauli matrices. In a
B-hBN/1L-G/B-hBN system, the perturbations from the top (t) and bottom
(b) interfaces superimpose and we obtain the combined
perturbation,\cite{Cosma_2014}
\begin{equation}
\delta\mathcal{H}=\delta\mathcal{H}^\text{t}+\delta\mathcal{H}^\text{b}=
\sum_{\alpha=\text{t},\text{b}}\sum_{n=1}^6\left[ U^{\alpha}_0+(-1)^n
  \left( i U^{\alpha}_3\sigma_3 +U^{\alpha}_1 \mathbf{e}_n\cdot {\bm
    \sigma} \right) \right] e^{i
  \mathbf{G}^{\alpha}_n\cdot\left(\mathbf{r}\mp \mathbf{R}/2\right)
}\, e^{i \mathbf{g}_n\cdot
  \mathbf{u}_\mathbf{R}(\mathbf{r})}, \label{sup_eq:SLHam}
\end{equation}
where $\mathbf{e}_n$ are unit vectors in the directions of the 1L-G
lattice points in the first star in real space, $\mathbf{R}$ is the
reference shift between the top and bottom B-hBN crystals, and
$\mathbf{u}_\mathbf{R}(\mathbf{r})=\mathbf{u}^\text{t}(\mathbf{r}-\mathbf{R}/2)
+\mathbf{u}^\text{b}(\mathbf{r}+\mathbf{R}/2)$. The parameters take
into account the 1L-G displacement caused by the top and bottom B-hBN
crystals.  The parameters
$U_0^\text{t/b}=16.4\frac{4\pi\sqrt{\delta^2+(\theta^\text{t/b})^2}}
{\sqrt{3} a_\text{G}}$ eV\,{\AA},
$U_1^\text{t/b}=-32.8\frac{4\pi\delta} {\sqrt{3}a_\text{G}}$ eV\,{\AA},
and
$U_3^\text{t/b}=-32.8\frac{2\pi\sqrt{\delta^2+(\theta^\text{t/b})^2}}
{a_\text{G}}$ eV\,{\AA} are taken from the earlier moir\'e SL model in
aligned 1L-G/B-hBN LMHs, and $\{\mathbf{G}_n^{t/b}\}$ are the first
star of reciprocal lattice vectors of the SLs corresponding to the
top/bottom interfaces.

We consider two mechanisms that give rise to supermoir\'{e}
perturbations to the electronic structure.  Firstly, the interference
between the top and bottom moir\'{e} patterns leads to electrons
scattering off both lattices with respective reciprocal-lattice
vectors $\mathbf{G}^\text{t/b}_m$. This can be described within
second-order perturbation theory in
$\delta\mathcal{H}^\text{t/b}(\mathbf{u}=\mathbf{0})$ as scattering
with the combined reciprocal-lattice vectors
$\mathbf{K}_{nm}=\mathbf{G}^\text{t}_{n}-\mathbf{G}^\text{b}_{m
}$. Secondly, mixing between top and bottom moir\'{e} patterns occurs
due to 1L-G lattice relaxation caused by the combined influences of
the top and bottom B-hBN\@. Using the Fourier representation of the
displacement field, we perform an expansion in \eq\ref{sup_eq:SLHam}
for small displacement fields (\eq\ref{sup_eq:small_field}). The
resulting expression features different pairs of the moir\'{e}
reciprocal-lattice vectors $\mathbf{G}^\text{t/b}_m$, which combine to
give the effective reciprocal-lattice vectors $\mathbf{K}_{nm}$, each
describing one of the emergent periodic supermoir\'{e} structures,
\textit{i.e.}\@ different beats between moir\'{e} patterns at the top
and bottom interfaces.

\begin{figure}[!htbp]
\centerline{\includegraphics[width=110mm]{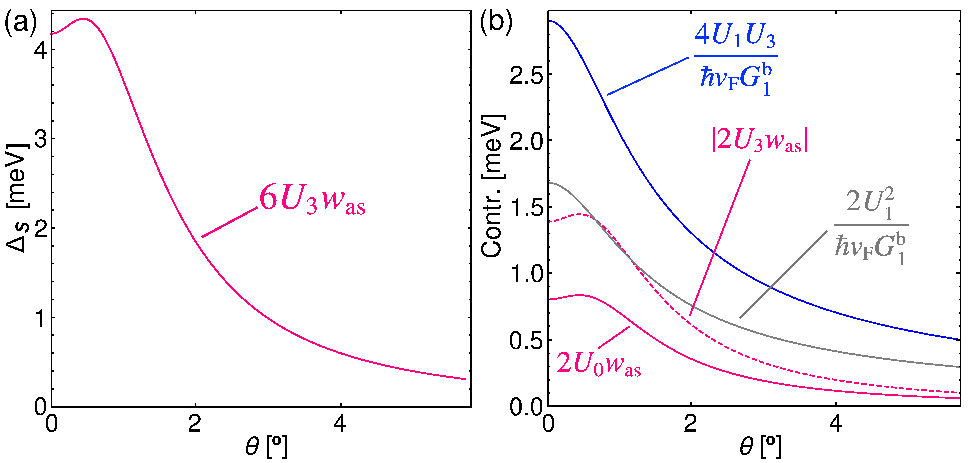}}
\caption{(a) Strain-induced energy gap $\Delta_\text{S}$ of a single
  interface between 1L-G and B-hBN in B-hBN/1L-G/B-hBN LMH (b)
  Comparison of the magnitudes of the different energy contributions
  in \eqs\ref{sup_eq:dnnSLHAMal} and
  \ref{sup_eq:dnnSLHAMrev}.\protect\label{sup_fig:GapSI}}
\end{figure}

If one of the top or bottom B-hBN layers is strongly misaligned with
respect to the 1L-G, while the other is almost aligned
(\textit{e.g.}\@ $\theta^\text{t}\gg\theta^\text{b}\sim0$), all the
supermoir\'{e} reciprocal-lattice vectors are large, so that there is
no effect of the supermoir\'{e} periodic structure at low
energies. However, strain from the near-aligned layer opens a gap:
\begin{equation}
\Delta_\text{S} (\theta^\text{b})=-6 U_3w_\text{as},
\end{equation}
where $w_\text{as}^\text{t/b}=-(-1)^m\text{Re}(\mathbf{g}_m\cdot
\mathbf{u}_{\pm\mathbf{G}^\text{t/b}_m})$. The gap is plotted against
misalignment angle in \fig\ref{sup_fig:GapSI}a.

For any supermoir\'{e} reciprocal-lattice vector $\mathbf{K}_{nm}$ the
real-space period of the corresponding supermoir\'{e} pattern is
$\Lambda_{nm}=\frac{4\pi}{\sqrt{3}}\frac{1}{K_{nm}}$. In the following
we focus on the longest-period effects, and hence keep only the
shortest nonzero $\mathbf{K}_{nm}$. As we illustrate in
\fig\ref{sup_fig:SLLattVecL}d, for certain misalignment angles of the
top and bottom B-hBN layers in B-hBN/1L-G/B-hBN, combinations of the
form $\mathbf{K}_{nm}=\mathbf{G}^\text{t}_{n}-\mathbf{G}^\text{b}_{m
}$ become very small and can even vanish completely. Hence these
combinations yield the shortest reciprocal-lattice vectors at those
misalignment angles. The most generic conditions for long-period beats
are for $m=n$, for which the period of the supermoir\'{e} lattice
diverges for any stacking configuration with
$\theta^\text{t}=\theta^\text{b}$. We now discuss the corresponding
low-energy Hamiltonians and properties of this configuration for the
relevant low-energy regime with
$\theta^\text{t}\approx\theta^\text{b}$. Since 1L-hBN does not
preserve inversion symmetry, the configurations in which the top and
bottom B-hBN layers are parallel and antiparallel yield distinct
cases.

The perturbative term in the electronic 1L-G Hamiltonian for the case
in which the top and bottom B-hBN are aligned
($\theta^\text{t}=\theta^\text{b}$) is
\begin{align}
\delta\mathcal{H}_\text{par} =& -12 {U}_3 w_\text{as}\sigma_3
\nonumber \\ & {} -\sum_{n=1}^6 \left[ 2 {U}_3 w_\text{as} \sigma_3+
  \frac{4U_1U_3} {\hbar v_\text{F}G^\text{b}_{n}}+i\frac{2
    U_1^2}{\hbar v_\text{F}G^\text{b}_{n}}\frac{(\mathbf{e}_z\times
    \mathbf{K}_{nn })\cdot {\bm \sigma}}{K_{nn}} \right]
e^{-\frac{i}{2}{\mathbf{R}}\cdot
  (\mathbf{G}^\text{t}_{n}+\mathbf{G}^\text{b}_{n })}\;
e^{i\mathbf{K}_{nn} \cdot \mathbf{r}},
\label{sup_eq:dnnSLHAMal}
\end{align}
where $\mathbf{e}_z$ is the unit vector in the $+z$-direction.  As the
inversion-symmetry breaking is enhanced by the two aligned B-hBN
acting jointly, the gap is twice as big as that induced by a single
interface, giving the first term in \eq\ref{sup_eq:dnnSLHAMal}. The
effect of both interfaces combined oscillates with the periodicity of
the supermoir\'{e} lattice, giving the second term in
\eq\ref{sup_eq:dnnSLHAMal}.

\begin{figure}[!htbp]
\centerline{\includegraphics[width=120mm]{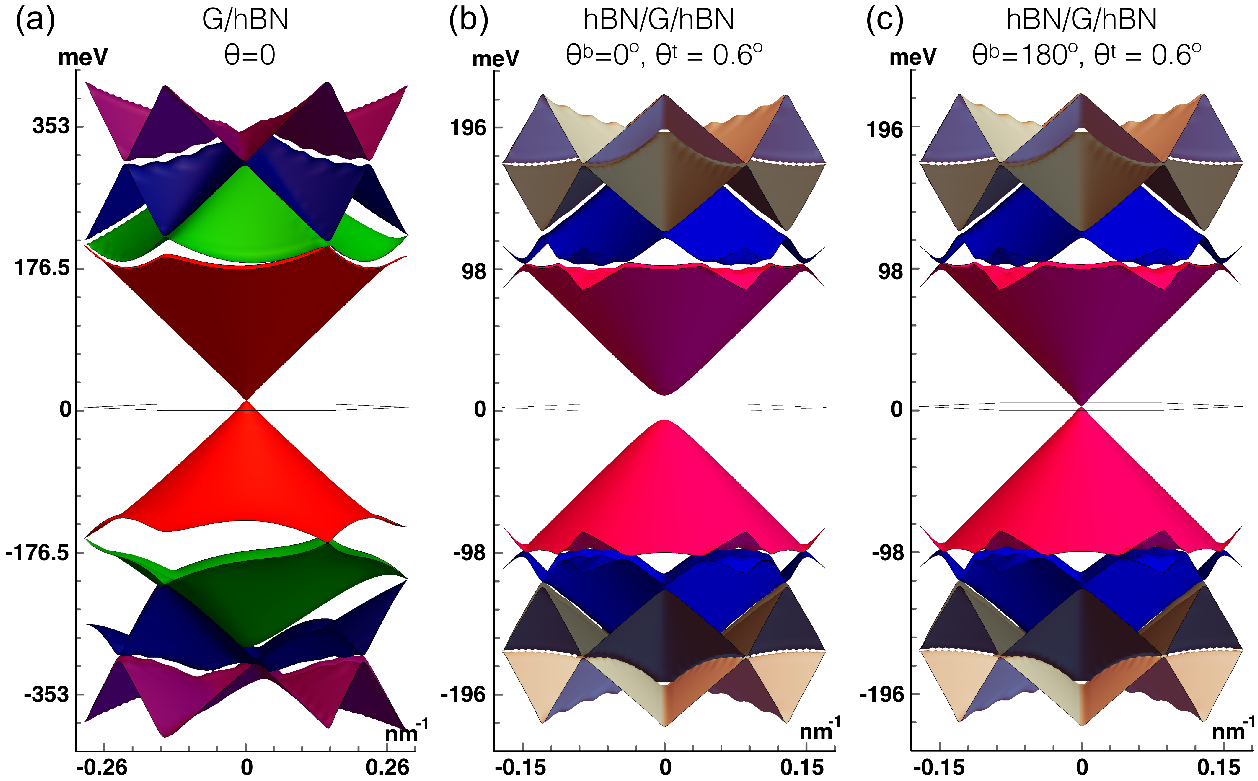}}
\caption{Band structures in the first supermoir\'{e} mini BZ for: (a)
  a single 1L-G/B-hBN interface, (b) $H_{mm}$ parallel alignment, and
  (c) antiparallel alignment. The dispersion is plotted in the
  hexagonal BZ\@. \protect\label{sup_fig:BSboth}}
\end{figure}

The perturbative term in the Hamiltonian for
$\theta^\text{t}=\theta^\text{b}+180^{\circ}$ is
\begin{equation}
\delta\mathcal{H}_\text{antipar}= \sum_{n=1}^6 \left[i(-1)^m 2{U}_0
  w_\text{as} - \frac{4U_1U_3 } {\hbar
    v_\text{F}G^\text{b}_{n}}-i\frac{2 U_1^2}{\hbar
    v_\text{F}G^\text{b}_{n}}\frac{(\mathbf{e}_z\times\mathbf{K}_{nn})
    \cdot{\bm\sigma}}{K_{nn}} \right] e^{-\frac{i}{2}{\mathbf{R}}
  \cdot(\mathbf{G}^\text{t}_{n}+\mathbf{G}^\text{b}_{n})}\;e^{i\mathbf{K}_{nn}
  \cdot \mathbf{r}}.
\label{sup_eq:dnnSLHAMrev}
\end{equation}
In this case, the large gap (\fig\ref{sup_fig:BSboth}b) is canceled by
reversing one of the top or bottom B-hBN layers against the other as
shown in \fig\ref{sup_fig:BSboth}c. Thus, only a small gap due to the
inversion-symmetry-breaking components is present. The magnitudes of
the different terms are compared in
\fig\ref{sup_fig:GapSI}b. \fig\ref{sup_fig:BSboth} shows examples of
band structures corresponding to the supermoir\'{e} Hamiltonians of
\eqs\ref{sup_eq:dnnSLHAMal} and \ref{sup_eq:dnnSLHAMrev}.

\section{Band-Structure Reconstruction of B-hBN/1L-G/B-hBN: \\
Derivation of the Hamiltonian of Shortest Period}

We take into account the combination of the top and bottom
single-interface moir\'{e} patterns via two mechanisms: (i)
quantum-mechanical interference; (ii) lattice reconstruction, where
the total 1L-G strain field is the sum of the strain fields from each
interface.

Within second-order perturbation theory, the Hamiltonian for the
middle SL generated by the shortest effective reciprocal-lattice
vectors
$\mathbf{K}_{n,m}=\mathbf{G}_{n}^\text{t}-\mathbf{G}_{m}^\text{b}$
that combine due to interference is $\Hh_{n,m}^\text{int} =
\delta\Hh^{(2)}_\text{tb} + \delta\Hh^{(2)}_\text{bt}$, with
\begin{footnotesize}
\begin{align}
\nonumber& \delta\Hh^{(2)}_{\substack{\text{tb}\\(\text{bt})}} =
\sum_{n,m}\frac{1}{\hbar v_\text{F} (G^\text{b}_m)^2}\Biggl( \bigg[
  U_0 U_1 (-1)^{m} \mathbf{G}_{m}^{\text{b(t)}} \cdot
  \mathbf{e}_{m}+U_0 U_1 (-1)^{n } \mathbf{G}_{m}^\text{b(t)} \cdot
  \mathbf{e}_{n }\\ \nonumber &+ U_1U_3 (-1)^{n+m}
  (\mathbf{e}_z\times\mathbf{G}_{m}^\text{b(t)}) \cdot
  (\mathbf{e}_{m}+\mathbf{e}_{n}) \bigg]e^{\mp
  i\frac{\mathbf{R}}{2}(\mathbf{G}_{m}^\text{b}+\mathbf{G}_{n}^\text{t})}
\; e^{i(\mathbf{G}_{m}^\text{b}-\mathbf{G}_{n}^\text{t}) \cdot
  \mathbf{r}}\\ \nonumber &+ i \bigg[ (-1)^{m} U_1 U_0
  (\mathbf{e}_z\times\mathbf{G}_{m}^\text{b(t)}) \cdot \mathbf{e}_{m}
  - (-1)^{n } U_1 U_0 (\mathbf{e}_z\times\mathbf{G}_{m}^\text{b(t)})
  \cdot \mathbf{e}_{n }\\ \nonumber &-   U_1 U_3 (-1)^{n+m}
  \mathbf{G}_{m}^\text{b(t)} \cdot (\mathbf{e}_{m}-\mathbf{e}_{n})
  \bigg] \sigma_3 \, e^{\mp
  i\frac{\mathbf{R}}{2}(\mathbf{G}_{m}^\text{b}+\mathbf{G}_{n}^\text{t})}
\;e^{i(\mathbf{G}_{m}^\text{b}-\mathbf{G}_{n}^\text{t}) \cdot
  \mathbf{r}}\\ \nonumber &+ \bigg[  i (-1)^{n+m} U_1^2 \big[
    (\mathbf{e}_{n^{ }}\cdot \mathbf{G}_{m}^\text{b(t)})
    (\mathbf{e}_{m} \cdot {\bm \sigma}) + [\mathbf{e}_{n} \cdot
      (\mathbf{e}_z\times\mathbf{G}_{m}^\text{b(t)})]
    [(\mathbf{e}_z\times\mathbf{e}_{m}) \cdot {\bm \sigma}] \big]
  \bigg]e^{\mp i\frac{\mathbf{R}}{2}\cdot
  (\mathbf{G}_{m}^\text{b}+\mathbf{G}_{n}^\text{t})}
\;e^{i(\mathbf{G}_{m}^\text{b}-\mathbf{G}_{n}^\text{t})\cdot
  \mathbf{r}}\\ &+\bigg[U_0^2 \mathbf{G}_{m}^\text{b(t)} -
  (-1)^{n+m} U_3^2 \mathbf{G}_{m}^\text{b(t)} +  U_0 U_3 \big(
  (-1)^{m} + (-1)^{n} \big)
  (\mathbf{e}_z\times\mathbf{G}_{m}^\text{b(t)}) \bigg] \cdot {\bm
  \sigma} e^{\mp i\frac{\mathbf{R}}{2} \cdot
  (\mathbf{G}_{m}^\text{b}+\mathbf{G}_{n}^\text{t})}
\;e^{i(\mathbf{G}_{m}^\text{b}-\mathbf{G}_{n}^\text{t})\cdot
  \mathbf{r}} \Biggr).
\end{align}
\end{footnotesize}

The terms in the SL Hamiltonian in which combinations
$\mathbf{K}_{n,m}=\mathbf{G}_{n}^\text{t}-\mathbf{G}_{m}^\text{b}$
arise due to lattice reconstruction can be written as
\begin{footnotesize}
\begin{align}
\nonumber\delta\Hh^\text{rec}_{n,m}=&\sum_{n,m} \Bigg( i
\Big[U_0^\text{b}\mathbf{g}_n \cdot
  \mathbf{u}_{\mathbf{G}_m^\text{t}}-U_0^\text{t}\mathbf{g}_n \cdot
  \mathbf{u}_{-\mathbf{G}_m^\text{b}}\Big] + i\sigma_3 \Big[ (-1)^n i
  U_3^\text{b}\mathbf{g}_n \cdot \mathbf{u}_{\mathbf{G}_m^\text{t}} +
  (-1)^m i U_3^\text{t}\mathbf{g}_n \cdot
  \mathbf{u}_{-\mathbf{G}_m^\text{b}}\Big] \\ &+ \Big[ (-1)^n i
  U_1^\text{b} \left( \mathbf{e}_n \cdot {\bm \sigma}\right)
  (\mathbf{g}_n \cdot \mathbf{u}_{\mathbf{G}_m^\text{t}}) - (-1)^m i
  U_1^\text{t} \left( \mathbf{e}_m \cdot {\bm \sigma}
  \right)(\mathbf{g}_n \cdot \mathbf{u}_{-\mathbf{G}_m^\text{b}})\Big]
\Bigg)\; e^{i
  (\mathbf{G}_{n}^\text{t}+\mathbf{G}_{m}^\text{b})\cdot\frac{\mathbf{R}}{2}
} \; e^{i \mathbf{K}_{nm}\cdot\mathbf{r}}.
\label{sup_eq:FulldnnHamSM}
\end{align}
\end{footnotesize}
For each emergent reciprocal-lattice vector of the supermoir\'{e}
pattern, the corresponding term in the Hamiltonian of
\eq\ref{sup_eq:FulldnnHamSM} is of equivalent form to
\eq\ref{sup_eq:SLHam} with $\mathbf{u}=\mathbf{0}$, where the
reciprocal-lattice vectors are replaced by those of the supermoir\'{e}
lattice, and the coefficients are rescaled:
\begin{align}
\nonumber&\delta\Hh^\text{rec}_{n,m}= C_{\sigma_0} + \Delta \sigma_3
+\mathbf{A}\cdot{\bm \sigma}\\ \nonumber &+a^{ (0)}_{nm} f^{(n,m)}_1
(\mathbf{r}) +a^{ (3)}_{nm}  f^{(n,m)}_2 (\mathbf{r}) \sigma_3
+a^{(12)}_{nm} [\mathbf{e}_z\times \nabla f^{(n,m)}_2 (\mathbf{r})
]\cdot {\bm \sigma} \\ &+b^{ (0)}_{nm} f^{(n,m)}_2 (\mathbf{r})  +b^{
  (3)}_{nm}  f^{(n,m)}_2 (\mathbf{r})  f^{(n,m)}_1 (\mathbf{r})
\sigma_3+ b^{(12)}_{nm} f^{(n,m)}_2 (\mathbf{r}) [\mathbf{e}_z\times
  \nabla f^{(n,m)}_1 (\mathbf{r}) ]\cdot {\bm \sigma} ,
\label{sup_eq:SLHamModel}
\end{align}
where $f^{(n,m)}_1(\mathbf{r})=\sum_m e^{i
  (\mathbf{G}^\text{b}_{n}+\mathbf{G}^\text{t}_{m})\cdot\mathbf{R}/2}
e^{i\mathbf{K}_{n^{ },m}\cdot\mathbf{r}}$ and
$f^{(n,m)}_2(\mathbf{r})=i\sum_m (-1)^m e^{i
  (\mathbf{G}^\text{b}_{n}+\mathbf{G}^\text{t}_{m})\cdot \mathbf{R}/2}
e^{i\mathbf{K}_{n^{ },m}\cdot\mathbf{r}}$ encode the periodicity of
the supermoir\'{e} potential. The first line in
\eq\ref{sup_eq:SLHamModel} preserves inversion symmetry, whereas the
second line does not. The coupling constants $a^{(i)}_{nm}$ and
$b^{(i)}_{nm}$ are given in \tabs\ref{sup_tab:hieroglyphics_aligned}
and \ref{sup_tab:hieroglyphics_antialigned}. The constants in
\eq\ref{sup_eq:SLHamModel} are:
\begin{align}
g^{n,n+1}_{\text{al}, \perp} &=-\frac{a_\text{G}\left(2 \sqrt{3}
  {\delta}-3 {\theta^\text{t}}+3 {\theta^{ \text{b}}}\right)}{8 \pi
  \left({\delta}^2+\sqrt{3} {\delta} ({\theta^{ \text{b}}}-{\theta^{
      \text{t}}})+{\theta^{ \text{b}}}^2-{\theta^{ \text{b}}}
       {\theta^{ \text{t}}}+{\theta^{ \text{t}}}^2\right)}
,\\ g^{n,n+1}_{\text{rev}, \perp} &= \frac{3a_\text{G} ({\theta^{
      \text{b}}}+{\theta^{ \text{t}}})}{8 \pi
  \left({\delta}^2+\sqrt{3} {\delta} ({\theta^{ \text{b}}}-{\theta^{
      \text{t}}})+{\theta^{ \text{b}}}^2-{\theta^{ \text{b}}}
       {\theta^{ \text{t}}}+{\theta^{ \text{t}}}^2\right)}
\\ g^{n,n+2}_{\text{al},\text{s}, \perp} &=
g^{n,n+2}_{\text{rev},\text{as}, \perp}= \frac{3a_\text{G} \left(
  \sqrt{3} {\delta}+ {\theta^\text{t}}+2 {\theta^{
      \text{b}}}\right)}{8 \pi \left(3{\delta}^2+\sqrt{3} {\delta}
  ({\theta^{ \text{b}}}-{\theta^{ \text{t}}})+{\theta^{
      \text{b}}}^2+{\theta^{ \text{b}}} {\theta^{ \text{t}}}+{\theta^{
      \text{t}}}^2\right)}\\ g^{n,n+2}_{\text{al},\text{as}, \perp}
&=g^{n,n+2}_{\text{rev},\text{s}, \perp} =-\frac{3a_\text{G} \left(2
  \sqrt{3} {\delta}+ {\theta^\text{b}}- {\theta^{ \text{t}}}\right)}{8
  \pi \left(3{\delta}^2+\sqrt{3} {\delta} ({\theta^{
      \text{b}}}-{\theta^{ \text{t}}})+{\theta^{
      \text{b}}}^2+{\theta^{ \text{b}}} {\theta^{ \text{t}}}+{\theta^{
      \text{t}}}^2\right)} \\ g^{n,n+3}_{\perp}
&=-\frac{\sqrt{3}a_\text{G}}{2\pi}
\frac{\delta}{4\delta^2+(\theta^\text{b}+\theta^\text{t})^2}\\ g^{n,n+4}_{\text{al},\text{s},
  \perp} & = g^{n,n+4}_{\text{rev},\text{as}, \perp} =- \frac{3
  a_\text{G} ({\theta^{ \text{b}}}+{\theta^{ \text{t}}})}{8 \pi
  \left(3{\delta}^2+\sqrt{3} {\delta} ({\theta^{ \text{t}}}-{\theta^{
      \text{b}}})+{\theta^{ \text{b}}}^2+{\theta^{ \text{b}}}
       {\theta^{ \text{t}}}+{\theta^{
           \text{t}}}^2\right)}\\ g^{n,n+4}_{\text{al},\text{as},
  \perp} & =g^{n,n+4}_{\text{rev},\text{s}, \perp} -\frac{3a_\text{G}
  \left(2 \sqrt{3} {\delta}+ {\theta^\text{t}}- {\theta^{
      \text{b}}}\right)}{8 \pi \left(3{\delta}^2+\sqrt{3} {\delta}
  ({\theta^{ \text{t}}}-{\theta^{ \text{b}}})+{\theta^{
      \text{b}}}^2+{\theta^{ \text{b}}} {\theta^{ \text{t}}}+{\theta^{
      \text{t}}}^2\right)}\\ g^{n,n+5}_{\text{al}, \perp}
&=-\frac{a_\text{G}\left(2 \sqrt{3} {\delta}-3 {\theta^\text{b}}+3
  {\theta^{ \text{t}}}\right)}{8 \pi  \left({\delta}^2+\sqrt{3}
  {\delta} ({\theta^{ \text{t}}}-{\theta^{ \text{b}}})+{\theta^{
      \text{b}}}^2-{\theta^{ \text{b}}} {\theta^{ \text{t}}}+{\theta^{
      \text{t}}}^2\right)} ,\\ g^{n,n+5}_{\text{rev}, \perp} &=
\frac{3a_\text{G} ({\theta^{ \text{b}}}+{\theta^{ \text{t}}})}{8 \pi
  \left({\delta}^2+\sqrt{3} {\delta} ({\theta^{ \text{t}}}-{\theta^{
      \text{b}}})+{\theta^{ \text{b}}}^2-{\theta^{ \text{b}}}
       {\theta^{ \text{t}}}+{\theta^{ \text{t}}}^2\right)}  .
\end{align}

\begin{table}[!htbp]
\centering
\caption{Coupling constants for B-hBN/1l-G/B-hBN with aligned B-hBN
  lattice vectors [$C_{\sigma_0}=-6 (U_0^\text{t}w^\text{t}_\text{s}+
    U_0^\text{b} w^\text{b}_\text{s}) $, $\Delta= -6(U_3^\text{t}
    w^\text{t}_\text{as}+ U_3^\text{b} w^\text{b}_\text{as}) $, $A_1
    =0$, and
    $A_2=0$]. $w_\text{as}^\text{t/b}=-(-1)^m\text{Re}(\mathbf{g}_m\cdot\mathbf{u}_{\pm\mathbf{G}^\text{t/b}_m})$
  and
  $w_\text{s}^\text{t/b}=\text{Im}(\mathbf{g}_m\cdot\mathbf{u}_{\pm\mathbf{G}^\text{t/b}_m})$
  \protect\label{sup_tab:hieroglyphics_aligned}}
\resizebox{1.\linewidth}{!}{$\displaystyle
\begin{array}{|c||c|c|c|c|c|c|}
    &a^{ (0)}_{nm}  & a^{ (3)}_{nm} &a^{(12)}_{nm} &b^{ (0)}_{nm}  &
  b^{ (3)}_{nm} &b^{(12)}_{nm} \\ \hline n,n & -
  (U_0^\text{b}w^\text{t}_\text{s}+ U_0^\text{t} w^\text{b}_\text{s})
  &  -(U^\text{b}_3 w^\text{t}_\text{s}-
  U^\text{t}_3w^\text{b}_\text{s})&0 &  (U_0^\text{b}
  w^\text{t}_\text{as}-U_0^\text{t} w^\text{b}_\text{as}) &
  -(U^\text{b}_3w^\text{t}_\text{as}+U^\text{t}_3
  w^\text{b}_\text{as})&0\\ n,n+1 & -\frac{1}{2}(U_0^\text{b}
  w^\text{t}_\text{s}+U_0^\text{t} w^\text{b}_\text{s}) &
  \frac{1}{2}(U_3^\text{b} w^\text{t}_\text{s}-U_3^\text{t}
  w^\text{b}_\text{s})&  -\frac{U_1}{2}w_\text{s}
  g^{n,n+1}_{\text{al},\perp} & \frac{1}{2}(U_0^\text{b}
  w^\text{t}_\text{as}+U_0^\text{t}
  w^\text{b}_\text{as})&-\frac{1}{2}(U_3^\text{b}
  w^\text{t}_\text{as}+U_3^\text{t} w^\text{b}_\text{as}) &
  \frac{U_1}{2}w_\text{as} g^{n,n+1}_{\text{al},\perp} \\ n,n+2 &
  -\frac{1}{2}(U_0^\text{b} w^\text{t}_\text{s}+U_0^\text{t}
  w^\text{b}_\text{s})  & -\frac{1}{2}(U_3^\text{b}
  w^\text{t}_\text{s}-U_3^\text{t} w^\text{b}_\text{s}) &-
  \frac{U_1}{2}w_{s} g^{n,n+2}_{\text{al},\text{s},\perp} &
  \frac{1}{2}(U_0^\text{b} w^\text{t}_\text{as}-U_0^\text{t}
  w^\text{b}_\text{as}) &-
  \frac{U_3}{2}(w^\text{t}_\text{as}+w^\text{b}_\text{as}) &
  \frac{U_1}{2}w_\text{as} g^{n,n+2}_{\text{al},\text{as},\perp}
  \\ n,n+3 & (U_0^\text{b} w^\text{t}_\text{s}+ U_0^\text{t}
  w^\text{b}_\text{s})  & -(U_3^\text{b}
  w^\text{t}_\text{s}+U_3^\text{t} w^\text{b}_\text{s})  &
  -(U_1^\text{b} w^\text{t}_\text{s}+U_1^\text{t}
  w^\text{b}_\text{s})g^{n,n+3}_{\perp} & -(U_0^\text{b}
  w^\text{t}_\text{s}+U_0^\text{t} w^\text{b}_\text{s}) &
  -(U_3^\text{b} w^\text{t}_\text{as}+ U_3^\text{t}
  w^\text{b}_\text{as}) & (U_1^\text{b}
  w^\text{t}_\text{as}+U_1^\text{t}
  w^\text{b}_\text{as})g^{n,n+3}_{\perp}    \\ n,n+4 &
  -\frac{1}{2}(U_0^\text{b} w^\text{t}_\text{s}+U_0^\text{t}
  w^\text{b}_\text{s})   &   -\frac{1}{2}(U_3^\text{b}
  w^\text{t}_\text{s}-U_3^\text{t} w^\text{b}_\text{s})&
  -\frac{U_1}{2}w_\text{s} g^{n,n+4}_{\text{al},\perp} &
  \frac{1}{2}(U_0^\text{b} w^\text{t}_\text{as}-U_0^\text{t}
  w^\text{b}_\text{as}) & -\frac{1}{2}(U_3^\text{b}
  w^\text{t}_\text{as}+U_3^\text{t} w^\text{b}_\text{as}) &
  \frac{U_1}{2}w_\text{as} g^{n,n+4}_{\text{al},\text{as},\perp}
  \\ n,n+5 &
  -\frac{1}{2}(U_0^\text{b}w^\text{t}_\text{s}+U_0^\text{t}w^\text{b}_\text{s})
  &  -\frac{1}{2}(U_3^\text{b} w^\text{t}_\text{s}+U_3^\text{t}
  w^\text{b}_\text{s}) & -\frac{U_1}{2}w_\text{s}
  g^{n,n+5}_{\text{al},\perp} &
  -\frac{U_0}{2}(w^\text{t}_\text{as}+w^\text{b}_\text{as}) &
  \frac{1}{2}(U_3^\text{b} w^\text{t}_\text{as}+U_3^\text{t}
  w^\text{b}_\text{as})  & -\frac{U_1}{2}w_\text{as}
  g^{n,n+5}_{\text{al},\perp}   \\
\end{array} $
}
\end{table}

\begin{table}[!htbp]
\caption{Coupling constants for B-hBN/1L-G/B-hBN with anti-aligned
  B-hBN lattice vectors [$C_{\sigma_0}=-6
    (U_0^\text{t}w^\text{t}_\text{s}- U_0^\text{b}
    w^\text{b}_\text{s})$, $\Delta= -6(U_3^\text{t}
    w^\text{t}_\text{as}- U_3^\text{b} w^\text{b}_\text{as})$, $A_1
    =0$, and
    $A_2=0$]. \protect\label{sup_tab:hieroglyphics_antialigned}}
\begin{center}
\resizebox{1.\linewidth}{!}{$\displaystyle
\begin{array}{|c||c|c|c|c|c|c|}
    &a^{ (0)}_{nm}  & a^{ (3)}_{nm} &a^{(12)}_{nm} &b^{ (0)}_{nm}  &
  b^{ (3)}_{nm} &b^{(12)}_{nm} \\ \hline n,n &  -(U_0^\text{b}
  w^\text{t}_\text{s}- U_0^\text{t} w^\text{b}_\text{s}) &
  -(U_3^\text{b} w^\text{t}_\text{s}+U_3^\text{t}
  w^\text{b}_\text{s})&0 &
  U_0(w^\text{t}_\text{as}+w^\text{b}_\text{as}) & -(U_3^\text{b}
  w^\text{t}_\text{as}- U_3^\text{t} w^\text{b}_\text{as})&0\\ n,n+1 &
  -\frac{1}{2}(U_0^\text{b} w^\text{t}_\text{s}-U_0^\text{t}
  w^\text{b}_\text{s}) &  \frac{1}{2}(U_3^\text{b}
  w^\text{t}_\text{s}+U_3^\text{t} w^\text{b}_\text{s})
  &-\frac{U_1}{2}w_\text{s} g^{n,n+1}_{\text{rev},\perp}   &
  \frac{1}{2}(U_0^\text{b} w^\text{t}_\text{as}-U_0^\text{t}
  w^\text{b}_\text{as})& -\frac{1}{2}(U_3^\text{b}
  w^\text{t}_\text{as}-U_3^\text{t} w^\text{b}_\text{as}) &
  \frac{U_1}{2}w_\text{as} g^{n,n+1}_{\text{rev},\perp} \\ n,n+2 &
  -\frac{1}{2}(U_0^\text{b} w^\text{t}_\text{s}-U_0^\text{t}
  w^\text{b}_\text{s}) &   -\frac{1}{2}(U_3^\text{b}
  w^\text{t}_\text{s}+U_3^\text{t} w^\text{b}_\text{s}) &-
  \frac{U_1}{2}w_{s} g^{n,n+2}_{\text{rev},\text{s},\perp}  &
  \frac{1}{2}(U_0^\text{b} w^\text{t}_\text{as}+U_0^\text{t}
  w^\text{b}_\text{as}) &-
  \frac{U_3}{2}(w^\text{t}_\text{as}-w^\text{b}_\text{as}) &
  \frac{U_1}{2}w_\text{as} g^{n,n+2}_{\text{rev},\text{as},\perp}
  \\ n,n+3 &  (U_0^\text{b} w^\text{t}_\text{s}- U_0^\text{t}
  w^\text{b}_\text{s}) & -(U_3^\text{b}
  w^\text{t}_\text{s}-U_3^\text{t} w^\text{b}_\text{s})  &
  -(U_1^\text{b} w^\text{t}_\text{s}-U_1^\text{t}
  w^\text{b}_\text{s})g^{n,n+3}_{\perp}  &  -(U_0^\text{b}
  w^\text{t}_\text{s}-U_0^\text{t} w^\text{b}_\text{s})  &
  -(U_3^\text{b} w^\text{t}_\text{as}- U_3^\text{t}
  w^\text{b}_\text{as}) &  (U_1^\text{b}
  w^\text{t}_\text{as}-U_1^\text{t}
  w^\text{b}_\text{as})g^{n,n+3}_{\perp}  \\ n,n+4 &
  -\frac{1}{2}(U_0^\text{b} w^\text{t}_\text{s}-U_0^\text{t}
  w^\text{b}_\text{s})    &    -\frac{1}{2}(U_3^\text{b}
  w^\text{t}_\text{s}+U_3^\text{t} w^\text{b}_\text{s}) &
  -\frac{U_1}{2}w_\text{s} g^{n,n+4}_{\text{rev},\perp} &
  \frac{1}{2}(U_0^\text{b} w^\text{t}_\text{as}+U_0^\text{t}
  w^\text{b}_\text{as}) & -\frac{1}{2}(U_3^\text{b}
  w^\text{t}_\text{as}-U_3^\text{t} w^\text{b}_\text{as}) &
  \frac{U_1}{2}w_\text{as} g^{n,n+4}_{\text{rev},\text{as},\perp}
  \\ n,n+5 &
  -\frac{1}{2}(U_0^\text{b}w^\text{t}_\text{s}-U_0^\text{t}w^\text{b}_\text{s})
  &  -\frac{1}{2}(U_3^\text{b} w^\text{t}_\text{s}-U_3^\text{t}
  w^\text{b}_\text{s}) & -\frac{U_1}{2}w_\text{s}
  g^{n,n+5}_{\text{rev},\perp} &
  -\frac{U_0}{2}(w^\text{t}_\text{as}-w^\text{b}_\text{as})
  &\frac{1}{2}(U_3^\text{b} w^\text{t}_\text{as}-U_3^\text{t}
  w^\text{b}_\text{as}) &  -\frac{U_1}{2}w_\text{as}
  g^{n,n+5}_{\text{rev},\perp} \\
\end{array}$ }
\end{center}
\end{table}

\section{Time-Step Errors in DMC Calculations}

\begin{figure}[!htbp]
\centerline{\includegraphics[width=90mm]{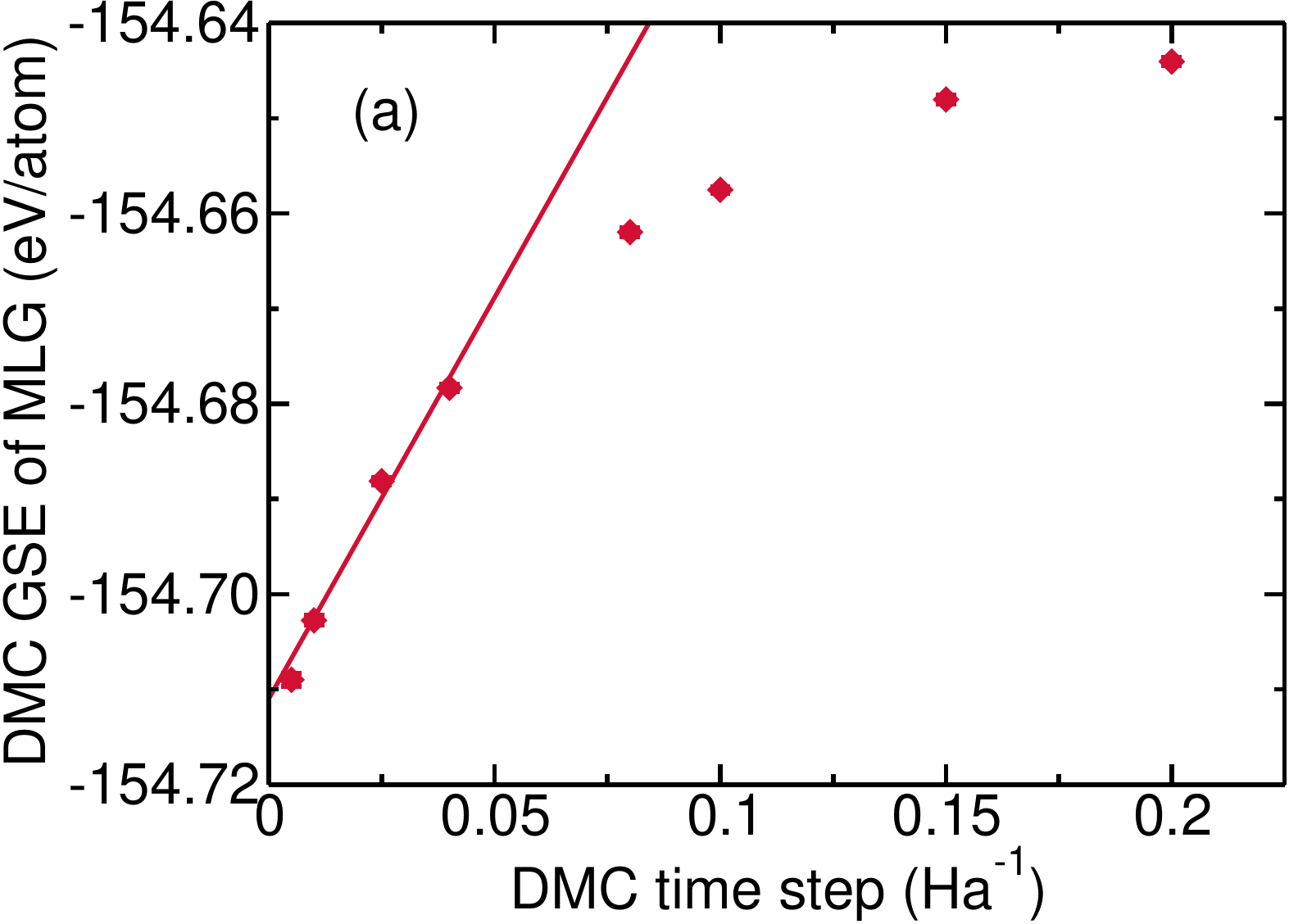}}
\centerline{\includegraphics[width=90mm]{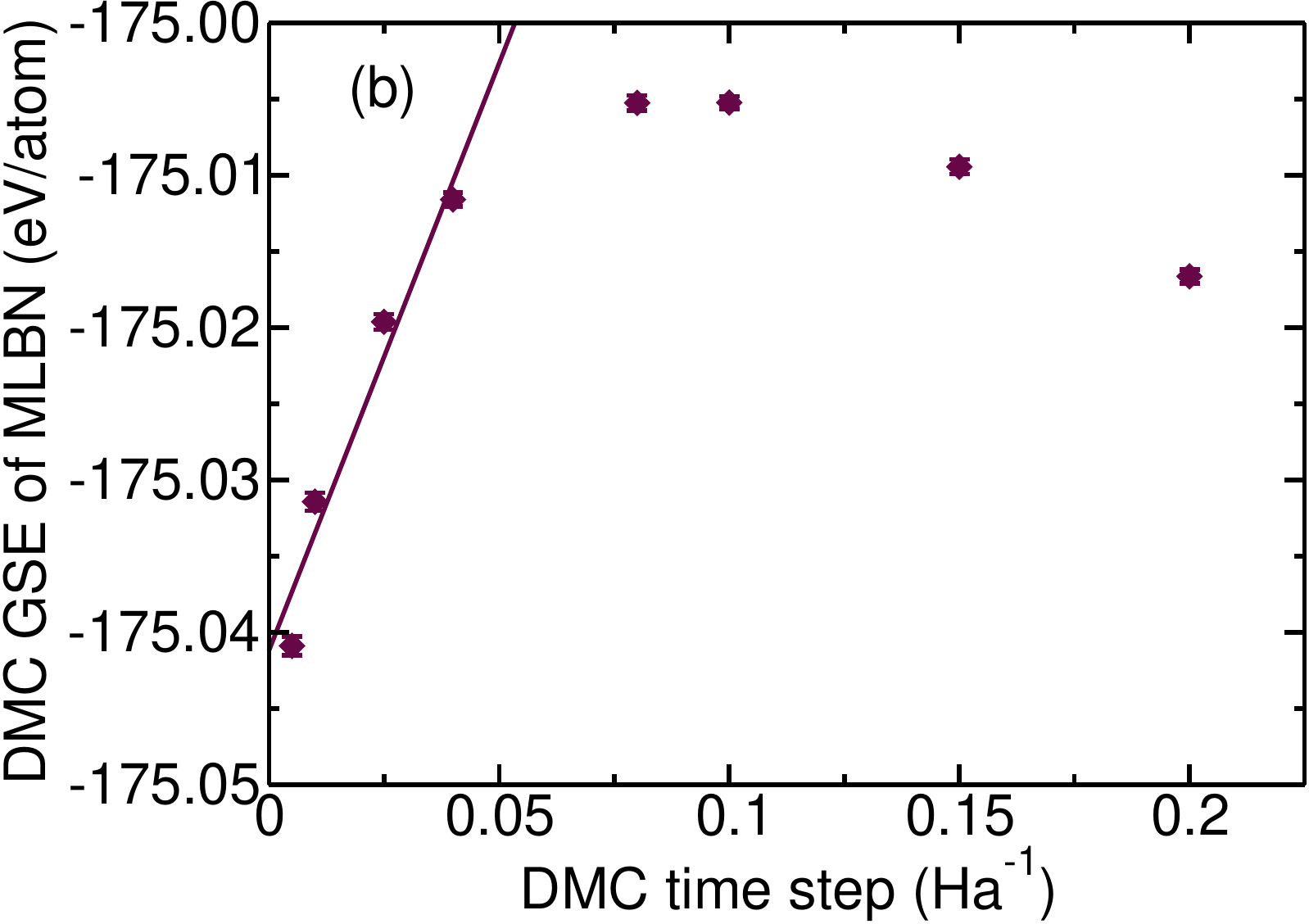}}
\centerline{\includegraphics[width=90mm]{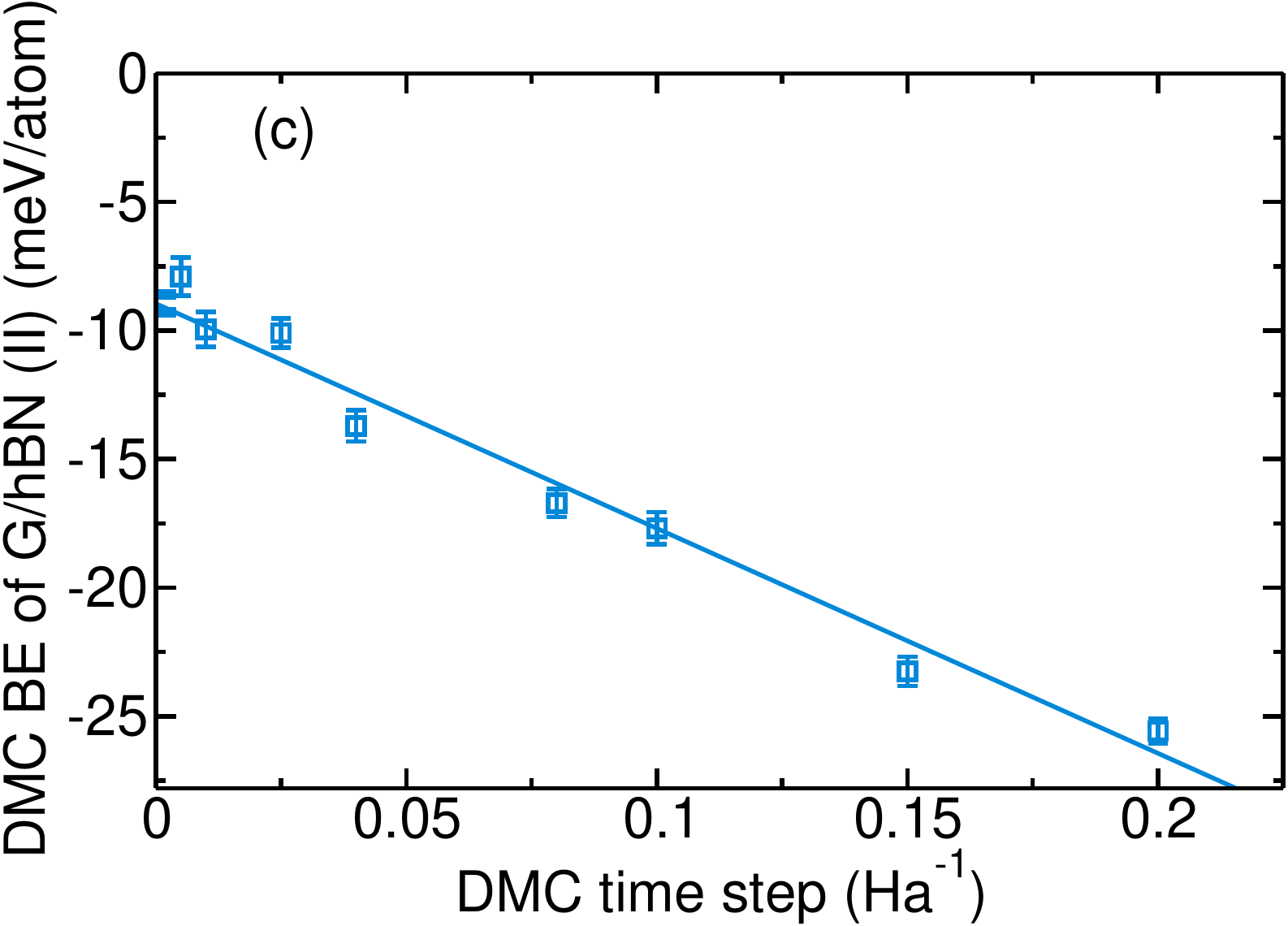}}
\caption{DMC ground-state energy (GSE) of (a) 1L-G and (b) 1L-hBN
  against time step for a supercell consisting of $3\times 3$
  primitive cells. (c) DMC BE for stacking configuration II of
  1L-G/1L-hBN against time step.  These DMC results are not twist
  averaged. \protect\label{sup_fig:dmc_dt_bias}}
\end{figure}

To investigate the finite-time-step errors in our DMC energies, we
calculated the non-twist-averaged ground-state DMC energy of 1L-G and
1L-hBN as well as the BE of 1L-G/1L-hBN (stacking configuration II)
for a supercell composed of $3\times 3$ primitive
cells. \fig\ref{sup_fig:dmc_dt_bias} shows that the time-step errors
in the total energies at a time step of $0.04$ Ha$^{-1}$ are typically
$\sim30$ meV/atom, but these errors partially cancel when the BE is
calculated. The DMC total energy of each 1L varies linearly with time
step, up to $\sim 0.04$ Ha$^{-1}$. The nonlinear contribution to the
time-step bias is primarily due to the atomic cores, and hence largely
cancels out of the BE, which exhibits linear time-step bias up to a
much larger time step of $\sim 0.2$ Ha$^{-1}$. We therefore
extrapolate all our final DMC BEs to zero time step using time steps
of $0.04$ and $0.1$ Ha$^{-1}$.

\bibliography{G_BN}